\documentclass[12pt]{article}
\let\ssection=\section
\renewcommand{\section}{\setcounter{equation}{0}\ssection}

\newcommand\mathC{\mkern1mu\raise2.2pt\hbox{$\scriptscriptstyle|$}
        {\mkern-7mu\rm C}} 
\newcommand{\mathR}{{\rm I\! R}}         

\newcommand\bi{\begin{itemize}}
\newcommand\ei{\end{itemize}}
\newcommand\be{\begin{equation}}
\newcommand\ee{\end{equation}}

\begin{document}

\begin{titlepage}

\begin{center}
{\large\bf On the Persistence of Homogeneous Matter}
\end{center}

\vspace{0.8 truecm}

\begin{center}
        J.~Butterfield\\[10pt] All Souls College, Oxford OX1 4AL\\
		email: jb56@cus.cam.ac.uk;
            jeremy.butterfield@all-souls.oxford.ac.uk
\end{center}

\begin{center}
        Saturday 22 May 2004
\end{center}

\vspace{0.8 truecm}

\begin{abstract}
Some recent philosophical debate about persistence has focussed on an 
argument against perdurantism that discusses rotating perfectly homogeneous 
discs (the `rotating discs argument'; RDA). The argument has been mostly 
discussed by metaphysicians, though it appeals to ideas from classical 
mechanics, especially about rotation. In contrast, I assess the RDA  from 
the perspective of the philosophy  of physics.

After introducing the argument and emphasizing the relevance of physics 
(Sections 1 to 3), I review some metaphysicians' replies to the argument, 
especially those by Callender, Lewis, Robinson and Sider (Section 4). 
Thereafter, I argue for three main conclusions. They all arise from the 
fact, emphasized in Section 2, that classical mechanics (non-relativistic as 
well as relativistic) is both more subtle, and more problematic, than 
philosophers generally realize. 

The first conclusion is that the RDA can be formulated more strongly than is 
usually recognized: it is not necessary to ``imagine away'' the dynamical 
effects of rotation (Section \ref{134Goodbadunnecy}). The second is that in 
general relativity, the RDA fails because of frame-dragging (Section 
\ref{432:RDAfailsGR}). 

The third  is that even setting aside general relativity, the strong 
formulation of  the RDA can after all be defeated (Section \ref{sec:soln}). 
Namely, by the perdurantist taking objects in  classical mechanics (whether 
point-particles or continuous bodies) to have only temporally extended, i.e. 
non-instantaneous, temporal  parts: which immediately blocks the RDA. 
Admittedly, this version of perdurantism defines persistence in a weaker 
sense of `definition' than {\em pointilliste} versions that aim to define 
persistence assuming only instantaneous temporal parts. But I argue that 
temporally extended temporal parts: (i) can do the jobs within the 
endurantism-perdurantism debate that the perdurantist wants temporal parts 
to do; and (ii) are supported by  both classical and quantum  mechanics.

 \end{abstract}

\end{titlepage}
\tableofcontents

\section{Introduction}\label{secIntro}
This paper is an attempt to connect what modern  physics, especially 
classical physics, says about matter, with the debate in  analytic 
metaphysics  whether an object persists over time by the selfsame object 
existing at different times (nowadays called  `{\em endurance}'), or by 
different temporal parts, or stages, existing at different times (called 
`{\em perdurance}'). This is a multi-faceted debate, with various 
connections to physics and the philosophy  of physics. This paper focusses 
on just one such connection. I will assess, from the perspective of the 
philosophy of physics, a metaphysical argument against perdurantism, which 
is based on the idea of rotating homogeneous matter, and is nowadays often 
called the `rotating discs argument' (RDA).  I will argue, against much of 
the literature, that the argument fails, because of some features of 
classical mechanics (including how it should be interpreted in the light of 
quantum mechanics). But my larger hope is to connect the philosophy of 
physics, and metaphysics, communities. In this hope, I will sometimes 
expound details familiar to one community or the other---at the cost of some 
length!

I begin by outlining the argument, the kinds of reply usually made to it, 
and my own preferred reply (Section \ref{secIntro}). This will lead to an 
discussion of how physics, and philosophy  of physics, is relevant to the 
argument (Section \ref{ssec;therelevcephys}). This will include a more 
detailed prospectus of the later Sections (Section 
\ref{ssec;RDAprospectus}). But in short, I will:\\
\indent (i): present some more details about the RDA's assumptions and scope 
(Section \ref{sec2;argtkindsreply}); \\
\indent (ii): discuss some replies made to it in the metaphysical literature 
(Section \ref{ssec;causevelrot}); \\
\indent (iii): present some details of how physics describes rotation, and 
thereby formulate a stronger version of the RDA than the usual one---albeit 
one that fails in general relativity (Section \ref{ssec;describerot});\\
\indent (iv): present my own two replies to the RDA (Sections 
\ref{sec:soln}-\ref{sssec:Appealbeyondclassl}); of which I favour the 
second. This reply involves some novel proposals about the 
intrinsic-extrinsic distinction among properties. It is also  supported by 
the way in which the objects of classical mechanics emerge from quantum 
theory.

This paper is a part of a larger project. My (2004, 2004a) describe some 
other connections between the endurantism-perdurantism debate and aspects of 
physics and its philosophy. In particular, my (2004)  presents in  more 
detail both the endurantism-perdurantism debate, and my arguments against 
philosophers' tendency to interpret classical mechanics in what I will call 
a {\em pointilliste} way (cf. Section \ref{sssec;vsStrtfwd}). For the 
moment, suffice it to say  that I conceive the endurantism-perdurantism 
debate in much the same way as Sider's fine recent survey, and defence of 
perdurantism (2001); cf. also Hawley (2001). Hawley and Sider also both 
discuss the RDA, at pp. 72-90, and 224-236, respectively. 

\subsection{The RDA}\label{ssec;theRDA} 
The argument envisages that the perdurantist with her ontology of temporal 
parts faces the project of defining  persistence: since persistence is not 
identity, she needs to tell us what it is. (This project  is  called 
`analyzing persistence', and `analyzing or defining the genidentity relation 
between temporal parts'.) In particular, she needs to define persistence so 
as to distinguish ``ordinary persisting objects'' (i.e. the referents of 
ordinary terms, and elements of ordinary domains of quantification) from the 
countless other ``spacetime worms'', i.e. mereological fusions of temporal 
parts. (Most perdurantists accept unrestricted mereological composition, so 
that they  also accept these worms as genuine objects.) On pain of 
circularity, the {\em definiens} is not to presuppose the notion of 
persistence.

 The argument urges that the perdurantist cannot succeed in this 
project.\footnote{The RDA arose in recent philosophy in 
Kripke (unpublished lectures) and Armstrong (1980).
Zimmerman (1998) reports how the argument goes back at least to Broad in 
1925. Sider (2001, p. 226) notes that Leibniz (1698, sect. 13) deploys 
essentially the same argument: but Leibniz's target is Descartes' doctrines 
about matter and motion.} It is based on two ideas:\\
\indent (i) {\em Homogeneous}:  In a continuum (i.e. a continuous body whose 
composing matter entirely fills its volume) that is utterly homogeneous 
throughout a time-interval containing two times $t_0,t_1$, a spatial part at 
the time $t_0$ is equally  qualitatively similar to any spatial part 
congruent to itself (i.e. of the same size and shape) at the later time 
$t_1$. (The properties of the continuum can change over time, but must not 
vary over space; e.g. the continuum could cool down, but must at each time 
have the same temperature everywhere.)\\
\indent (ii) {\em Follow}: The perdurantist will presumably try to define 
persistence in terms of suitable relations of qualitative similarity between 
temporal parts. The obvious tactic is to have the {\em definiens}  
``follow'' the curves in spacetime that are timelike and track maximum 
qualitative similarity.

\indent The tactic of {\em Follow} seems to work well   when applied to 
point-particles moving  in a void each with a continuous spacetime 
trajectory (worldline). For however exactly we define `maximum qualitative 
similarity', there will no doubt be,  starting at a point-particle at $t_0$, 
a unique timelike curve of qualitative  similarity passing through it: the 
worldline of the particle. (Indeed, for this case we could dispense with 
qualitative similarity, and have the {\em definiens} refer just to spacetime 
points' property  of being occupied by matter.)  Similarly for 
point-particles moving, not in a void, but in a continuous fluid with 
suitably different properties---a different ``colour'', or made of different 
``stuff'', than the point-particle. (Again, we could have the {\em 
definiens} refer just to spacetime points' property of being occupied by 
matter with the ``colour'', or made of the ``stuff'', of the given 
point-particle.)\footnote{Agreed, one can object that: (i) any such {\em 
definiens} is too weak, i.e. not sufficient for persistence, since a god 
could instantaneously destroy a point-particle and immediately replace it 
with a qualitative replica---suggesting that the {\em definiens} must invoke 
causal notions; and-or that (ii) any such {\em definiens} is too strong, 
i.e. not necessary  for persistence, since a point-particle could jump about 
discontinuously. I address these objections in Sections 4.1-4.2.1 of my 
(2004a). In short: as to (i) I am sceptical of the appeal to causation---a 
topic I will return to in Section \ref{sssec;appealcause} below; and as to 
(ii), I suggest we restrict the {\em definiens} to point-particles assumed 
to have continuous worldlines. But the details of my replies are not needed 
for this paper:  for they are no help to the perdurantist in facing the 
trouble made by the RDA.}

\indent  But {\em Homogeneous} implies that {\em Follow}'s strategy stumbles 
when applied to a homogeneous continuum. There are altogether too many 
spatial parts at $t_1$ that are tied-first-equal as regards qualitative 
similarity to the given spatial part at $t_0$: any congruent spatial part 
will do. In other words: the curves of qualitative similarity run ``every 
which way''.

\indent This problem is  made vivid by urging that the perdurantist  cannot 
distinguish two  cases that, the argument alleges, must be distinguished. 
One main example, on which I will focus, is the case of a perfectly circular 
and rigid disc of homogeneous matter that is stationary; and a duplicate 
disc  (rigid and congruent to, made of the same homogeneous material as, at 
the same temperature as {\em etc.} the first) that is rotating about the 
axis through its centre. It will be convenient to have labels for two such 
possibilities: call them `(Stat)' and `(Rot)'.\\
\indent Hence the argument  is nowadays often called the `rotating discs 
argument' (RDA). (In some discussions, both discs are rotating, but with 
different velocities, maybe even in different senses.) But all agree that 
countless other examples would serve just as well as a disc: e.g. a sphere; 
or a body of fluid, like a river, that can be either stationary or flowing 
(or flowing with different speeds, or in different directions).

It seems that the endurantist can easily distinguish the two possibilities, 
according to whether the very same non-circularly-symmetric part, e.g. a 
segment, is in the same place at two times. Later (especially Sections 
\ref{sssec;tuquoque}, \ref{132connectmetrot}), I will pursue the question 
whether this is really so: can the endurantist legitimately use the notion 
of being in the same place at two times, i.e. the notion of persisting 
spatial points? (This question is almost entirely ignored in the 
metaphysical literature: authors often appeal without further discussion to 
the idea of ``the same place'' (e.g. Hawley 2001, p. 85).) But for the 
moment, I just assume, in order to give the RDA as good a run as possible, 
that the answer is Yes. 

On the other hand, it seems the perdurantist has a problem. Surely she  must 
say that all the relations (and therefore, all her proffered ``suitable 
relations'' for analysing persistence) between two  temporal parts of the 
disc (say, second-long parts at noon and 12.01) are the same---whether the 
disc is rotating or not? And similarly  for temporal parts at the two times 
of any spatial part of the disc, such as a segment.

The rest of this Section clarifies the scope of this argument, and the kinds 
of reply the perdurantist can give to it. This will yield a statement of a 
consensus which is widespread in the literature---and an announcement of how 
the remainder of the paper will argue against that consensus.

\subsection{Intrinsic properties  and velocities}\label{ssec;intrpropvelies}
So far I have expressed the RDA's main idea as the inadequacy, for defining 
persistence, of qualitative similarity. But in some versions of the 
argument, the emphasis is instead on the inadequacy of intrinsic properties. 
The intrinsic-extrinsic distinction among properties is controversial, but 
the rough idea is that possession of an intrinsic  property implies nothing 
about the possessor's environment, i.e. about matters of fact beyond the 
instance.  So in these versions, the target is a perdurantist who seeks a 
{\em definiens} using intrinsic properties of temporal parts. And in some 
versions, the target is a yet stronger neo-Humean doctrine to the effect 
that (roughly speaking) {\em all} facts---not just facts of 
persistence---are determined by all the various intrinsic properties of all 
the points of spacetime. The most influential version of this sort of 
extreme {\em pointilliste} doctrine is {\em Humean supervenience}, as 
formulated and defended by Lewis (1986, p. ix-x; 1994, p. 474; 1999).

This emphasis on intrinsic properties goes along with an objection to the 
obvious suggestion that what distinguishes the two cases is the direction of 
the instantaneous velocity of the disc's (or sphere's, or river's) 
constituent parts. Thus for the stationary disc, all the disc's parts have 
zero velocity; while for the rotating disc, the parts have various 
velocities (and for a perfectly rigid disc, a common angular velocity); and 
similarly for the sphere or river. But, says the RDA, the notion of  
velocity presupposes the persistence of the object concerned. For average 
velocity is a quotient, whose numerator must be the distance traversed by 
the given persisting object: otherwise you could give me a superluminal 
velocity by dividing the distance between me and the Sun by a time less than 
eight minutes. So average velocity's limit, instantaneous velocity, surely 
also presupposes the notion of persistence. Accordingly, says the RDA, the 
perdurantist cannot adopt the obvious suggestion, of distinguishing the 
cases in terms of  instantaneous velocity (or angular velocity)---on pain of 
circularity.

The notion of presupposition, like the intrinsic-extrinsic distinction,  is 
controversial. Besides, Tooley (1988) proposes a heterodox account of 
instantaneous velocity as an intrinsic property of an object at a time; (and 
Bigelow and Pargetter (1989, 1990 Section 2.6) propose a similar account). 
Though these authors are not concerned with the debate over persistence, 
their account of velocity has been discussed in the context of the RDA.  So 
I shall later return to the idea of appealing to velocity, and to this 
heterodox account of it (Section \ref{sssec;appealvel}).  But for the most 
part, I will concede the above objection. That is, I will concede that both 
average and instantaneous  velocity presuppose the notion of persistence, 
and are extrinsic properties. Nevertheless, my favoured reply to the RDA 
(Section \ref{psa;perdmwouttears}) will be that a perdurantist who accepts 
only non-instantaneous temporal parts (a version of perdurantism which, I 
contend, is supported by physics) can endorse the obvious suggestion, i.e. 
can appeal to velocities to distinguish the two cases.

\subsection{``Naturalism''}\label{ssec;naturalism}
So far, my description of the perdurantist project of defining persistence, 
and of the RDA against it, might well be read within the tradition of 
conceptual analysis. By this I mean that the perdurantist's definition would  
be both finite in length, and formulated using everyday concepts. But 
nowadays, the literature also considers a ``naturalized'' perdurantist 
project of\\
\indent \indent  (a): providing  only a supervenience basis for persistence 
(i.e. allowing infinitely long definitions), rather than a finite definition 
or analysis of it; and-or \\
\indent \indent (b): appealing to technical notions, and contingent bodies 
of doctrine, in particular the laws of dynamics.  Also, some authors  
combine  (b) with use of the Ramsey-Lewis technique for simultaneous 
functional definition; (in particular, Sider (2001, 224-236)---details in 
Section \ref{sssec;Sider}).\\
\indent Accordingly, the RDA is nowadays sometimes formulated as targeting 
even: (a) the supervenience of persistence on qualitative similarity among, 
and-or intrinsic properties of, the perdurantist's temporal parts; where (b) 
such supervenience may even be contingent, say relative to the laws of a 
dynamical theory.\\
\indent This situation prompts two comments: the first relates mostly to 
(a), the second mostly to (b). 

(1): {\em Non-reductive perdurantism}:--- There is also an even more 
naturalistic conception of perdurantism, which might well avoid the RDA. On 
this conception, the perdurantist seeks a theory of perdurance and related 
concepts, that can  appeal to scientific technicalities, that can revise 
rather than describe our concepts---and that does {\em not} have to define  
persistence in terms that do not presuppose it. Of course, analogous 
``non-reductive'' conceptions are nowadays commonplace in the philosophical 
study of many concepts, such as  causation, perception and action. So just 
as a philosophical theory of causation might decline to define causation 
(even infinitarily, even by Ramsey-Lewis functional definition), a 
perdurantist might decline to define persistence (even in these liberalized 
senses), on the grounds that she nevertheless says enough to adequately 
distinguish ``ordinary persisting objects'' from other ``spacetime worms''. 
I shall return to this modest (because non-reductive) perdurantism in 
Section \ref{psa;perdmwouttears}. But until then I shall  consider the more 
ambitious perdurantist, who aspires to define persistence, and so faces the 
RDA.

(2): {\em How many worlds?}:--- Once we allow that a perdurantist theory of 
persistence might appeal to a contingent body of doctrine, such as a 
physical theory, the discussion of the RDA (or even the whole 
endurantism-perdurantism debate) is liable to become relative to a theory. 
There are two aspects to this; the first leads on to the second.

\indent (2.A): The RDA might hold good in one theory, and fail in another. 
Thus it is a familiar thought that any consistent theory lays out a set of 
possibilities: in philosophical jargon, possible worlds according to the 
theory; in physics  jargon, a space of solutions.   So relative to any 
consistent theory  about matter and rotation (describing them no doubt 
partially rather than completely---and perhaps falsely), the two cases 
(Stat) and (Rot) are either two distinct possibilities: or they are not, 
either because at least one is not possible (since e.g. the theory  denies 
that matter is homogeneous), or because they are the {\em same} possibility.

\indent  This point is independent of whether to accept the notion of a law 
of nature, not relativized to a specific scientific theory. Authors (on 
either side of the endurantism-perdurantism debate) who accept this notion 
can consider contingent  theses of reduction or supervenience cast in such 
terms, e.g. supervenience across a class of possible worlds that each make 
true all the actual laws of nature. For example, Armstrong and Lewis (to 
whom I will return in Section \ref{ssec;causevelrot}) are two such authors, 
and both perdurantists: though  they disagree about how to understand laws 
of nature, and how the perdurantist should respond to the RDA.\\
\indent On the other hand, authors who reject the notion will construe 
contingent supervenience theses, and so perhaps their discussion of the RDA 
(or even the whole endurantism-perdurantism debate), as relative to a given 
theory.\\
\indent I myself will not need the notion of a law of nature; (indeed, I am 
wary of it). And we will later see the RDA holding good, and failing, in 
different theories. But I will of course allow for evaluation of the RDA 
which is not wholly relative to a theory. In particular, special interest 
will of course attach to the case of true theories: or to put it from our 
epistemic perspective, theories that are our best guess for truth. Such an 
interest does not presuppose ``scientific realism'', which concerns whether 
we should {\em believe} the theoretical claims of our best theory to be at 
least approximately true. Any ``naturalist'', whether or not they are a 
``scientific realist'',  will of course  be especially interested in whether 
the RDA holds good in our best theory of matter and rotation.\\
\indent (I will in fact argue that the RDA fails, not only in general 
relativity and quantum theory---our best guesses about space, time and 
gravity, on the one hand, and about matter, on the other---but even in 
classical mechanics, under an interpretation I favour.)    

\indent (2.B): But we should beware of just dismissing the RDA on the ground 
that according to our best theories, matter is in fact made of atoms and so 
not homogeneous. For presumably:\\
\indent (i): A continuous, rigid and utterly homogeneous form of matter 
could exist and be formed into a disc that either rotates or is stationary. 
And:\\
\indent (ii): No philosopher of persistence  is ``so far gone'' in 
naturalism  as to be interested only in how objects persist, given all the 
contingencies of the actual world.

\indent In what follows, I will agree with these presumptions, so as to give 
the RDA against perdurantism as good a run as possible. But it is worth 
drawing attention to them since, as we shall see:\\
\indent (i'): The sort of continuous and homogeneous matter the RDA needs is 
a much subtler and more problematic affair than the RDA literature typically 
recognizes; (cf. Section \ref{ssec;therelevcephys}). This leads in to 
(ii'):---\\
\indent (ii'): Some perdurantists reply to the RDA by saying that for the 
possibilities (Stat) and (Rot) to exhibit no difference to which the 
perdurantist  can appeal,  the advocate of the RDA needs to ``imagine away'' 
so many actual laws, technical and-or everyday, which describe various 
causes and effects of rotation, that the RDA's possibilities (Stat) and 
(Rot) are, though logically or metaphysically possible, very arcane. Indeed, 
they are so arcane that a naturalist perdurantist need feel no shame in 
being unable to accommodate them.

 To put the reply (ii') in the jargon of possible worlds: the perdurantist 
claims their theory of persistence, though contingent and unable to 
discriminate the possibilities (Stat) and (Rot), is true in so broad a class 
of possible worlds that excluding (at least one of) (Stat) and (Rot) is a 
small price---and worth paying. (Examples of this reply include: Lewis 
(1986, p.xiii, 1994, p. 475), Callender (2001), and (less explicitly) Sider 
(2001, 230-236).) This leads to the next Subsection.

\subsection{The accompaniments of rotation}\label{sssec;accompanimtsrotn}
 Rotation has countless typical causes and effects; or if one is wary of 
causal talk: countless typical accompaniments. Typically, a rotating object  
was previously set in motion, say by being pushed by someone, and exhibits 
distinctive dynamical effects: for example, a solid object tends to become 
oblate, and a fluid,  like water in a whirlpool, develops a concave surface.  
These accompaniments do not depend on matter being in fact atomistic (or on 
the laws of physics being relativistic and quantum). So in a possible  world 
that contained continuous and homogeneous matter but was otherwise ``like 
the actual world'', these accompaniments---even the ``technical'' ones, like 
oblateness and concavity---would occur. In which case, the RDA needs to 
block the perdurantist appealing to them so as to distinguish the cases.

True to the tradition of conceptual analysis, the  literature on the RDA 
almost entirely sets aside the technical accompaniments, and concentrates on 
the everyday ones, like having been pushed in the past; and on related 
everyday counterfactuals, such as `were I to spray a spot of paint on the 
disc, I would see it move', or `were I to  grasp the disc, I would feel 
friction'. More specifically, the literature tends to assume that the RDA 
can legitimately set aside all the technical accompaniments by just 
stipulating that the rotating disc is not only solid but perfectly rigid, so 
that it does not become oblate; (hence Section \ref{ssec;theRDA}'s mention 
of rigidity).  The philosophical battle can then be joined on two 
battlefields familiar to metaphysicians; as follows.

First, there is debate about whether the RDA can legitimately ``imagine 
away'' the everyday accompaniments of rotation, so that the perdurantist 
cannot appeal to them. In particular: if (as usual) the RDA stipulates that 
the  present and ``occurrent'' everyday accompaniments are absent, can the 
perdurantist appeal to past or future accompaniments, or perhaps to 
counterfactuals about them? For example:\\
\indent (i) Can the perdurantist make the distinction by appealing to  a 
past cause, such as a push, or to a present counterfactual about seeing a 
paint-spot move?\\
\indent (ii) Or would appealing to a past cause amount to postulating an 
unacceptable ``temporal action-at-a-distance'' (e.g. Robinson 1989 p. 
405-406; Hawley 2001 p. 81)?\\
\indent (iii) And would appealing to a present counterfactual amount to 
postulating unacceptably ``ungrounded'' counterfactual truths (Robinson 1989 
p. 403; Hawley 2001 p. 74-75)?

Second, there is debate about whether the perdurantist can appeal to 
 differences between (Stat) and (Rot) that are distinctively metaphysical 
(neither everyday nor technical-physical).  For example: Can the 
perdurantist appeal to:\\
\indent (i) a special (non-Humean) relation of immanent causation between 
temporal parts that subvenes (or even yields an analysis of) persistence 
(Armstrong 1980, 1997, pp. 73-74); or\\
\indent (ii) special vectorial properties that are numerically equal to, yet 
different from, velocities (Robinson 1989 pp. 406-408, Lewis 1999: 
incidentally, this idea echoes Leibniz's proposal against Descartes (1698, 
sect. 13)); or\\
\indent (iii) non-causal relations between temporal  parts that are not 
supervenient on the intrinsic natures of the parts that are the relata, and 
yet are not just spatiotemporal relations (Hawley: 1999, p. 63-66; 2001, p. 
85-90)?

For my reply to the RDA, I do not {\em need} to enter either of these 
battlefields; (fortunately, since they remain well-populated, despite the 
crossfire!). As to the first, I can set aside the ``everyday 
accompaniments''. For I shall argue (especially in Sections 
\ref{ssec;therelevcephys} and \ref{ssec;describerot}) that the RDA should 
not just set aside technicalities, in particular the technical 
accompaniments of rotation; and that in any case, it {\em cannot} do so just 
by stipulating perfect rigidity. As to the second, my reply to the RDA (in 
Section \ref{sec:soln}) does not need controversial metaphysical proposals 
like immanent causation, special vectorial properties etc. (of which I am in 
any case wary). However, I will make some points about these proposals, from 
the perspective of the philosophy of physics (Section 
\ref{ssec;causevelrot}).

\subsection{Two kinds of reply: Against the consensus}\label{psa; against 
the consensus}
We can sum up ``the story so far'' in two stages. First, there are two main 
ways perdurantists can reply (and have replied) to the RDA. They can 
either:\\
\indent\indent (`Appealing Differences'): argue that there are differences 
between the discs to which they can appeal; whether everyday (e.g. `someone 
pushed it'), technical (e.g. `it's oblate') or metaphysical (e.g. `the 
timelike curves  of immanent causation are helical, not straight'); or\\
\indent\indent (`No Difference'): argue that possible  worlds in which the 
discs show no such difference are too arcane to matter: i.e. they do not 
fall within the scope of their ``naturalist'' account of persistence.

Second: In the literature on the RDA, considerations of metaphysics, and in 
particular conceptual analysis, tend to dominate. This dominance has led to 
a widespread consensus on four points: two in support of the RDA, and two 
against the perdurantist. Namely:\\
\indent (I): The RDA  {\em can} legitimately\\
\indent\indent (a) imagine away the usual accompaniments of rotation: both 
the everyday ones; and the technical ones such as discs tending to be become 
oblate (in the latter case, by requiring the discs to be rigid); \\
\indent\indent (b) assume the intuitive notion of rotation, with its idea of 
persisting spatial points.

On the other hand:\\
\indent (II): the perdurantist {\em cannot} legitimately\\
\indent\indent (c) appeal to differences of velocity, since velocity   
presupposes persistence; nor can they \\
\indent\indent (d) appeal to the atomic, indeed quantum-theoretic, nature of 
matter, since the topic of debate is our common-sense conception of 
persistence---which surely allows continuous matter.      
 
Turning to this paper: I shall argue against the consensus (a)-(d). Section 
\ref{ssec;describerot} argues against (a) and (b); Sections \ref{sec:soln} 
to \ref{sssec:Appealbeyondclassl} against (c) and (d). (Sections 
\ref{ssec;therelevcephys} to \ref{ssec;causevelrot} will set the stage for 
these arguments.) The overall effect will be twofold. As to (a) and (b): I 
will concede that there are sound versions of the RDA. Indeed, the RDA can 
be formulated more strongly than usual (i.e. than Section 
\ref{ssec;theRDA}'s formulation): for it does not need to imagine away the 
usual accompaniments of rotation. But as to (c) and (d): a certain sort of 
perdurantist---roughly speaking, one who accepts only non-instantaneous 
temporal  parts---{\em can} both appeal to differences of velocity, and 
garner support for their position from quantum theory.  

\section{The relevance of physics}\label{ssec;therelevcephys} 
So much by way of introducing the RDA. 
We have already seen that it raises issues in the philosophy of physics as 
much as in metaphysics. There is of course a spectrum here, from ``common 
sense'' doctrines about persistence to ``folk physics'' to technical 
physics. And I agree that  it is in part a matter of intellectual judgment 
and-or interest: (i) how far along the spectrum to move; and if one 
considers technical physics, (ii)  which physical theories to consider, 
classical or quantum, relativistic or non-relativistic. But only in part! I 
shall argue that the philosophy of persistence needs to go further towards 
technical physics than the  literature on the RDA tends to.

By and large, the RDA literature engages a bit with ``folk physics'', but 
not technical physics. There are two connected aspects to this restriction. 
First, the literature sets aside the fact that matter is in fact made of 
atoms.  Almost all authors maintain that our ``common-sense'' notions of 
matter and its persistence are surely compatible with matter's being 
continuous in its composition; so that a perdurantist seeking an account of 
these notions faces the RDA.\footnote{I say `surely compatible' since some 
authors toy with the view that the RDA  shows that the compatibility is an 
illusion: our notion of matter and its persistence requires atomistic 
matter. Robinson (1989, p.404, reporting Lewis) portrays this as an example 
of the traditional ``paradox of analysis'': roughly, that philosophical 
analysis can reveal surprising truths.} And most authors, especially those 
closer to traditional conceptual analysis, maintain that these notions  form 
a framework sufficiently widespread, and cognitively central, for such an 
account to be not parochial, but worthwhile; and worthwhile even if 
continuous matter is ``science fiction physics''  (e.g. Robinson 1989, pp. 
396-398; Zimmerman 1999, p. 213).

Second, although  many authors in this literature discuss velocity, even 
instantaneous velocity---and several briefly discuss allied concepts from 
elementary mechanics, like force and momentum---almost all set aside 
technical physics: not just the modern theories which are our best guesses, 
viz. relativity theory and  quantum theory, but also the details of the 
classical mechanical description  of rotation and of continua (i.e 
continuous bodies).\footnote{So far as I know, only one article about the 
RDA engages with technical physics: viz. Callender (2001), who discusses the 
classical physics of rotation; I discuss it below. Oppy (2000) surveys 
various threats from physics, including quantum physics,  to Lewis' Humean 
supervenience; but without focussing on persistence.}

I believe these two aspects---taking common sense to encompass continua, and 
setting aside technical physics---arise from two mutually related, and 
widespread, assumptions. But these assumptions are in fact {\em false}---and 
correcting these assumptions will be the main ingredient in my rebuttal of 
the RDA. In short, the assumptions are:\\
\indent ({\em Straightforward}): The ontology of classical mechanics, 
including the classical mechanics of continua, is straightforward, i.e. 
unproblematic.\\
\indent ({\em Bracket}): Although the world is in fact relativistic and 
quantum, we can ``bracket'' this fact when we investigate persistence. That 
is: classical mechanics, or at least classical physics as a whole, forms a 
coherent whole, which can be safely assumed to provide the supervenience 
basis on which facts about the persistence of macro-objects supervene.  

 So in the next two Subsections, I shall spell out the errors of these two 
assumptions, and so urge that the philosophy of physics is relevant to the 
RDA. That will serve to introduce Section \ref{ssec;RDAprospectus}'s 
Section-by-Section prospectus.\footnote{For a more detailed discussion of 
the two aspects above, and these two assumptions, cf. Sections 2.2 and 2.3 
of my 2004.} 

\subsection{Classical mechanics is subtle and 
problematic}\label{sssec;vsStrtfwd}
({\em Straightforward}) says that the ontology of classical mechanics, 
including the classical mechanics of continua, is unproblematic. More 
precisely, I think the literature assumes a  conception of the ontology of 
classical mechanics, which I call the {\em particles-in-motion  picture}. 
This analyses matter into extensionless particles: either point-particles 
separated from each other by a vacuum, or the extensionless infinitesimal 
constituents of a continuum (i.e. continuous body), ``cheek by jowl'' with 
each other. In either case, the composition and behaviour of matter is 
analysed in terms of  extensionless particles,  interacting by 
particle-to-particle forces such as gravity (with their motions through 
Euclidean space determined by the forces, according to Newton's second law). 
Furthermore, the literature assumes that this particles-in-motion picture is 
unproblematic, as regards {\em matter}; i.e. once one sets aside the various 
familiar philosophical problems about space and time (e.g. ``absolute'' or 
``relational''?).

 This assumption, ({\em Straightforward}), leads to the first aspect of the 
RDA literature's restriction, i.e. its taking common sense to encompass 
continua.  For the assumption implies that: (i) philosophical discussions of 
persistence have no need to tangle with the details of mechanics, either of 
point-particles or of continua; and (ii) since continua are both 
countenanced by common sense and   unproblematic, an account of persistence 
needs to allow for matter being continuous, even though  matter is in fact 
made of atoms---so the perdurantist faces the RDA.

But the particles-in-motion picture is wrong. There are in fact considerable 
conceptual tensions in classical mechanics' description  of matter, whether 
conceived as point-particles or as continua. Besides, classical continua 
cannot be treated in the ``{\em pointilliste}'', i.e. particle-by-particle, 
way envisaged by the particles-in-motion picture.

 Obviously I cannot here enter into details about the foundations of 
classical mechanics. I will only present two points that bear directly on my 
concern with the RDA. The first illustrates classical mechanics' subtlety, 
and will be directly relevant below: in Sections \ref{sssec;appealvel} and 
\ref{psa;perdmwouttears} it will give the perdurantist a reply to the RDA. 
The second illustrates how classical mechanics is problematic, and will lead 
in to the next Subsection (against assumption ({\em Bracket})).\footnote{My 
2004 gives a more detailed critique of the particles-in-motion picture, 
especially of its  {\em pointillisme}. For classical mechanics' anti-{\em 
pointillisme}, cf. e.g. Truesdell (1991, especially Sections II.2, III.1, 
III.5). For a philosopher's general introduction to the conceptual tensions 
in classical mechanics' description  of matter, I recommend Wilson's papers, 
e.g. his (1997) and (2000).}

(1): {\em Against pointillisme}:  The first point is that classical 
mechanics does not in fact describe continua in the {\em pointilliste} way 
that the particles-in-motion picture envisages.  Instead, the classical 
mechanics of continua has to be formulated in terms of spatially extended 
regions and their properties and relations. In particular, one cannot 
understand the forces operating in continua (whether solids or fluids) as 
particle-to-particle. Rather, one needs to conceive of a force being exerted 
on the entirety of an arbitrary finite (i.e. not infinitesimal) portion of 
matter, and of a force being exerted at the surface of such an portion. In 
Sections \ref{sssec;appealvel} and \ref{psa;perdmwouttears}, this anti-{\em 
pointillisme} will be extended to include {\em temporal} extension 
(motivated in part by quantum theory). It will thereby give the perdurantist 
the right to have only temporally  extended, i.e. non-instantaneous, parts: 
and this will secure  a reply to the RDA. 

  This need to take extended regions as primitives is worth stressing, even 
apart from the debate about RDA; for two reasons. First,  I admit that {\em 
prima facie} the particles-in-motion picture's strategy for analysing 
continuous matter is more attractive. For the alternative strategy, of 
describing the states of all the countless overlapping extended sub-regions 
of a continuum, is highly redundant: each sub-region is described countless 
times, viz. as a part of the description of a larger region in which it is 
included. Nevertheless, classical mechanics adopts---and needs to 
adopt---this alternative strategy. In short: a redundancy worth remarking. 

Second, {\em pointillisme} has been a prevalent theme  in recent analytic 
metaphysics: witness the recent interest in Lewis' {\em pointilliste} 
doctrine of Humean supervenience. But classical mechanics's anti-{\em 
pointillisme}  seems not to have been noticed in metaphysics; though the 
relevant physics goes back to Euler.  

(2): {\em Problems about point-particles}: Even if we set aside continua and 
consider only point-particles, i.e. extensionless point-masses, there are 
conceptual problems. One main group of problems arises once we add to the 
idea of a point-particle the notion of a field: I will postpone them to 
Section \ref{sssec;vsbracket}. Here, I will just mention two obvious 
problems, independent of the notion of a field.

The first problem is: how can we describe in terms of point-particles, {\em 
contact} between objects? As I see it: this problem divides into two 
sub-problems---both of them hard. First: how can we reconcile 
point-particles with solid objects' impenetrability? Even if we postulate, 
as Boscovitch did,  that when point-particles are very close some repulsive 
force dominates the attractive force of gravity, questions abound. For 
example: how can we describe the difference between solids and fluids? 
Second: there is the problem of describing (or else somehow prohibiting!) 
collisions of point-particles. Such collisions are clearly problematic, not 
just kinematically but dynamically. In particular,  under Newtonian gravity 
(or any interaction described by an infinite potential well around each 
particle), two colliding point-particles each have infinite kinetic energy 
at the instant of collision.

The second problem arises from the fact that, barring collisions, 
point-particles require that all forces act at a distance. Newton famously 
``deduced from the phenomena'' that gravity acted at a distance. In 
particular, it acts instantaneously: according to his theory, if the Sun  as 
a whole were now to move by, say, a metre, the direction of its 
gravitational pull on the Earth would now change, albeit by a minuscule 
angle. On the other hand, since light takes eight minutes to travel from Sun 
to Earth, the minuscule change in the visual direction, from our standpoint, 
of the Sun  would take eight minutes to occur. But Newton also agreed that 
it is `inconceivable that inanimate brute matter should, without the 
mediation of something else which is not material, operate upon and affect 
other matter without mutual contact'; so that  he `contrived no hypotheses' 
about `the reason for these properties of gravity'. Though the outstanding 
successes of Newtonian gravitational theory during the next two centuries 
accustomed people to action at a distance, the advent of general relativity, 
in which gravity propagates at the same speed as light, has now revived the 
natural suspicion of it---which Newton shared.\footnote{The quotations are 
from a letter to Bentley of 1693, and the General Scholium added to the {\em 
Principia} in 1713: for discussion and references, cf. Torretti 1999, p. 78. 
I also stress that it is general, not special, relativity, that militates 
against action at a distance: Lorentz invariance does not prohibit action at 
a distance, whether along the light cone or across spacelike intervals; (cf. 
Earman 1989, p. 156, who cites Kerner 1972).}   

In this Subsection, I have argued against the particles-in-motion  picture. 
I end with a conjectural, but humdrum, reason  why this wrong picture is so 
widespread in philosophy. I think it is a result of the educational 
curriculum's inevitable limitations.  In the elementary mechanics that most 
of us learn in high school, extended bodies are assumed to be small and 
rigid enough to be treated as point-particles. One never faces the 
subtleties of classical mechanics' treatment of continua.  Philosophers 
often augment high school mechanics with some  seventeenth-century 
mechanics, through studying such great natural philosophers as Descartes, 
Hobbes and Leibniz. But there ends most philosophers' acquaintance with 
mechanics. About 1700, natural philosophy divided into physics and 
philosophy, so that   few philosophers know about mechanics' later 
development. In particular, as regards the eighteenth century: philosophers 
read Berkeley, Hume and Kant, not such figures as Euler and Lagrange---whose 
monumental achievements in developing mechanics, and in particular its 
treatment of continua, changed the subject out of all recognition. 

{\em A fortiori}, philosophers also tend not to know about  relativity 
theory  and quantum theory. So this conjectural, but humdrum, reason also 
helps explain the second aspect of the RDA literature's restriction, viz. 
its setting aside these theories.

But I think there is also another explanation. Namely, the literature tends 
to make the  assumption I labelled ({\em Bracket}): that classical 
mechanics, or at least classical physics as a whole, forms a coherent whole, 
which can be safely assumed to provide the supervenience basis on which 
facts about the persistence of macro-objects supervenes. In the next 
Section, I argue that ({\em Bracket}) is false.

\subsection{Classical physics leads to relativity theory and  quantum 
theory}\label{sssec;vsbracket} ({\em Bracket}) says that, although the world 
is in fact relativistic and quantum, we can ``bracket'' this fact when we 
investigate persistence. More precisely:  All agree that the everyday 
macroscopic world ``emerges'' somehow or other from the relativistic  
quantum realm; and that in describing that world, classical physics, in 
particular  classical mechanics, is outstandingly successful. This suggests 
that some enquiries, even some in the foundations of physics or in 
metaphysics, will be able to take the classical mechanical description of 
the world as the physical ``given'', ignoring the fact that it is emergent 
and approximate.  ({\em Bracket}) proposes that enquiries, physical or 
metaphysical, about the persistence of macroscopic objects are among them. 
In more philosophical jargon: the classical mechanical description of the 
world provides the supervenience basis on which facts about the persistence 
of macro-objects supervenes.

 The tendency of the metaphysical literature on persistence  to invoke 
``folk physics'' and classical mechanics, but to set aside relativity theory  
and quantum theory, suggests that ({\em Bracket}) is widespread. But I claim 
that it is false. Besides, its falsity is central to my own position about 
the RDA. For the moment, I will just argue that ({\em Bracket}) is false, in 
three stages. The first two are brief and general; the third is an 
illustration.  (1): First, I will urge that interpreting classical mechanics 
leads one to the vast landscape of all of classical physics. (2): Then I 
will  describe how classical physics leads to relativity theory and quantum 
theory. (3): I will illustrate (2) with the example of self-interaction.

(1): {\em From classical mechanics to classical physics}\\
 There is vastly more to classical physics than is contained in classical 
mechanics: for example, thermodynamics, optics and electromagnetism. 
Furthermore, classical mechanics conceptually depends on these other fields 
in an open-ended way that is even today not wholly and rigorously mapped 
out. That is, classical mechanics  {\em cannot} be assumed to have some 
consistent and unproblematic ontology; (whether along the {\em pointilliste} 
lines of Section \ref{sssec;vsStrtfwd}'s particles-in-motion picture, or in 
my preferred non-{\em pointilliste} terms, using extended regions). Even for 
the special case of  point-particles in a void, we saw that one can raise 
worries, about collisions and the comprehensibility of action-at-a-distance 
forces. But in any case, the mechanics of continua (even if conceived in a 
non-{\em pointilliste} way) leads out into these other fields of physics in 
so open-ended a way as to raise many questions of ontology, or more 
generally, of interpretation.

\indent Obviously I cannot go into detail: it must suffice to make one basic 
point. Energy's role as a grand unifying concept in physics (as discovered 
in the nineteenth century) means that the classical mechanics of continua 
needs to be unified with thermodynamics: how else could  we understand 
rigorously such phenomena as the expansion of a (classical!) tarmac road in 
the heat of the day? For a glimpse of such a  unified theory, cf. Truesdell 
(1991, pp. 79-83, and references therein). But we can hardly stop there. 
Since the sunlight heats the road, we are led to optics; and thermodynamics 
leads us to statistical mechanics and the atomic constitution of matter. And 
so it goes: it would be a brave, nay a foolhardy, person who claimed to 
descry hereabouts a consistent and unproblematic ontology even for classical 
mechanics, let alone for all of classical physics. 

(2): {\em From classical physics to relativity and the quantum}\\
 Not only does classical mechanics lead out into the unsurveyably vast 
landscape of classical physics. Also, that landscape has---as Lord Kelvin 
famously put it in 1900---clouds on the horizon. Kelvin was in fact 
referring to the failures of the equipartition theorem in statistical 
mechanics, and of attempts to detect the motion of the earth through  the 
ether: failures  which in due course led to quantum theory, and relativity 
theory, respectively. But what matters for us is the general point: that 
classical physics' description of the microscopic structure of matter, and 
of matter's interaction with the electromagnetic field, turns out to be 
embroiled in paradox. This means that the would-be ontologist of classical 
mechanics faces, not just the problem of open-endedness discussed under (1), 
but an in-principle difficulty---of paradox.  In short, classical mechanics, 
together with the rest of classical physics, turns out to be a house built 
on sand. 

We know now that it is quantum sand---and that it somehow keeps the house 
up. But it remains pretty darned mysterious how it does so. By this I do not 
just mean that the interpretation of quantum theory (especially the 
resolution of its measurement problem) remains mysterious. Also, some 
aspects of how ``the house manages to stay up'' are current research 
projects in theoretical, not foundational, physics. One obvious example is 
the physics of decoherence; which also, all agree, will play an important 
role in solving the measurement problem. But I postpone this example till 
Section \ref{sssec:Appealbeyondclassl} and the Appendix, where it will be 
important in replying to the RDA.

Another example, closely related to the paradoxes of classical physics' 
description of matter's interaction with the electromagnetic field,  is the 
stability of matter. Classically, atoms would be unstable, since the 
orbiting electrons would radiate, lose energy and so tumble down to the 
nucleus. But quantum theory promises to secure stable atoms. The main idea 
here is the Pauli exclusion principle: it prevents an atom's electrons all 
tumbling down to be together in the atom's lowest electronic energy levels; 
(and similarly the principle prevents a cascade down nuclear energy levels). 
But the details are very complicated; and though they have been attacked 
successfully, especially in work from the 1960s, they remain an active 
research area (Levy-Leblond (1995), Lieb (1997)). 

(3): {\em An illustration: self-interaction}\\
 Even apart from atoms, there are paradoxes about the interaction of a 
classical charged particle with the  electromagnetic field. These paradoxes  
will illustrate (2). They also illustrate the spectrum from conceptual 
analysis to technical physics, with which I began this Section. For they 
show that any philosophical theory of persistence (whether endurantist or 
perdurantist), even one that considered only  the apparently straightforward 
case of point-particles, is liable to get led along this spectrum,  even as 
far as relativity theory and quantum theory. ({\em A fortiori}, there is 
good  reason to think technical physics is relevant to the more complex case 
of continua, considered by the RDA.) 

The paradoxes arise as soon as we accept that the  electromagnetic field 
carries energy, and that energy is conserved. Here it must suffice to sketch 
the main idea; there is not space for a proper discussion.\footnote{Rohrlich 
(2000) is a  philosophically and historically oriented entry into this large 
subject. It also covers treatments of charged particles as extended rather 
than point-like; indeed Lorentz's original statement (1892) of the Lorentz 
force law assumed an extended charge.} Classical electrodynamics says that 
an accelerating charge emits radiation, and thereby energy; and the 
conservation of energy then requires that it slow down. If the charge were a 
point-particle and was the only particle in the universe, the only force 
present that could cause it to slow down is that derived from its own 
electromagnetic field. So it seems that we must take a classical charged 
point-particle to ``feel'' its own field, even though in ordinary 
calculations we do not do so---e.g. in applying the Lorentz force formula 
${\bf F} = e({\bf E} + {\bf v} \wedge {\bf B})$ we ignore the charge's 
(infinite!) contribution to ${\bf E}$.

This predicament suggests two possible strategies.\\
\indent (i): To try to formulate a consistent interaction of a charge with 
its own field which will both (a) give some sensible result for the solitary 
accelerating charge, and (b) vindicate as approximately correct our usual 
calculational practice of ignoring each charge's self-interaction. (We need 
both (a) and (b) since the point-particle can hardly ``know'' whether it is 
lonely or not.)\\
\indent (ii): To revise classical electrodynamics so as to avoid the 
solitary-charge argument for a self-interaction; e.g. by postulating that 
energy is radiated only if, later on, something will absorb it.

Each of these strategies has had very distinguished proponents: from Lorentz 
and Dirac, for (i), to Feynman and Wheeler, for (ii). But the details of 
their work, and that of others, do not matter here. Anyway, a proper 
discussion of how to reply to these paradoxes must nowadays include quantum 
theory's description of matter (and specifically topics like renormalization 
in quantum electrodynamics): which lies far beyond this paper's scope. Here 
it is enough to have shown that the concept of a point-particle, which at 
first seems part of ``educated common sense'', is in fact problematic: as we 
saw in Section \ref{sssec;vsStrtfwd}, point-particles with 
action-at-a-distance are problematic, and now we see that problems remain 
when we  combine point-particles with the concept of a field.

\subsection{Prospectus}\label{ssec;RDAprospectus}
In this Section, I have argued: that  classical mechanics is both more 
subtle, and more problematic, than philosophers generally recognize (Section 
\ref{sssec;vsStrtfwd}); and that in addressing its problems, one is led to 
the rest of classical physics, and even to relativity theory and quantum 
theory---though it is still  unclear, in various ways, how the everyday 
macroscopic world ``emerges'' from the relativistic  quantum realm (Section 
\ref{sssec;vsbracket}). I can now describe how these views yield my main 
claims about the RDA.

My overall position is that the perdurantist can rebut the RDA; but physics 
also shows how the RDA can be formulated more  strongly than it has been. 
More specifically,  I will argue for three main conclusions. The first two 
are in Section \ref{ssec;describerot}, which focusses on the details of 
physics' description  of rotation, especially for continuous matter. 
(Quantum theory is set aside until Section \ref{sssec:Appealbeyondclassl}.) 
The first conclusion is that the RDA can be formulated more strongly than is 
usually recognized. For it is not necessary to ``imagine away'' the 
dynamical effects of rotation (e.g. the tendency of a spinning sphere to be 
oblate), as advocates of the RDA usually do; (Section 
\ref{134Goodbadunnecy}). The second conclusion is that in general 
relativity, the RDA (even in its stronger formulation) fails, because an 
(amazing but well-established) physical effect called `frame-dragging'  
implies that there are differences between rotation and non-rotation which 
the perdurantist can appeal to (Section 
\ref{432:RDAfailsGR}).\footnote{These two conclusions are not original to 
me. (1): The strengthened formulation of the RDA is due to Paul Mainwood and 
David Wallace, in an Oxford seminar, autumn 2003; and is hinted at by 
Zimmerman (1998, p. 268-269). (2): Callender (2001, p. 38) mentions 
frame-dragging as one of many differences between rotation and non-rotation 
the perdurantist can appeal to. So to all four, my thanks: my only 
contribution is to set these conclusions in a broader landscape than did 
their originators.}

The third conclusion is in Section \ref{sec:soln}. It is that even setting 
aside general relativity, the strong formulation of  the RDA can after all 
be defeated. I argue that the subtleties and problems of classical mechanics 
(including the way it ``emerges'' from quantum theory) mean the perdurantist 
can take objects in  classical mechanics (whether point-particles or 
continuous bodies) to have only temporally extended, i.e. non-instantaneous, 
temporal  parts (stages): which blocks the RDA.

I stress that I will {\em not} claim that considerations of physics show 
perdurantism superior to endurantism; though, as I will discuss in Section 
\ref{133Allowboth}, endurantism and perdurantism  have traditionally been 
associated with conceptions of metaphysics as {\em a priori} conceptual 
analysis, and as {\em a posteriori} theory-construction, respectively. I do 
not even claim this, when one considers quantum theory's description of 
atomic particles as ``wave-like'' and evanescent. My reason is that (as I 
said in Section \ref{sssec;vsbracket}) it remains  mysterious how the 
macroscopic world, with its persisting objects ``emerges'' from the quantum 
realm---despite impressive recent  progress in understanding the physics of 
decoherence (Section \ref{sssec:Appealbeyondclassl}). This mystery means 
that it is, at least {\em today}, impossible for anyone to state precisely 
what is the  ``supervenience basis'' for macroscopic objects' persistence. 
In short, the jury is still out, scientifically as well as philosophically. 
The most we can now claim is that the RDA fails, and that perdurantism is, 
so far, tenable.

I will prepare the ground for these three conclusions, by first discussing: 
(i) the RDA in more detail (Section \ref{sec2;argtkindsreply}); (ii) some 
metaphysicians' replies to it (Section \ref{ssec;causevelrot}). Besides: 
though Sections \ref{sec2;argtkindsreply} and \ref{ssec;causevelrot} are 
written from the perspective of the philosophy of physics, they will exclude 
technical physics, especially relativity and  quantum theory. (Indeed, so 
will Sections \ref{ssec;describerot} onwards,  to a large extent.)

 More specifically, nothing in  Sections \ref{sec2;argtkindsreply} and 
\ref{ssec;causevelrot} contradicts the metaphysicians' prevalent assumption 
which I labelled ({\em Bracket}): that classical mechanics, or perhaps 
instead the whole of classical physics, provides a supervenience basis for 
persistence. (But of course, nothing I say depends on this assumption, which 
I reject.) To that extent, these two Sections should be of interest to 
friends of that assumption. In effect, these Sections report some of the 
themes and arguments of the RDA literature, from the perspective of the 
philosophy of physics. So it is only in Sections \ref{ssec;describerot} 
onwards that the claims of Sections \ref{sssec;vsStrtfwd} and 
\ref{sssec;vsbracket} come to the fore: in Section \ref{ssec;describerot}, I 
will ``re-admit'' relativity theory; and in Section 
\ref{sssec:Appealbeyondclassl}, I re-admit quantum theory. 

Finally, by way of prospectus: it may help to announce what my answers will 
be to the following questions  about rotation and continuous matter, which 
are  obviously relevant to evaluating the RDA. As to rotation, one naturally 
asks:\\
\indent (1) How exactly does physics, or more specifically a given physical  
theory, describe rotation?\\
\indent (2) Do endurantist and perdurantist have equal rights to that 
description?\\
\indent (3) Does that description supply some differences between rotation 
and non-rotation which the endurantist's RDA has ignored, but which the the 
perdurantist can appeal to? Or does it strengthen the RDA?\\
\indent \indent  And about continuous matter, one naturally asks:\\
\indent (4) Must a theory of persistence allow that continuous matter is 
possible, and so address the RDA's distinction between the two discs? Or 
could it legitimately set aside continuous matter, and so duck out of 
discussing the  RDA? (As we have seen: the metaphysical literature says Yes 
to the first question;  and tends to explicitly set aside atomism and 
especially  quantum theory.)

My answers to these questions will be broadly as follows.\\
\indent (1): I will report in Section \ref{ssec;describerot} how physics 
describes rotation, including some peculiarities of rotation in our best 
theory of space and time, viz. general relativity.\\
\indent (2) I will allow in Section \ref{133Allowboth} that endurantist and 
perdurantist have equal rights to this description; though this is largely 
for the tactical reason of giving the RDA as good a run as possible.\\
\indent (3) This description has both a positive and a negative  implication 
for  the RDA. The positive implication is that  the RDA can be formulated 
more strongly than is usually recognized; (my first conclusion, Section 
\ref{134Goodbadunnecy}). But within general relativity, the RDA fails: 
frame-dragging implies that there are differences between rotation and 
non-rotation which the perdurantist can appeal to; (my second conclusion, 
Section \ref{432:RDAfailsGR}).\\
\indent (4) Though I maintain that classical continua are subtler and more 
problematic than usually recognized, I agree that they are indeed logically 
possible. And I therefore agree that a philosophical  theory of persistence 
should if possible allow for continuous matter, and so not duck out of 
discussing the RDA. But as stressed, I will go on to maintain (Section 
\ref{sec:soln} onwards) that in fact a perdurantist theory can reply to the 
RDA, by endorsing its distinction between the discs.

\section{The RDA and kinds of reply---in detail}\label{sec2;argtkindsreply}
In this Section, I first present the RDA more fully than in Section 
\ref{ssec;theRDA}; (Sections \ref{sssec;track} and \ref{sssec;whycts}). Then 
I consider what the RDA implies for endurantism (Section 
\ref{sssec;tuquoque}).  Finally, in Section \ref{sssec;tworeplies}, I 
distinguish the two kinds of perdurantist  reply to it, more fully than in 
Section \ref{psa; against the consensus}.

\subsection{The argument: keeping track of homogeneous 
matter}\label{sssec;track}
Here are two distinct possibilities for a perfectly circular disc made of 
homogeneous matter: where `homogeneous' means that the properties of the 
matter do not vary across space even on the smallest length scales: \\
\indent (Stat): that it is stationary, and in particular  not rotating about 
an axis perpendicular to the plane of the disc; (of course this possibility 
can be subdivided as regards the disc's size, and the properties of its 
matter, e.g. its mass-density):\\
\indent (Rot): that it rotates about this axis; (of course this possibility 
can be further subdivided, apart from the subdivisions in (Stat), viz. as 
regards the angular velocity  of the disc.)

It seems that  the endurantist can easily recognize and describe the two 
possibilities, according to whether the very same non-circularly-symmetric 
part, e.g. a segment, is in the same place at two times. But I postpone 
considering the endurantist until Section \ref{sssec;tuquoque}.  For the 
moment, consider the perdurantist. It seems she has a problem: surely she  
must say that all the relations (and therefore, all her proffered ``suitable 
relations'' for analysing, or at least subvening, persistence) between two 
stages (temporal parts) of the disc, say at noon and 12.01, are the 
same---whether the disc is rotating or not? And similarly  for stages at the 
two times of any spatial part of the disc, such as a segment: surely 
perdurantism  must say that all the relations are the same?

More precisely: the phrase `any spatial part of the disc, such as a segment' 
seems to presuppose persistence---which is precisely in question here. So a 
better way to put the second rhetorical question is to say: Similarly  for 
any spatial part of the disc-at-noon, such as a segment, and {\em any} 
congruent subvolume of the disc-at-12.01: surely the perdurantist  must say 
that all the relations are the same, whichever of the many congruent 
subvolumes of the disc-at-12.01 are chosen?

In the sequel, it will sometimes be useful to have mnemonic labels for the 
temporal parts being compared. So let me express perdurantism's apparent 
problem by using some memorably ugly labels, as follows. For {\em any} four 
choices of spatially congruent temporal parts of the discs:\\
 \indent  \indent a spatial segment of the stationary disc at noon, call it 
StatNoon;\\
 \indent  \indent a congruent spatial segment of the stationary disc at 
12:01, call it StatMin;\\
 \indent  \indent a congruent spatial segment of the rotating disc at noon, 
call it RotNoon;\\
 \indent  \indent a congruent spatial segment of the rotating disc at 12:01, 
call it RotMin;\\
StatNoon and RotNoon match in their  properties; as do StatMin and RotMin; 
and StatNoon bears to StatMin exactly the same relations as RotNoon does to 
RotMin.

\indent  I have phrased this argument  so as to allow:\\
\indent \indent (i) the properties of spatial parts of the discs to vary, 
provided they vary in a circularly symmetric way, e.g. by each disc being 
decorated with circles of colour, centred on the centre of the disc;\\
\indent \indent  (ii) the discs' properties to change over time, provided 
that the two discs always match, e.g. the discs could have a temperature, 
even a circularly symmetric distribution of tempearture, and could cool 
down, provided temperatures always match.

\subsection{Why continuous and homogeneous matter?}\label{sssec;whycts}
Let us first ask: Why does the RDA use a disc composed of continuous and 
homogeneous, rather than atomistic, matter? Most of the philosophical 
discussions do not say why. But  the implicit  answer is: 
\begin{quote}
Because if the disc is atomistic (i.e. a swarm of point-particles), the set 
of spatial points occupied by matter varies over time, or stays constant, 
according as the disc rotates or not; so that the perdurantist  can  
distinguish rotation from non-rotation by ``following'' which spatial points 
are occupied by matter at which times. Similarly, if the disc's matter is 
continuous but inhomogeneous, the perdurantist  can distinguish the cases by 
following lines of qualitative similarity. But for continuous homogeneous 
matter, the perdurantist is stuck: she cannot distinguish the cases. 
\end{quote}
This answer raises three issues. The first is straightforward, independent 
of the distinction between the discs, and is largely a matter of setting 
some matters aside. The second and third are important for us, and will need 
more attention later on.

\subsubsection{Tracking matter}\label{sssec;track2}
The idea of the answer  is twofold. First, the answer concedes that the 
perdurantist can provide an account of persistence (in other jargon: a  
diachronic criterion of identity) for a point-particle in  a void. The 
criterion is just to follow the continuous  curve of the presence of mass 
(or perhaps of charge; or more generally,  of qualitative similarity). The 
ambient void means that starting from a point-particle at a time, there is a 
unique way to go forward or backward in time. Similarly for a point-particle 
moving, not in a void, but in a continuous fluid with suitably different 
properties---a different ``colour'', or made of different ``stuff'', than 
the point-particle; (cf. the discussion of {\em Follow} in Section 
\ref{ssec;theRDA}). \\
\indent But second: in continuous matter, there is no void---the lines of 
matter-occupation run ``every which way''. And it seems that if the matter 
is also homogeneous, then even the lines of qualitative similarity, however 
they are exactly defined, run every which way---leading to the argument's 
challenge to the perdurantist.

This second point will of course preoccupy us in what follows. Here I just 
make three ancillary remarks about the first point. Though straightforward, 
they have the merit of showing the scope of the sort of criterion that says 
``follow the lines of matter-occupation or qualitative similarity''. (Recall 
also from footnote 2 in Section \ref{ssec;theRDA} that this sort of 
criterion  can be disputed; but that dispute is not directly relevant to the 
RDA, and I set it aside.)\\
\indent (i): This sort of criterion will also work for extended objects 
moving through a void, or through  a suitably different continuous  
fluid---provided it is understood as applying only to such an object as a 
(spatial) whole. For of course, applying it to the spatial  parts of an 
extended object (e.g. parts of a rigid homogeneous sphere) just resets, on a 
smaller scale, the problem first posed by the RDA. (So I  agree that the 
perdurantist who considers extended  objects will have to  add to this sort 
of criterion some account of the objects' parts.)\\
\indent (ii): This sort of criterion even works for a homogeneous  rotating 
disc as a whole, if it is not {\em perfectly} circularly symmetric. Imagine 
a disc made of Lego---of {\em rectangular} Lego blocks: it is approximately 
circular, and may be treated as circular for certain purposes, e.g. if 
looked at from a sufficient distance, and-or if sufficiently larger than the 
individual blocks that the edge can be treated as smooth. Now imagine a disc 
of exactly the same shape, but which is homogeneous (no bricks!). Since the 
disc's edge is in fact rough, our ``follow the lines'' sort of criterion 
works:  exactly tracking the lines of matter-occupation or qualitative 
similarity at the edge reveals whether or not the disc rotates.  Similarly 
of course for all actual rotating objects: they are not perfectly circularly 
symmetric, and so the spatially varying qualitative  features, such as a 
roulette wheel's numerals, suggest the correct way the  ``identify'' spatial  
parts across time (i.e. to define persistence)---so that the challenge of 
the RDA does not arise.\\
\indent (iii): On the other hand, exact homogeneity is not needed for the 
RDA. In my version above, I allowed the spatial properties to vary in a 
circularly symmetric way. If they do, the lines of qualitative similarity 
will have to be  circularly symmetric: but the challenge to the perdurantist  
will remain, since there nevertheless seems to be an abundance---a 
continuous infinity---of such lines. How can the perdurantist specify those 
that define persistence?

\subsubsection{The persistence of spatial 
points}\label{sssec;persesptlraised}
The second issue  is that the  answer above glosses the distinction between 
rotation and non-rotation intuitively. It presupposes that there are 
persisting spatial points, so that it can say: only in the rotating disc do 
the point-sized bits of matter occupy different spatial points at different 
times. This prompts the question: What account is to be given of the 
persistence of spatial points?  This question is very important to 
evaluating the RDA, but is almost entirely ignored in the metaphysical 
literature: even in the best discussions, authors often appeal without 
further analysis to the idea of `the same place' (e.g. Hawley 2001, p. 85). 
This question will take centre-stage in Section \ref{ssec;describerot}, 
where I leave metaphysics for the philosophy  of space and time and the 
physics of rotation.

\subsubsection{The accompaniments of 
rotation---again}\label{sssec;accompanimtsrotnagain}
The third issue arises from the last sentence of the answer. That sentence 
is contentious: and (unlike the assumption of persisting spatial points) it 
{\em is} contested in the metaphysical literature---as I reported in Section 
\ref{sssec;accompanimtsrotn}. And as I also announced there, my own view 
will be that:---\\
\indent (i): the RDA should not just set aside the technical accompaniments 
(this goes along with the importance of the technical description of 
rotation, just announced in Section \ref{sssec;persesptlraised}); \\
\indent (ii): it cannot do so just by stipulating perfect rigidity; and\\
\indent (iii): the perdurantist can reply to the RDA without resorting to 
distinctively metaphysical proposals such as immanent causation, or special 
vectorial properties. (I shall also join some perdurantists such as Sider in 
accepting appeal to everyday causes, effects and counterfactuals. I discuss 
Sider's position in Section \ref{sssec;Sider}; and develop an analogue of 
his position in Section \ref{sssec:detailclassl}.)

So, to sum up this presentation of the RDA:--- Its strategy is clear. It 
needs to ``imagine away'' enough of the usual accompaniments of rotation 
(and-or non-rotation) to make it plausible that the perdurantist (or Humean) 
has trouble making distinquishing the discs.  This Subsection has developed 
the first main example of this strategy: the RDA imagines a perfectly 
circular and perfectly rigid disc, made of continuous and perfectly 
homogeneous matter; thereby aiming to block the perdurantist from  
``tracking'' matter through the void, or ``following'' lines of qualitative 
similarity.

\subsection{Tu quoque?}\label{sssec;tuquoque}
So far as I know,  the RDA literature never considers whether the rotating 
discs harbour any problems or projects for the endurantist (or more 
generally, non-Humean). I think this is a mistake. Surely the endurantist 
owes us a discussion of diachronic criteria of identity for the spatial 
parts of a  piece of  homogeneous matter, such as the disc: a discussion 
that will secure the distinction. Of course, it is not my brief here to 
develop endurantism. But as a preliminary to considering perdurantist 
replies to the RDA, it is worth discussing the factors that combine to make 
us forget that the endurantist owes us such a discussion.

First, we easily slip into relying on intuitive judgments of sameness of 
place. But as we saw in Section \ref{sssec;persesptlraised}, it is not 
enough for the endurantist to  say just that  the difference between the 
possibilities is a matter of whether the worldlines of the enduring pieces 
of matter are straight or helical. The intuitive contrast ``straight vs. 
helical'' depends on the idea of persisting spatial points. The endurantist 
owes us an account of this idea, just as much as the perdurantist does. 
Either this idea must be vindicated, or some other (maybe more technical) 
notions that describe rigorously the distinction between rotation and 
non-rotation must be invoked and justified.   This is an endeavour which 
leads  into issues in the philosophy of space and time, and the physics of 
rotation. I will discuss these issues in Section \ref{ssec;describerot}. In 
fact, I will there allow that as regards these issues, the honours are even, 
or roughly even, between the endurantist and perdurantist. That is, I will 
allow that both sides have equal right to the notion of persisting spatial 
points, or to whatever  notions are needed to describe rigorously the 
rotation/non-rotation distinction.  I say `allow', because my reason will be 
that since I want to give endurantism and the RDA  as good a run as 
possible,  I  give them the benefit of the doubt about their rights to these 
notions.  

On that assumption, it is tempting to think, as the discussion in Section 
\ref{sssec;track} implicitly did, that only the perdurantist ``has work to 
do''. More precisely: it is tempting to think that:\\
\indent (i): the perdurantist needs to say  what, in terms of qualitative 
similarity or causation or whatever, distinguishes the correct worldlines 
from all the other  spacetime worms (mereological fusions of stages) they 
accept as objects;\footnote{How difficult this is, how much work there is to 
do, will of course depend on their other views. In particular, perdurantists 
who are Humean about causation, like Lewis, will presumably have more work 
to do, in distinguishing the discs, than perdurantists who are not, like 
Armstrong; cf Section \ref{sssec;appealcause}.} while\\
\indent (ii): the endurantist can take the distinction between the correct 
and incorrect worldlines (straight vs. helical) as ``bedrock'':  no more can 
be said, and besides, no more needs to be said.

\indent  I think this temptation arises from a widespread  belief that only 
for the endurantist does a persisting object remain self-identical over 
time. This belief leads to the idea that---at least for the spatial parts of 
a  piece of  homogeneous matter---diachronic criteria of identity are 
unnecessary, or even unintelligible: i.e. the idea that the identity over 
time of such parts is just ``good old identity'', and is both unanalysable, 
and in no need of analysis---it is as clear as crystal!

\indent  But this belief is {\em false}. Sider (2001, p. 54-55) exposes the 
error: also for the perdurantist, the persisting object is genuinely 
self-identical over time. (Sider seems to forget this insight on p. 226, 
para 2, when he endorses assumption (i) above, i.e. says the endurantist has 
an `easy answer' about how to distinguish the discs.) 

So I think a good case can be made that the endurantist also has work to do 
(even after persisting spatial points, or whichever  notions are used to 
describe the rotation/non-rotation distinction, are in play). I agree that 
it is unclear exactly what sort of account of matter's identity over time, 
the endurantist is to give. But in what follows, I shall not go further into 
this: developing endurantism is not my brief. Suffice it to make three 
remarks:\\
\indent \indent (i): I think part of the reason for the obscurity is that it 
is unclear exactly how to formulate endurantism (Sider 2001, p. 63-68).\\
\indent \indent (ii): Some endurantists agree that some such account is 
needed. For example, Zimmerman, after developing a detailed account of 
immanent causation for the perdurantist (1997, p. 449-456), argues that the 
endurantist should accept some parts of it; (roughly speaking, for histories 
of enduring objects: p. 456-459). For more on immanent causation, cf. 
Section \ref{sssec;appealcause}. \\
\indent \indent (iii): I think endurantists are likely to see what 
perdurantists say in order to distinguish the discs as so complex, as to 
amount to a serious disadvantage for perdurantism. In particular, they are 
likely to accuse Sider's position (and mine) of this. My response will be, 
in effect, that endurantists will themselves need them to {\em say} much of 
what the perdurantist says: though for them it may well be collateral 
information about the distinction between the discs, rather than (as for the 
perdurantist) ``constitutive'' of the distinction. 

\subsection{Kinds of reply}\label{sssec;tworeplies}
In this last Subsection,  I will develop Section \ref{psa; against the 
consensus}'s two kinds of reply to the RDA. This  will also bring out a 
complaint implicit in Section \ref{sssec;persesptlraised}'s question about 
the account of space and rotation. Namely, that as Callender (2001, p.30) 
puts it: `analysing RDA is frustrating because the possible worlds described 
are left so vague'.

A preliminary remark. I admit that my distinction is not exhaustive: there 
are other possible replies.  In particular, Teller (2002, p. 207-208) gives 
a reply which, though at a glance similar to my first kind, is much more 
radical. He suggests the perdurantist, or at least the advocate of Humean 
supervenience, should say that even for a inhomogeneous disc like a roulette 
wheel, there is nothing objectively right about defining persistence in 
terms of ``tracking'' spatially varying qualitative  features such as one of 
the wheel's numerals. Teller agrees that the perdurantist can and should 
accept that:\\
\indent  (i) it is convenient to ``identify'' parts across time (i.e. to 
define persistence) on the basis of such features; and \\
\indent (ii) this is convenient because of its association with what Teller 
calls  `rotational phenomena': i.e. what I called `accompaniments of 
rotation', e.g. having been pushed by someone, and the tendency of rotating 
solid objects to become oblate.\\
\indent But, according to Teller, perdurantists should not accept that any 
disc, even an inhomogeneous one, ever has any `literal rotation ... for them 
there is only rotation by courtesy' (p. 208). (So though the perdurantist 
might call being pushed, oblateness etc. `rotational phenomena', she should 
not  call them, as we normally do, `typical causes and effects of rotation': 
for that suggests there is literal rotation.)\\
\indent Teller's reply certainly blocks the RDA: but, by my lights, at {\em 
far} too high a price.  Its denial that there are any facts of persistence 
in the countless unproblematic cases (e.g. of inhomogeneous discs), facts 
which perdurantism must accept, amounts to a sort of nihilism about 
persistence---which I find incredible: but I will not argue against it 
here.\footnote{Note that it is much more radical than both the non-reductive 
perdurantism mentioned in (1) of Section \ref{ssec;naturalism}, and the (No 
Difference) reply discussed below. Both these positions accept the facts of 
persistence in the countless unproblematic cases.}

Though my distinction between replies is not exhaustive, it is natural. We 
saw in Section \ref{sssec;whycts} that the strategy of the RDA is to imagine 
away enough of the usual accompaniments of rotation to make it plausible 
that the perdurantist (or Humean) has trouble distinguishing the discs.  
So in reply, the perdurantist and Humean can either\\
\indent (i) say  there is no difference: too much has been imagined away for 
there to be a difference remaining; or\\
\indent (ii) say that in fact, they have the wherewithal to describe the 
difference.\\
These are my two kinds of reply, which I label `(No Difference)' and 
`(Appealing Differences)'. Of course, a perdurantist  can in a sense combine 
them. For in the back-and-forth of debate, she might move from one  reply to 
the other: `well, if you imagine away {\em all of those} accompaniments of 
rotation, I will then reply that there is after all no difference'. I turn 
to stating the two replies in more detail.

{\bf (No Difference)}: There is no good reason to distinguish the 
possibilities (Stat) and (Rot). More precisely: though there is of course a 
distinction between rotating and non-rotating discs, a distinction 
manifested in various differences between the discs---recall the usual 
accompaniments of rotation---the RDA needs to assume that {\em its} discs, 
in the possibilities (Stat) and (Rot), do not manifest {\em any} of these 
differences. And {\em then} there is  no reason to distinguish the discs.\\
\indent Note that this reply does {\em not} need the differences to in some 
sense ``ground'' the rotation/non-rotation distinction. It is enough that 
the differences exist, and so could be mentioned in the perdurantist's 
prospective definition of persistence in such a way that the definition 
yields for each disc its correct (straight or helical) lines of 
persistence/worldlines. That is: this is enough, as regards replying to the 
RDA. Of course, knowing these differences does not by itself tell us how to 
frame the definition.\\
\indent  Similar remarks apply, when the RDA's target is not perdurantism, 
but some other neoHumean doctrine such as Lewis' Humean supervenience (1986, 
p. ix-x; 1994, p. 474). I will discuss this in more detail in Section 
\ref{sssec;LewRob}. Here I only note that also with this target, the RDA 
needs to imagine away any differences between the discs to which the 
neoHumean could appeal.

\indent So far as I know, Callender (2001) is the main example of this 
reply; (he focusses on Humean supervenience, rather than perdurantism, but 
he also uses the label `no-difference'). Lewis  also gives essentially the 
(No Difference) reply, again taking the RDA to have as its target Humean 
supervenience not just perdurantism   (1986, p.xiii, 1994, p. 475). But he 
later changed his mind, endorsing a proposal of Robinson (1989).  And since 
I want to discuss that proposal only after discussing velocities, I shall 
postpone Lewis' views, and the comparison of Lewis with Callender,  till 
then (Section \ref{sssec;LewRob}).\\
\indent For the moment, I just bring out the flavour of the (No Difference) 
reply by reporting Callender's analogy between the discs and the up/down 
distinction. He says the distinction between (Stat) and (Rot) is as spurious 
as the distinction between\\
\indent (Up): an arrow in an otherwise empty world pointing up; and\\
\indent (Down): an arrow in an otherwise empty world pointing down:\\
which all   agree to be  a distinction without a difference, since there is 
no up/down distinction except with reference to some other direction, in 
particular the direction of the local gravitational force. (At least: 
nowadays, if not in Aristotle's day, all agree to this.)\\
\indent  Callender makes the analogy between the disc-worlds and the 
arrow-worlds closer, by:\\
\indent \indent (i): imagining the discs to be each alone in its world; 
and\\
\indent \indent (ii) saying `Assuming Newtonian spacetime with its absolute 
standard of rest is not crucial to the argument, a harmless change of 
coordinates will change our case [viz.: one disc rotating, the other not] 
into one with one disc rotating clockwise, and the other rotating 
counter-clockwise [i.e. with equal speeds]' (p. 32).\\
\indent So: since the clockwise/counter-clockwise distinction depends on a 
choice of direction (a clock-dial moves counter-clockwise when seen from 
behind!), this really is a distinction without a difference, just like (Up) 
vs. (Down).

{\bf (Appealing Differences)}: According to this kind of reply, the 
perdurantist (or Humean) {\em can} distinguish the possibilities. That is: 
even supposing that the RDA stipulates that its discs do not manifest the 
{\em usual} differences between rotation and non-rotation---so that the 
argument seems to get a grip---there are differences the perdurantist (or 
neoHumean) will find acceptable---even appealing!---and can appeal to. In 
short: there are more things in the (perdurantist or Humean) heaven and 
earth than are dreamt of by the RDA's advocates.

\indent  This reply is much more common than (No Difference). Indeed, so far 
as I know, Hawley (1999, p. 55-56; 2001, p. 74-76) is the only author, apart 
from Callender and Lewis, who considers the (No Difference) reply at any 
length: (but she believes it defective, and advocates a version of 
(Appealing Differences)).

\indent I think the reason (Appealing Differences) is more common lies in 
the facts noted in Section \ref{sssec;accompanimtsrotn}. Namely:\\
\indent  (i): The metaphysical literature  concentrates on the everyday, not 
technical physical, accompaniments of rotation. And: \\
\indent (ii): The usual everyday differences that the RDA stipulates to be 
absent involve present and ``occurrent'' accompaniments of rotation, such as 
a roulette wheel's numeral, or a spot of paint, moving relative to the 
disc's environment.\\
\indent (iii): This leads the metaphysical literature to focus on whether 
the perdurantist can legitimately appeal to:\\
\indent \indent (a): differences in past or future or counterfactual 
everyday accompaniments of rotation; such as having been pushed in the past; 
or that if a spot of paint were sprayed on the disc, it would move; and-or 
\\
\indent \indent (b): differences that are distinctively metaphysical 
(neither everyday nor technical physical), such as a special relation of 
immanent causation, or special vectorial properties that are numerically 
equal to, yet different from, velocities.\\
\indent (iv): The issues raised in (a) and (b) are familiar to 
metaphysicians.
 
\indent  In the next Section, I shall discuss various examples of this 
reply. But by no means all. I shall concentrate on distinctively 
metaphysical differences, i.e. (iii) (b). But even there I will omit some 
views, e.g. Hawley's proposal there are relations between temporal  parts 
that are not supervenient on the intrinsic natures of the parts that are the 
relata, and yet are not just spatiotemporal relations (the paradigm case of 
such non-supervenient relations: 1999, p. 60, 63-66; 2001, p. 85-90).

\section{Some metaphysical replies}\label{ssec;causevelrot}
In this Section, I give a survey of some metaphysicians' replies to the RDA, 
emphasising points that will be important later on (or that I think 
important in themselves!). For the most part, these replies are examples of 
(Appealing Differences). I begin with two such examples: appealing to 
causation (Section \ref{sssec;appealcause}), and appealing to velocities 
(Section \ref{sssec;appealvel}). I reject the first, but am more sympathetic 
to the second. This leads me to consider a third example of (Appealing 
Differences): Lewis' and Robinson's appeal to a quantity analogous to (but 
different from!) velocity; and Zimmerman's reply to that proposal (Section 
\ref{sssec;LewRob}). Finally (Section \ref{sssec;Sider}), I discuss Sider's 
reply, which combines (Appealing Differences) and (No Difference). I discuss 
Sider in some detail, since his reply is in effect my ``fallback position'': 
if my own reply failed, I would endorse an analogue of his (Section 
\ref{sssec:detailclassl}).

Much of this Section involves the distinction between intrinsic and 
extrinsic properties---a distinction which I have so far just mentioned. 
Though intuitively compelling, this distinction is controversial, and in 
particular hard to analyse. By and large, I will not need contentious claims 
about it. But I will argue (in Section 4.2.2.C) that to assess the RDA, it 
is worth distinguishing (though the literature has not done so):\\
\indent (i): degrees of extrinsicality;\\
\indent (ii): whether predicating an extrinsic property has implications for 
other times, or for other places (which I will call `temporal 
extrinsicality' and `spatial extrinsicality' respectively).    \\
(It will also be obvious that proposals (i) and (ii) might also be useful 
for other problems in metaphysics.)

\subsection{Appealing to causation}\label{sssec;appealcause}
 Much discussion of the RDA as an argument against perdurantism concerns the 
relation between persistence and causation within the persisting object. I 
shall note four  points, in (a)-(d), and then in (e) express my scepticism 
about appealing to causation.
 
\indent (a): {\em The Idea}$\;\;$ Quite apart from the RDA, many 
philosophers take causation to somehow underpin persistence. For in some 
puzzle cases, causation seems to be what is needed for persistence: instead 
of, or in addition to, qualitative similarity. A simple oft-cited example is 
the imaginary case in which a god destroys an object and immediately 
replaces it with a qualitative replica: it seems that what is missing in 
such a case of non-persistence are causal relations between the (states of 
the) destroyed object and the replica. Some philosophers call such causal 
relations (perhaps together with special doctrines claimed about them) 
`immanent causation''. (For details and references, cf. the discussion of 
Armstrong after (d) below, and e.g. Zimmerman (1997, p. 435-437).)

\indent (b): {\em An obvious reply?}$\;\;$ If so, the obvious reply to RDA 
is to appeal to whether or not there is (an appropriate sort of) causation 
between the given stages of spatial parts of the disc. Using the ugly labels 
of Section \ref{sssec;track}: only if StatMin is  chosen so as to comprise 
the same matter as does StatNoon will there be (the appropriate sort of) 
causation between them; and similarly for RotMin and RotNoon. In short: the 
obvious reply is to deny the RDA's claim that for {\em any} choices of the 
four stages, StatNoon bears to StatMin exactly the same relations as RotNoon 
does to RotMin: the causal relations {\em are} sensitive to the choices. \\
\indent But this reply is ``a bit quick'', for two reasons. That is: there 
are two problems about appealing to causation to subvene, or even yield an 
analysis of, persistence. The first problem, in (c), is much more often 
discussed, and so presumably thought more important; but I shall later 
develop ideas from the second problem, presented in (d).

\indent  (c): {\em Trouble for Humeans}$\;\;$  Some perdurantists want to 
endorse some broadly Humean account of causation. (Indeed, for some, 
Humeanism is a leading motivation for their perdurantism.) For such 
perdurantists, the RDA still threatens. For  surely, once we restrict 
attention to properties and relations that are intrinsic (or 
``qualitative'', or ``occurrent''---or whatever the Humean regards as 
characterizing their supervenience basis for causation and so persistence), 
the properties and relations within the two pairs, \{StatNoon,StatMin\}  and 
\{RotNoon,RotMin\}, do match for {\em any} choices of the four stages---as 
the RDA alleged. So for a Humean, the causal relations should also match: so 
the RDA seems to show that  perdurantism is incompatible with a broadly 
Humean account of causation. (For a fuller exposition, cf. Zimmerman's 
discussion of `Humean supervenience of the Causal Relation' (called `(HS)'); 
1998, p. 271.)

(d): {\em Causation and motion}$\;\;$ The second problem is a threat of 
circularity, arising from connections between the notions of causation and 
motion. As Shoemaker (1979, p. 328) puts it: `it seems very unlikely that we 
can specify the relevant causal relationships without invoking the notion of 
motion and with it the notion of cross-temporal identity (i.e. persistence) 
... which we are trying to analyse.'\\
\indent But (as Shoemaker goes on (p. 329-330) to describe) there seems to 
be a way in which this {\em can} be done. Namely, one adopts the following 
two-stage procedure.\\
\indent \indent (i): One can apply the concept of motion---and its 
associated quantitative concepts like average and instantaneous  velocity or 
acceleration etc.---to an arbitrary spatiotemporally continuous series of 
momentary stages.\\
\indent \indent (ii): Only then does one appeal to causation to underpin 
persistence. That is: One now assumes that the worldlines or worldtubes of  
persisting objects are distinguished by the stages of each of them having 
(maybe: the ancestral of) some suitable relaton of causation (maybe of 
immanent causation)).\\
\indent Shoemaker's two-stage procedure is rarely discussed; but (so far as 
I know) is endorsed by those who do discuss it; for example,
Zimmerman (1998, p. 279-280). But Zimmerman goes on to emphasise the first 
problem, (c) above. That is, in the context of the RDA, with its two pairs 
of spatial parts of discs, one pair causally related and the other not:  the 
appeal in stage (ii) to causal relations surely requires non-Humeanism about 
causation.

\indent I myself agree that Shoemaker's stage (i) works. One can certainly 
apply the  concept of motion and its associated quantitative concepts, as 
made precise by differential geometry, to:\\
\indent \indent (a) any worldline, or to a foliated worldtube (in the latter 
case, one would  assign the velocity and acceleration vector to, say, the 
centroid of each of the foliation's leaves); and even to\\
\indent \indent  (b) spacelike curves and tubes (though for these cases, one 
might resist using the usual language of motion, e.g. calling the tangent 
vector of a spacelike curve a `velocity').\\
 \indent Indeed, textbooks of modern spacetime theories contain countless 
examples of  (a) and (b). Agreed,  the notions of  velocity, acceleration 
etc., as usually understood,  presuppose the notion of persistence; (as I 
conceded in Section \ref{ssec;intrpropvelies} and will discuss in Section 
\ref{sssec;appealvel}). But Shoemaker's procedure does {\em not} conflict 
with that understanding. His stage (i) applies such notions, stripped of 
that presupposition; (how this works will become clearer in Section 
4.2.2.C). Then, after stage (ii),  notions like the instantaneous  velocity 
of spatial parts of the discs are to be reinstated, ``piggy-backing'' on the 
relation of causation. That is: the  notions as usually understood are 
applied just to the worldlines or worldtubes that stage (ii), i.e. 
causation, picks out.\\
\indent But on the other hand, turning to Shoemaker's stage (ii):  I fear 
that it may fail, because causation  might presuppose persistence in a 
different way than that Shoemaker alerts us to, i.e. in a way  independent 
of the notion of motion---details below.

\indent (e): {\em Armstrong, Lewis and a warning}$\;\;$ The issues raised in 
(a)-(d), especially (a)-(c), are illustrated in  many perdurantists' 
discussions of the RDA, e.g. Armstrong's and Lewis'. Armstrong is not a 
Humean about causation and so endorses the reply in (b): he believes the 
perdurantist can and should reply to the RDA by appealing to causal 
relations between stages, and spatial parts of stages (1980, 1997, 
pp.73-74). More precisely: he thinks that,  quite apart from the RDA, the 
perdurantist should take persistence---the ``suitable relations'' between 
stages of a persisting object---to be a matter of what (following Broad and 
W.E. Johnson) he calls `immanent causation': `a form of causality which 
remains confined to a single particular and that, further, does not proceed 
by interaction between sub-particulars' and which involves `the actual {\em 
bringing into existence} of later by earlier temporal parts' (1997, 
pp.73-74). So for Armstrong, the moral of the RDA is simply that immanent 
causation does {\em not} supervene on the intrinsic natures of the relata: a 
conclusion which, as a non-Humean about causation in general, Armstrong is 
happy to endorse.\footnote{So far as I know, the fullest account of immanent 
causation is Zimmerman (1997), which builds on his (1995).} 

\indent On the other hand, Lewis is a Humean about causation. (More 
precisely: he defends a counterfactual analysis of causation, which makes 
the truth-values  of counterfactuals, and so of causal statements, supervene 
on the qualitative nature of the world (1979, 1986a, p.22).) But he also 
agrees with (a) above that causation is crucial to understanding 
persistence: as he says in his last discussion of the RDA: `the most 
important sort of glue that unites the successive stages of a persisting 
thing is causal glue' (Lewis 1999, p. 210). So Lewis cannot endorse the 
reply  in (b) above, and has to reply to the RDA in another way; cf. Section 
\ref{sssec;LewRob}.

But I myself am wary about appealing to causation to solve a metaphysical 
problem, since, like many philosophers of science, I think the notion is too 
problematic to be relied on. (Recent discussion of its problems include 
Hitchcock 2003, Norton 2003.)  More specifically, analytic metaphysicians  
should take heed when their doctrines, e.g. about persistence, carry 
contentious commitments about causation. For example: Dowe  (2000) argues 
for a process theory  of causation which explicitly assumes the notion of a 
persisting object. So a metaphysician who appeals to causation to analyse or 
at least subvene persistence is commited to Dowe's theory being  wrong: not 
a commitment to be entered into lightly! (This warning is not a universal 
accusation: some metaphysicians are admirably explicit about their 
commitments about causation; for example Zimmerman (1997, p. 444-449, 
464-465).)

\subsection{Appealing to velocities}\label{sssec;appealvel}
On first meeting the RDA, most people's response is that since the two discs 
differ in their instantaneous angular velocity (similarly: corresponding 
spatial parts of them differ in instantaneous velocity), the perdurantist 
should  reply to the argument by attributing instantaneous angular velocity 
to the stages (or similarly: instantaneous velocity to spatial parts of 
stages).\\
\indent But as I said in Section \ref{ssec;intrpropvelies},  there is a 
consensus in the RDA literature against this tactic. The consensus urges 
that the notion of  velocity  presupposes the notion of persistence, so that 
appealing to velocity brings circularity; and that this is so, both for the 
usually notion of velocity, and a heterodox notion advocated by Tooley and 
others.\\
\indent I shall first report this consensus (Section 
\ref{sssec;againstvel}); and then present three replies to it, in ascending 
order of  importance (Section \ref{sssec;againstconsensus}). The second and 
third replies will foreshadow Sider's and my own replies to the RDA. And the 
third reply will develop my denial of {\em pointillisme} (announced in 
Section \ref{sssec;vsStrtfwd}).

\subsubsection{The consensus against velocities}\label{sssec;againstvel}
The consensus against appealing to velocities  relates to metaphysics rather 
than technicalities of physics. It uses only ``naive'' notions of average 
and instantaneous velocity, thereby implicitly assuming a space of 
persisting spatial points. So as I did in Section \ref{sssec;tuquoque}, I 
will postpone till Section \ref{ssec;describerot} the question how to 
describe rigorously the distinction between rotation and non-rotation: 
whether by invoking  a space of persisting spatial points, or by invoking 
some other  notions. 

The first, and main, point of the consensus is that  if velocity is 
understood, as usual, in terms of spatial separations of the places occupied 
at different times by one and the same object, then the notion of velocity  
{\em assumes} the idea of persistence. (This goes with the so-called 
`Russellian theory of motion', also called the `at-at theory of motion'.) 
This is obvious for the elementary definition of average velocity as the 
quotient of distance traversed and time elapsed. And the notion of 
instantaneous velocity is usually understood ``just'' as a limit of such 
quotients, so that it also assumes the idea of persistence.\\
\indent Many authors make this point, taking it to show: either that\\
\indent (i) it would be circular for the perdurantist to reply to the RDA by 
appealing to velocity; or \\
\indent (ii) velocity is not an intrinsic property of an object at a time, 
or of a temporal part; or \\
\indent (iii) both (i) and (ii).\\
For example, cf. Shoemaker (1979, p.327), Zimmerman  (1998, p.268), Sider 
(2001, p. 34) and Hawley (2001, p. 77-79).

\indent But perhaps  velocity should not be understood as usual. Various 
authors have sketched a rival, heterodox account of velocity, based on the 
idea that velocity should be an intrinsic property of an object at a time; 
for example, Tooley (1988, p. 236f.), Bigelow and Pargetter (1989, 
especially pp. 290-294; 1990, pp. 62-82) and Arntzenius (2000: pp. 189, 
197-201). These proposals seem to be mutually independent: the three later 
authors do not cite the previous work. But in what follows, I shall 
concentrate (as the RDA literature does) on Tooley; and so speak of 
`Tooleyan velocities'.\\
\indent Tooley denies the Russellian theory of motion on the grounds just 
mentioned: that it makes instantaneous velocity extrinsic to the moving 
object (at the time). He sketches an alternative, which aims to have 
velocity be an intrinsic property of the object: a property  that causes and 
explains its position at (shortly) later times---whereas the usual notion is  
a ``logical construction'' out of the object's positions at those times (and 
shortly earlier ones). In developing this rival notion, Tooley's strategy  
is to Ramsify the accepted laws of motion: i.e. to adapt Lewis' (1970) 
tactic for functional definition of theoretical terms. So, roughly speaking: 
Tooley says that the velocity of object $o$ at time $t$ is that unique 
intrinsic property of $o$ at $t$ that is thus-and-thus related to other 
concepts, as spelled out in the usual formulas of kinematics and dynamics. 
It follows, in particular, that velocity is  equal to the time-derivative of 
position, only as a matter of physical  law, and not as a matter of logic or 
conceptual analysis.\\
\indent Similarly, Bigelow and Pargetter (ibid.) propose that velocity 
should be an intrinsic  property  that causes and explains later positions: 
but they develop this idea in terms, not of Ramsification, but of the 
metaphysics of universals.   Finally, Arntzenius' recent survey  of three 
possible answers to Zeno's arrow argument takes one answer to be that the 
velocity of an object at an instant is an intrinsic property of it: a 
property that causes and explains change of position, on account of a law of 
nature ({\em not} a definition!) stating the value of velocity  to be equal 
to the time-derivative of position (ibid.). (Both Bigelow and Pargetter, and 
Arntzenius,  suggest their view is a descendant of  medieval views, in 
particular impetus theory.)

\indent At first sight, this heterodox view of velocity  has an obvious 
merit and an obvious defect. The merit is that velocity causes and explains 
later position: which sounds right. The defect is that on this view it is 
logically, though not nomically, possible that the velocity should point in 
the ``wrong direction''. It is logically possible that an object move to the 
right, while all the while its velocity vector pointed to the left---which 
sounds wrong!

\indent But to weigh this {\em pro} and {\em con}---and other {\em pros} 
suggested by Tooley, Bigelow and Pargetter, and {\em cons} suggested by 
Arntzenius---would take us too far afield.\footnote{For more discussion, cf. 
e.g.: Zimmerman (1998, pp. 275-278) who adds some discussion of the notion 
of intrinsicality (277-278); Sider (2001, pp. 35, 39, 228), who cites only 
Tooley; and Smith (2003) who defends the orthodox account of velocity, 
mostly against Arntzenius.} For  present purposes, I need only add to the 
above sketch that:\\
\indent (i): In my opinion, the  main motivation for this view is to secure 
a  ``{\em pointilliste}'' interpretation of mechanics; (as these authors say 
or hint: e.g. Arntzenius (2000, p. 200)). But there are good  reasons 
against such {\em pointillisme}; cf. Section \ref{sssec;vsStrtfwd} and my 
(2004).\\
\indent (ii): I will return to this view, in both a positive and a negative 
way, when I state my favoured reply to the RDA (in Sections 
\ref{psa;perdmwouttears} and \ref{sssec:Appealbeyondclassl}. Positively: the 
view will fare better in quantum theory than in the  context considered by 
the RDA, viz. classical mechanics. But negatively: my anti-{\em 
pointillisme}, which militates against the view, provides the best reply to 
the RDA.

\indent But it seems that ``Tooleyan velocities'' do {\em not} circumvent 
the consensus above, that velocity presupposes persistence.  For the laws of 
motion that Tooley Ramsifies make constant use of the notion of persistence. 
So even though a Tooleyan velocity is an intrinsic property of the object, 
the {\em concept} of velocity urged on us by Tooley involves the  notion of 
persistence no less than does the usual Russellian concept. Accordingly, 
Zimmerman (1998, p. 282-284) reiterates, for Tooleyan velocities, the 
consensus above; (Hawley (2001, p. 79) makes what is apparently the same 
point):
\begin{quote}
...the friends of temporal parts cannot appeal to [Tooleyan] velocities as 
theoretical properties implicitly defined by the laws of motion [to answer 
the RDA] ... [For it is] part of the definition of instantaneous  velocity  
that it be that property of an object which is such that its possession by 
an object at each instant of an interval, together with its location at the 
beginning of an interval and the length of the interval, determines where 
{\em that very same object} will be at the end of the interval. [Zimmerman 
1998, p. 282. Side-remark: Zimmerman adds a  footnote which sets aside 
forces acting during the interval, and refers to Tooley (1988, p. 238) for 
discussion of so doing.]   
\end{quote}  
Agreed, this consensus is no worries for Tooley. He is not concerned with 
the RDA, or in any way with the metaphysics of persistence: his {\em 
desideratum} for velocity is only that it should be an intrinsic property 
that causes and explains later positions.\\
\indent  As to the other advocates of heterodox velocities:---\\
\indent \indent (1): Arntzenius is also unconcerned with the debate about 
persistence.\\
\indent \indent (2): Bigelow and Pargetter briefly discuss the RDA, and deny 
the consensus above. They claim both that: (i) a portion of matter has a 
``non-qualitative identity'' across time; and, more directly against the 
consensus above, (ii) velocity, as understood by them, grounds such 
identities, and associated causal powers (1989, p. 297; 1990, pp. 72-74). 
But I shall not try to evaluate these claims, since:\\
\indent \indent (a): I shall later criticize a more developed version of 
claim (ii), due to Robinson and Lewis ((4) of Section 
\ref{sssec:LRproposal}); and\\
\indent \indent (b): being no friend of heterodox velocities, I think there 
are better ways than this to reply to the consensus ...

\subsubsection{Against the consensus}\label{sssec;againstconsensus}
I now present three replies to this consensus, in order of what I take to be 
ascending importance. The first two are objections to the verdict just 
given, that Tooleyan velocities are no help to a perdurantist who wants to 
distinguish the discs in terms of velocity. The third is more substantial: 
it develops the idea that the presupposition of persistence by velocity 
(understood either as usual, or {\em a la} Tooley) is mild and innocuous. 
All three replies will connect with other positions or topics in the 
debate---which I will return to.

\paragraph{4.2.2.A Functionally defining rotation?}\label{322AFuncDefRotn}   
Just as Tooley specifies velocity, in Ramsey-Lewis style, by its 
functional-causal role, i.e. the collection of its nomological or causal 
accompaniments, one might claim that the perdurantist should specify 
rotation (and associated quantitative measures) in terms of {\em its} 
functional-causal role, i.e. the accompaniments of rotation. And one might 
claim that this yields an (Appealing Differences) reply to the RDA.\\
\indent So the idea is to admit that Tooleyan velocity presupposes 
persistence: but to urge that rotation etc. can be functionally defined {\em 
without} such a presupposition.   So roughly speaking: the rotating disc is 
rotating ``in virtue of'' having one or other of the accompaniments of 
rotation.

I think the reply is coherent, but not  attractive.  (So far as I know, it 
has not been articulated in the literature; I learnt it from David Wallace 
in conversation---who also does not advocate it.) I say `not  attractive' 
because, as an example of the (Appealing Differences) reply, it will have to 
face the usual trouble for this reply: that an advocate of the RDA will 
argue that the accompaniments of rotation (i.e. the conjuncts within the 
functional-causal role) that it invokes can be ``imagined away'', while the 
disc nevertheless rotates. Agreed, the reply may well be able to outface 
this trouble, e.g. by the commonly discussed tactic of appealing to past or 
future of counterfactual differences between the discs. But I do not think 
the use of a functional definition adds much to the general strategy of the 
(Appealing Differences) reply: for two reasons.\\
\indent \indent (a): However exactly the intrinsic-extrinsic distinction is 
made precise, it seems that rotation (and its quantitative measures) should 
be intrinsic to an object; and that it needs to be intrinsic, if we are to 
answer the RDA, since the RDA can consider discs that are alone in their 
worlds. But there is no reason to think that when rotation is functionally 
defined in the proposed way, it will be intrinsic. For in general, a 
property that is functionally defined by its occupying a certain role need 
not be intrinsic.\footnote{So Tooley, keen to have velocity cause and 
explain (together with forces) the object's later position, needs his 
functional definition of velocity to require that the role-occupant be 
intrinsic to the object: intrinsicality does not come for free.}\\
\indent \indent (b): Functionally defining rotation without making a 
presupposition of persistence runs the risk that rotation, so defined, will 
not mesh appropriately with one's other doctrines about persistence, be they 
general philosophical doctrines (even analyses) or physical doctrines (as in 
the laws of mechanics). For presumably, both endurantist and perdurantist  
want rotation to involve, as a matter of conceptual analysis (not just 
scientific law), an object's (the disc's) persisting parts having circular 
(or approximately circular) orbits in space (on some appropriate  account of 
``space''---cf. Section \ref{ssec;describerot}). But if rotation is 
functionally {\em defined} by accompaniments none of which presuppose 
persistence, such as a tendency to oblateness, rotation will have only a 
nomological connection to persisting parts having circular orbits in 
space.\footnote{Point (b) brings out that this reply is similar to Teller's 
(Section \ref{sssec;tworeplies}), though more  ``positive'' than Teller's in 
that it accepts there are facts of rotation. But I will not pursue the 
comparison.}

\paragraph{4.2.2.B Functionally defining velocity and 
persistence?}\label{322BFuncDefVelyPerse}
 One might propose that even though Tooley's own version of Tooleyan 
velocities presupposes persistence, one could {\em extend} Tooley's appeal 
to Ramsey-Lewis style functional definition so as to simultaneously define 
both velocity {\em and} persistence. As we shall see,  this is very close to 
Sider's position (Section \ref{sssec;Sider}); which is my own ``fallback 
position''.\\
\indent For the moment, I note only that since this tactic functionally 
defines---not rotation alone---but both persistence and  velocity, objection 
(b) at the end of Section 4.2.2.A will {\em not} apply. For rotation will be 
defined in the usual way, after velocity and persistence have been defined. 
So as usual, and as desired, rotation will involve, as a matter of 
conceptual analysis, an object's  persisting parts having circular orbits in 
space.   (Again, to say `usual' is true but casual: a proper account of 
space is still needed---cf. Section \ref{ssec;describerot}). 

\paragraph{4.2.2.C Instantaneous  velocity is hardly 
extrinsic}\label{322CDegreeExtry} My third reply to the consensus is the 
most important of the three, for three reasons:\\
\indent (i): it introduces new ideas about the intrinsic-extrinsic 
distinction: specifically about the need to subdivide the distinction by 
admitting degrees of extrinsicality;\\
\indent (ii): it bears on the reply to the RDA proposed by Robinson and 
Lewis; which I will discuss in Section \ref{sssec;LewRob};   \\
\indent (iii): most important, it supports my favoured reply to the RDA (in 
Section \ref{psa;perdmwouttears}).\\
\indent This Subsection develops the reply's main ideas. The next Subsection 
adapts these ideas to ascriptions of specific values of velocity, as a 
preparation for the comparison with Robinson's and Lewis's proposal in 
Section \ref{sssec;LewRob}.   

My leading idea is that the consensus that velocity presupposes persistence 
is, though correct ``in the letter'', wrong ``in spirit''. Although velocity 
{\em does} presuppose persistence, the presupposition is milder than the 
literature allows. For ``most'' of the content of an ascription  of velocity 
to an object is free of this presupposition: though this ``most'' is about 
the object at other times, it does not imply that the object exists at any 
such times, since it is hypothetical (conditional) in content. This leading 
idea will also apply to acceleration and higher derivatives of position. In 
(1) and (2) below, I present two closely related ways of making this  idea 
precise. But I should first make three general points, (A)-(C).

(A): {\em Temporal intrinsicality and extrinsicality}:--- Our topic  prompts 
some terminology.  Since here and later, I will be focussing on whether the 
possession of a property $P$ by an object $o$ at a time implies propositions 
concerning matters of fact, especially about $o$, at other times, it will be 
convenient to use the phrase `{\em temporally intrinsic} property'. By this 
I mean ``intrinsic as regards time'': i.e.  roughly, a property whose 
possession by $o$ at a time implies nothing about matters of fact 
(especially about $o$) at other times (though it may imply propositions 
about other places).  Similarly, I shall talk of temporally extrinsic 
properties; and of spatially intrinsic and extrinsic properties.\\
\indent Two warnings about this terminology. (1): I agree that my 
explanation is vague, not least because the general intrinsic-extrinsic 
distinction on which it rides is itself vague;  (indeed probably 
ambiguous---cf. Humberstone 1996, Weatherson 2002). But my vague explanation 
will be enough for this paper. (2): Note that a property could be temporally 
extrinsic for one instance and not for another. Velocity itself provides 
examples of this. Imagine a non-instantaneous temporal part. That one of the 
part's constituent pieces of matter $o$ has a certain instantaneous  
velocity at a time $t$ ``within'' the part  surely corresponds to an 
intrinsic property of the part. But it is temporally  extrinsic for $o$ at 
the {\em instant} $t$. Humberstone (1996, p. 206, 227) notes that a similar 
phenomenon---extrinsic for one instance, but intrinsic for another---occurs 
for extrinsicality and intrinsicality {\em simpliciter}.\footnote{My (2004) 
further discusses temporal and spatial extrinsicality.}

(B): {\em Degrees of extrinsicality}:---  Extrinsicality is usually 
considered an all-or-nothing affair. But it is natural to suggest that it 
comes in degrees. Intuitively, a property is more extrinsic, the more that 
its ascription implies about the world beyond the property's instance: 
(compare the philosophy of mind's jargon of `wide' and `narrow' mental 
states---some are wider than others).\footnote{It is all the more natural 
when we consider how few properties are intrinsic: the extrinsic properties 
form so large a class as to merit being sub-divided.} 
That is rough speaking; and all the rougher because of controversies about 
the intrinsic-extrinsic distinction. But I expect that in many sufficiently 
limited contexts, the idea could  be  made precise in a natural way. In any 
case, I shall only consider the temporal extrinsicality at an instant of the 
properties of position and its time-derivatives (velocity, acceleration 
etc.), in the classical description of motion. This is certainly a 
sufficiently limited context for the idea to be made precise. 

(C): {\em Other conceptions of velocity}:--- My claims in (1) and (2) below 
could be carried over, with appropriate changes of wording, to Tooleyan 
velocities, accelerations etc. But  I shall develop my claims only for the 
orthodox view of velocity as the time-derivative of position, since as I 
said in Section \ref{sssec;againstvel} I am not convinced by the {\em 
pointilliste} motivation for Tooleyan velocities. (Indeed, I am not 
convinced in good part because of the present idea that orthodox velocity is 
``almost intrinsic''.)\\
\indent But here I should also admit that, from the perspective of physics 
rather than metaphysics, my discussion is limited; (and will remain so until 
Section \ref{511 Unitarity}). Namely: I go along with the RDA literature's 
assumption that, Tooleyan velocities apart,  velocity is defined as the 
time-derivative of position, so that position is  conceptually prior to 
velocity, and momentum is defined as mass times velocity. But I admit that 
even apart from Tooleyan velocities, this assumption is questionable. In 
particular, one can develop classical mechanics by taking momentum as 
primitive, together with position and mass, and defining velocity as 
momentum divided by mass. And in such a presentation, momentum does not need 
to be ``secretly understood'' as mass times velocity: one can introduce it 
abstractly, and without reference to time, as the generator of spatial 
translations. (Thanks to Gerard Emch for this point.) 

So in (1) and (2) below, I  present two closely related ways of making 
precise the idea that instantaneous velocity is ``hardly extrinsic'', i.e. 
hardly temporally extrinsic, since its ascription to an object $o$ at $t$ 
implies ``little'' about matters of fact at other times.\footnote{It is also  
hardly temporally  extrinsic, on a third construal of that notion discussed 
in  my 2004. Of course none of this is to deny that instantaneous velocity 
{\em is} temporally extrinsic at an instant, since it presupposes 
persistence.} Both ways are based on the obvious point that the only 
``categorical'' proposition that an ascription of a velocity (or indeed, of 
a higher derivative of position) to $o$ at $t$ implies about other times is 
that the $o$ exists for some open interval $(a,b)$ containing $t$: all the 
other implications are hypothetical.\\
\indent (The difference between the two ways will be that according to the 
first, which is ``read off'' the calculus, successively higher 
time-derivatives of position are more extrinsic; while on the second way, 
which is more logical and less mathematical, velocity acceleration and all 
higher derivatives are equally---and only mildly---extrinsic.) \\
\indent In what follows, we can think of $o$ as a point-particle; but it 
could equally well be a point-sized piece of matter in a continuum, or an 
extended body small and rigid enough to be treated as a point-particle. It 
will also be clear that the temporal extrinsicality of average velocity, 
acceleration etc. is mild for essentially the same reasons as for 
instantaneous velocity, acceleration etc. But to save space, I will focus on 
the instantaneous quantities. 

(1): {\em The sequence of time-derivatives}:---  The discussion will be 
tidier if we consider ascriptions, not of specific values of position, 
velocity, acceleration etc. to  $o$ at time $t$, but of some or other value. 
Then successive ascriptions are of increasing logical strength: having a 
velocity implies having a position, having an acceleration implies having a 
velocity etc. \\
\indent So consider a sequence of ascriptions to $o$ at time $t$: viz.\\
\indent \indent (Pos): an ascription of a position, i.e. a proposition 
saying that $o$ has some or other position at $t$;\\
\indent \indent (Vel): an ascription of an (i.e. some or other) 
instantaneous  velocity at $t$;\\
\indent\indent (Acc): an ascription of an instantaneous acceleration at 
$t$.\\
\indent These ascriptions are of course the first three members of an 
infinite sequence of ascriptions stating the existence of  higher 
time-derivatives of $o$'s position.  This gives an obvious sense in which 
instantaneous velocity is only mildly extrinsic. Each ascription is 
logically stronger than its predecessor; so (Vel), being almost at the start 
of the sequence, implies little in comparison with later members.\\
\indent In more detail: if a real function $f$ has a derivative at a point 
$t \in \mathR$, it must be defined on a neighbourhood of $t$ and be 
continuous at $t$. So the existence of $f''(t)$ requires the existence of 
$f'$ in a neighbourhood of $t$ and its (i.e. $f'$'s) continuity at $t$; and 
this in turn requires the continuity of $f$ in that {\em same} neighbourhood 
of $t$. And so on. In short: the existence of the $n$th derivative gives 
more information about times other than $t$ than does the existence of the 
$(n-1)$th derivative.
   
(2): {\em The ``only categorical implication''}:---  But there is also 
another sense in which velocity and the higher derivatives  of position are 
only mildly temporally extrinsic. This sense is more directly tied to the 
basic idea that the only categorical proposition that an ascription of such 
a quantity to $o$ at $t$ implies about other times is that the $o$ exists 
for some open interval $(a,b)$ containing $t$.\\
\indent In more detail. Let us ask what exactly is implied about other times 
by the ascriptions in the sequence; starting with (Pos). The metaphysical 
literature invariably assumes position to be temporally intrinsic: why?  The 
answer seems clear: `because (Pos), or even an ascription of a specific 
value `$o$ is at $\bf x$ at $t$', implies nothing about $o$'s position at 
other times'.\footnote{As discussed e.g. in Section 
\ref{sssec;persesptlraised}: I here set aside (i) the absolute-relational 
debate about space, and thereby (ii) possible  implications about other 
objects' positions, at $t$ or other times.} 

But to be more precise about `implying nothing' (apart of course from 
necessary or analytic propositions), we need:\\
\indent (a) to decide whether to allow that the object $o$ might  exist only 
for an instant; (as many metaphysical discussions of persistence do: true to 
the tradition of conceptual analysis, they allow all metaphysical or logical 
possibilities, not just the nomic ones); and \\
\indent (b) to distinguish categorical from hypothetical propositions.\\
\indent Although the categorical-hypothetical distinction is vague and 
contentious (because `logical form' is), I will not need to be precise or 
partisan about this: for it will be obvious from the calculus' definition of 
a limit which propositions  implied by ascriptions such as (Pos)-(Acc) to 
count as hypothetical.

If we allow $o$ to exist only for an instant (if we say `Yes' in (a)), then 
indeed (Pos) implies no categorical proposition about $o$'s positions at 
other times: there may be no such positions! But consider a hypothetical 
proposition along the lines: `if $o$ exists at a later time $t'$, and some 
value (or upper limit) is assumed about its average speed (defined in the 
usual way as distance traversed divided by time elapsed) over $[t,t']$, then 
$o$ is at $t'$  within a sphere of a certain radius, centred on $\bf x$'. 
Such a hypothetical proposition is of course not analytic; but it  follows 
by just definitions and logic  from `$o$ is at $\bf x$ at $t$'.

When we turn to the next member of the sequence, (Vel), we of course  get 
many more implications. $o$ must exist throughout some open interval, maybe 
tiny, around $t$; and since differentiability implies continuity, $o$'s 
position at a time $t'$ in the interval tends, as $t'$ tends to $t$, to 
$o$'s position $\bf x$ at $t$; and so on. But these implied propositions 
are, with one exception, hypothetical. The hypothetical propositions include 
those about average velocity discussed in the previous paragraph, and 
various others one can spell out by applying the definitions of continuity 
and differentiability. The exception is of course the categorical 
proposition  that $o$ exists throughout some open interval of times around 
$t$; (and, to be precise: {\em its} analytic consequences, like $o$'s 
existing at some time $t'$ not equal to $t$). In particular, (Vel) is 
compatible with $o$ being anywhere at any other time $t'$, no matter how 
close $t'$ is to $t$.\footnote{I here assume there is no limiting velocity, 
as in relativity.}  

Similarly again, for (Acc). There are again more implications, but they are 
almost all complicated hypotheticals: the only categorical proposition about 
other times that the ascription implies is the same one again: that $o$ 
exists throughout some open interval of times around $t$. And so on along 
the infinite sequence of ascriptions.

To sum up this discussion of (1) and (2): the temporal extrinsicality of  
velocity and higher derivatives of position is mild.  For almost all of the 
implied propositions  are hypothetical; and even the temporally intrinsic 
ascription (Pos) implies countless such propositions. Besides, the 
categorical propositions implied by an ascription of velocity, or of any 
higher derivative, are all just consequences of the one proposition  that 
$o$ exists throughout some open interval of times around $t$. So as regards 
categorical implications about other times, the temporal extrinsicality gets 
already at stage (Vel) as ``bad'' as it ever gets along the sequence: and 
that, I submit, hardly deserves the name `bad'---it is mild.\footnote{This 
view is reflected in the jargon of mathematics and physics. For example, 
mathematicians call not only (Pos), but also the ascriptions (Vel) etc., 
`local'; and physicists call equations of motion that determine the object's 
motion at $t$ in terms of its position and some of its derivatives then (but 
without reference to facts a finite temporal interval from $t$) `local in 
time'. For  more discussion, cf. Arntzenius 2000 pp. 192-195, Smith 2003 and 
my 2004.}

\paragraph{4.2.2.D Instantaneous velocity without presuppositions: 
``welocity''}\label{322Dwelocity}
Finally, I turn to ascriptions of specific values of velocity, acceleration 
etc. The first point to make is that the discussion above can of course  be  
carried over straightforwardly. For example, an ascription of velocity $\bf 
v$ to $o$ at $t$ simply adds to the ascription (Vel) information about what 
is the limit to which the countless average velocities tend for smaller 
time-intervals, viz. $\bf v$; and similarly for acceleration etc.
 
\indent But for my purposes, it is more important to notice that there is a 
way of representing my conclusion, that velocity etc. are hardly extrinsic, 
in terms of a novel vector-valued quantity that is like velocity---but lacks 
its presupposition of persistence (mild though that presupposition is).\\
\indent I will call this new-fangled quantity {\em welocity}, the `w' being 
a mnemonic for `(logically) weak' and-or `without (presuppositions)'. The 
benefit of introducing welocity will be clear in Section 
\ref{sssec;LewRob}'s comparison of it with Robinson and Lewis' proposed 
reply to the RDA.\\
\indent So the idea is that welocity is to reflect, in the way its values 
are defined, this lack of presupposition. That is: the values are to be 
defined in such a way that it is impossible to infer from the value of the 
welocity of the object $o$ at time $t$ that $o$ in fact exists and has a 
differentiable worldline in some neighbourhood of $t$: an inference which, 
as we have just seen, {\em can} be made from the value of velocity (at least 
as orthodoxly understood!).

\indent Developing this idea takes us to familiar philosophical territory, 
viz. rival proposals for the semantics of empty referring terms. In our 
case, the empty terms will be expressions for $o$'s instantaneous velocity 
at $t$; and, as we have seen, they can be empty either because:\\
\indent \indent (NotEx): $o$ does not exist for an open interval around $t$, 
or\\
\indent \indent (NotDiff): $o$ does exist for an open interval around $t$, 
but its position ${\bf x}$ is not differentiable at $t$; (roughly: there is 
a ``sharp corner'' in the worldline).\\
\indent (And similarly for acceleration and higher derivatives; but I shall 
discuss only velocity---tempting though words like `wacceleration' are!)

\indent In fact, it will be clearest to lead up to my proposal for welocity  
by first considering a more familiar one, which is modelled on Frege's 
proposal that (to prevent truth-value gaps) empty terms should be assigned 
some ``dustbin-referent'', such as the empty set $\emptyset$. Thus if one 
sets out to define a quantity that is like velocity but somehow avoids its 
presupposition of persistence, one naturally first thinks of a quantity, 
call it $\bf u$, defined to be\\
\indent \indent (a): equal to the (instantaneous)  velocity $\bf v$ for  
those times $t$ at which $o$ {\em has} a velocity; and\\
\indent \indent  (b):  equal to some dustbin-referent, say the empty set 
$\emptyset$, at other times $t$; i.e. times such that either (NotEx): $o$ 
does not exist for an open interval around $t$; or (NotDiff): $o$ does exist 
for an open interval around $t$, but its position ${\bf x}$ is not 
differentiable at $t$.\\
\indent Of course,  variations on (b) are possible. One could select 
different dustbin-referents for the two cases, (NotEx) and (NotDiff), (say, 
$\emptyset$ and $\{\emptyset \}$) so that $\bf u$'s value registered the 
different ways in which an instantaneous  velocity could fail to exist. And 
instead of using a dustbin-referent, one could say that the empty term just 
has no ``semantic value'', or ``is undefined'': (a contrast with 
dustbin-referents which would presumably show up in truth-value gaps, and 
logical behaviour in general).

Agreed, this definition is natural. But it does not do the intended job. For 
this quantity $\bf u$, whether defined using (b) or using the variations 
mentioned, does not avoid, in the way intended, the presupposition of 
persistence. For $\bf u$'s value (or lack of it, if we take the 
no-semantic-value option) registers whether or not the presupposed 
persistence holds true. That is: we {\em can} infer from the value of $\bf 
u$ (or its lack of value) whether (a) $o$ has a velocity in the ordinary  
sense, or (b) the presupposition has failed in that (NotEx) or (NotDiff) is 
true. In short: $\bf u$'s individual values tell us too much.

But there {\em is} an appropriate way of assigning semantic values to empty 
terms, i.e. a way of defining a quantity, {\em welocity}, that is like 
velocity but avoids its presupposition of persistence, in that welocity's 
values do not give the game away about whether the presupposition has 
failed, i.e. about whether (NotEx) or (NotDiff) is true. In order not to 
give the game away, welocity must  obviously take ordinary values, i.e. 
triples of real numbers, when the presupposition has failed. But how to 
assign them?\\
\indent The short answer is: arbitrarily. The long answer is: we can adapt 
schemes devised by logicians in which a definite description, whose 
predicate has more than one instance, is assigned as a referent {\em any 
one} of the objects in the predicate's extension. (The first such scheme was 
devised by Hilbert and Bernays; but we will only need the general idea.) 
Such a scheme applies to our case, because we can write the definition of 
welocity in such a way that when the presuppositions fail (i.e. (NotEx) or 
(NotDiff) is true), the   predicate (of triples of reals numbers) in the 
definition  is vacuously satisfied by all such triples; so that forming a 
definite description, and applying semantic rules like Hilbert-Bernays', 
welocity  is assigned an arbitrary triple of real numbers as value. Thus we 
get the desired result: if you are told that the value of welocity  for $o$ 
at $t$ is some vector in $\mathR^3$, say (1,10,3) relative to some axes and 
choice of a time-unit, you cannot tell whether:\\
\indent (a): (NotEx) and (NotDiff) are both false (i.e. the presuppositions 
hold), and $o$ has velocity (1,10,3); or\\
\indent (b): (NotEx) or (NotDiff) is true, the predicate is vacuously 
satisfied by all triples, and (1,10,3) just happens to be the triple 
assigned by semantic rules taken from Hilbert-Bernays' (or some similar) 
scheme.

The details are as follows. (1): Hilbert and Bernays introduced the notation 
$(\varepsilon x)(Fx)$ for the definite description `the $F$', with the rule 
that if $F$ had more than one instance, then $(\varepsilon x)(Fx)$ was 
assigned as referent any such instance, i.e. any element of $F$'s extension. 
(We need not consider their other rules, nor their rules' consequences for 
the semantics and syntax of singular terms.)

\indent (2): Next, we observe that intuitively the {\em velocity} of an 
object $o$ at time $t$ can be defined with a definite description containing 
a material conditional whose antecedents are the presuppositions of 
persistence and differentiability. That is: it seems velocity can be defined 
along the following lines:---

\noindent The velocity of $o$ at time $t$ is the triple of real numbers 
${\bf v}$ such that:
\begin{quote}
for some (and so any smaller) open interval $I$ around $t$:\\
\{[$o$ exists throughout $I$]  and [$o$'s position $\bf x(t)$ is 
differentiable in $I$]\} $\;\;\; \supset \;\;\;\;$ [$\bf v$ is the common 
limit of average velocities for times $t' \in I$, compared with $t$, as $t' 
\rightarrow t$ from above or below].
\end{quote}
This {\em definiens} uses a material conditional. So it will be vacuously 
true for all triples $\bf v$, if the antecedent is false for all open 
intervals $I$ around $t$, i.e. if (NotEx) or (NotDiff) is true: in other 
words, if velocity's presuppositions of continued existence and 
differentiability fail.

\indent (3): Now we put points (1) and (2) together. Let us abbreviate the 
displayed {\em definiens}, i.e. the open sentence with $\bf v$ as its only 
free variable, as $F({\bf v})$. Then I propose to define the {\em welocity}  
of $o$ at $t$ by the singular term $(\varepsilon {\bf v})(F{\bf v})$: which 
is, by Hilbert-Bernays' semantic rule:\\
\indent \indent (a): equal to the (instantaneous)  velocity of $o$ for  
those times $t$ at which $o$ {\em has} a velocity; and\\
\indent \indent  (b):  equal to some arbitrary triple of real numbers, at 
other times $t$; i.e. times such that either (NotEx): $o$ does not exist for 
an open interval around $t$; or (NotDiff): $o$ does exist for an open 
interval around $t$, but its position ${\bf x}$ is not differentiable at 
$t$.\\
Welocity, so defined, has the desired features: its values do not give the 
game away about whether (NotEx) or (NotDiff) is true.

That is all I need to say about welocity, for this paper's purposes; and in 
particular,  for Section \ref{sssec;LewRob}'s comparison with Robinson's and 
Lewis' proposal. But I end this Subsection by noting that there are of 
course various technical questions hereabouts, even apart from the logical 
questions about $\varepsilon$ (which are of course addressed by the masters, 
Hilbert and Bernays!).\\
\indent In a discussion of the RDA and so of continuous matter, a natural 
question arises from letting $o$ be a point-sized bit of matter in a 
continuum, and letting the presuppositions of velocity fail for various such 
point-sized bits of matter: some such bits may fail to exist, and some may 
have a non-differentiable worldline. One then faces the question: how widely 
across space, and in how arbitrary and gerry-mandered a spatial 
distribution, can these bits fail to exist, or have a non-differentiable 
worldline---i.e. how widely and arbitrarily can the presuppositions of 
velocity fail---while yet the welocity field might not give the game away, 
in that the arbitrary values {\em can} be assigned at all the points  where 
the presuppositions of velocity fail, in such a way as to give a smooth 
(e.g. continuous or even differentiable) welocity field?\\
\indent This is in effect a question about the scope and limits of 
``regularization'' of ``singularities'' in real vector fields: a good 
question---but not one for this paper!

To sum up this Subsection, i.e. Section \ref{sssec;againstconsensus}: I hope 
here---and especially in the last two parts, Subsections 4.2.2.C and 
4.2.2.D---to have ``set the cat among the pigeons'', to have ``upset the 
applecart'', about the literature's consensus that velocity presupposes 
persistence.\\
\indent These scattered pigeons and upset apples will be important for my 
reply to the RDA in Section \ref{psa;perdmwouttears}. But in the meantime, 
they will not much affect the discussion, except for the next Subsection's 
comparison of welocity with Robinson's and Lewis' proposal (Section 
\ref{sssec;LewRob}). So they can be set aside for:\\
\indent (i) most of this Section's discussion of metaphysical replies to the 
RDA; and for\\
\indent (ii) the next Section's discussion of the technical description of 
rotation, and of what it entails about the RDA.

\subsection{Velocities on the cheap? Lewis and Robinson}\label{sssec;LewRob}
I turn to Lewis' and Robinson's version of (Appealing Differences). They 
propose that a moving object has a vectorial property (i.e. a property 
represented by a vector) which is intrinsic to the object, and whose vector 
is equal to the velocity vector. But this property is not itself velocity, 
since velocity presupposes persistence, and this property is to be 
intrinsic, not merely ``almost intrinsic''.\\
\indent I will first present the proposal, in Section 
\ref{sssec:LRproposal}. I will emphasize: (i) Lewis' doctrine of Humean 
supervenience; and (ii) how Lewis came around to this defence of Humean 
supervenience (ca. 1998), after espousing for a while (ca. 1986-1994) a (No 
Difference) reply. Then in Section \ref{sssec;LRassessed} I will assess the 
proposal.

\subsubsection{The proposal}\label{sssec:LRproposal}
(1) {\em Humean supervenience}\\
We saw in Section \ref{sssec;appealcause} that Armstrong's appeal to 
causation, in response to the RDA, is not available to someone like Lewis 
who advocates both perdurantism, and a Humean view of causation. Indeed, as 
I mentioned, Lewis advocates a much stronger doctrine, Humean supervenience, 
which has become the paradigm in contemporary metaphysics for what I have 
called {\em pointillisme} (cf. Section \ref{sssec;vsStrtfwd}). He holds that 
{\em all} truths supervene on truths about  matters of local particular 
fact: where `matters of local particular fact' is to be understood in terms 
of Lewis' metaphysics of natural properties, with the properties having 
spacetime points, or perhaps point-sized bits of matter, as instances. He 
writes:
\begin{quote} 
\ldots all there is to the world is a vast mosaic of local matters of 
particular fact, just one little thing and then another ... We have ... 
relations of spatio-temporal distance between points ... And at those points 
we have local qualities ... For short: we have an arrangement of qualities. 
And that is all. There is no difference without a difference in the 
arrangement of qualities. All else supervenes on that. (1986, p. ix-x.) 
\end{quote}
Or in other words: Humean supervenience 
\begin{quote} 
\ldots says that in a world like ours, the fundamental relations are exactly 
the spatiotemporal relations: distance relations, both spacelike and 
timelike, and perhaps also occupancy relations between point-sized things 
and spacetime points. And it says that in a world like ours, the fundamental 
properties are local qualities: perfectly natural intrinsic properties of 
points, or of point-sized occupants of points. Therefore it says that all 
else supervenes on the spatiotemporal arrangement of local qualities 
throughout all of history, past and present and future. (1994, p. 474.) 
\end{quote}

(2) {\em Lewis' (No Difference) Reply}\\
So Lewis addresses the RDA as an objection to his Humean supervenience 
thesis. His reply changed over time; we can distinguish three phases (1986 
p.xiii, 1994 p. 475, 1999). At first (1986 p.xiii), he appealed to the fact 
(clear from the second quotation) that he advocates Humean supervenience as 
a contingent thesis, true at some worlds (including, he hopes, ours) but not 
at others. That is, he advocated Humean supervenience for an ``inner 
sphere'' consisting of the non-alien worlds---defined (in his quiddistic 
theory of natural properties) as the worlds where any instantiated natural 
property is not alien to the actual world. So he replied to the RDA, taken 
as putting its differing discs (Stat) and (Rot) each in its own world, by 
saying that one or both of the worlds must be outside the inner sphere.\\
\indent Of course, essentially the same reply can be given using other 
definitions of the ``inner sphere'' across which Humean supervenience is to 
hold as a contingent supervenience thesis. For example, as I mentioned in 
Section \ref{ssec;naturalism}: one might claim the supervenience to hold 
across all the worlds that each make true all the actual laws of nature.\\
\indent Similarly, Robinson (1989, p.404: crediting Lewis):\\
\indent \indent (i): toys with replying to the RDA that it shows that the 
common sense notion of homogeneous matter and its persistence requires 
atomistic matter; and\\
\indent \indent (ii): suggests this would be an example of the traditional 
``paradox of analysis'': roughly, that philosophical analysis can reveal 
surprising truths.

Later, Lewis adjusted this reply; (1994 p. 475---this is the second phase). 
He agreed that this reply had not given a reason  for thinking that enduring 
objects were different in their fundamental nature from perduring objects, 
so that one or both of the disc worlds had to be alien; (in response to 
Haslanger (1994); cf. also Robinson 1989, p. 403-404). But his preferred 
reply to the RDA remained a version of (No Difference): that one or both of 
the disc worlds were not ``worlds like ours''; i.e. they fell outside the 
class of worlds (now only vaguely specified) for which Humean supervenience 
was claimed.  

(3) {\em Comparison with Callender}\\
 It is worth  briefly comparing this reply with Callender's (No Difference) 
reply (cf. Section \ref{sssec;tworeplies}). The main difference is 
that:---\\
\indent Lewis admits that there are some possible worlds very `unlike ours' 
(roughly: outside the inner sphere of possibility)  which sustain the 
distinction between (Stat) and (Rot), in the sense that at least one of 
these two  worlds must be outside the inner sphere.\\
\indent  On the other hand, Callender (at least as I read him) takes a 
tougher stance.
He does not define a limited class of worlds in which there is no difference 
(i.e. from which the (Stat)/(Rot) distinction is banished). He apparently 
believes the (Stat)/(Rot) distinction is yet ``worse off''. For good 
metaphysical arguments can be given that it is as spurious as the up/down 
distinction (his Section 2, p. 30-35). And even if these arguments fail, 
there are methodological reasons (roughly: Occam's razor) to deny the 
distinction (his Section 3, p. 35-40). Thus he says that the discs that the 
RDA needs are `the metaphysical equivalent of fairies, ghosts and vital 
spirits' (p. 26). By this he means that they must be uncoupled (i.e causally 
isolated) from any of the fields (electromagnetism, gravity etc.) that exist 
in our world, and from all the usual causes and effects of rotation. `Such a 
disc is no different from a ghost, and is not something Humeans or 
non-Humeans ought to posit' (p. 37).\\
\indent (But maybe Callender is closer to Lewis than this summary suggests: 
perhaps he thinks the (Stat)/(Rot) distinction makes some kind of sense, but 
is  less committed than Lewis to the framework of possible worlds and so to 
defining some kind of ``inner sphere'' from which the distinction is 
banished.)
  
(4) {\em Lewis' (Appealing Differences) Reply}\\
But in a final short paper (replying to Zimmerman 1998), Lewis endorsed 
(1999, p. 211) a proposal that had been floated  by Robinson (1989; p. 405 
para 2, p. 406 para 2---p. 408 para 1). Roughly speaking, the proposal is 
that:\\
\indent (i) a vectorial property at a point can be an intrinsic property of 
that point;\\
\indent (ii) the propagation of continuous matter through spacetime  
involves such a property at every spacetime point; and \\
\indent (iii) these properties distinguish the rotating and non-rotating 
discs, since the vector that represents the property at a point is timelike, 
and  points in the same direction as the instantaneous four-dimensional 
velocity vector at that point; \\
\indent (iv) the distribution of these properties, from point to point, 
determines (subvenes) the relations of qualitative similarity between 
points, and especially the relations of causal dependence between events at 
those points; and\\
\indent (v) the distribution of these properties, by determining the lines 
of causal dependence, determines the lines of persistence.

Three side-remarks: (a) So this proposal takes causal dependence to underpin 
persistence: as I noted  at the start of Section \ref{sssec;appealcause}, 
many philosophers endorse this.\\
\indent (b) In fact, Lewis already agreed to (i) in his (1994, p. 474); but 
lacking (ii) and (iii), and so also (iv) and (v), he there retained his 
``not like ours'' (No Difference) reply.\\
\indent (c) Robinson's (i)-(iv) are clearly similar in spirit to Tooley's 
and Bigelow and Pargetter's heterodox understanding of velocity as an 
intrinsic property, discussed in Section \ref{sssec;againstvel}. Robinson 
does not refer to their papers which were of course  contemporaneous. But 
Zimmerman (1998, p. 281, p. 284) and Sider (2001, p. 228) both see the 
similarity to Tooley's proposal (1988). Zimmerman first discusses reading 
Robinson's proposal as the same as Tooley's (p. 281), and then discusses 
reading it as just similar (p. 284, note 65). Sider reads the proposals as  
similar. More specifically, Sider and Zimmerman's second reading both see 
Robinson's proposal as going with a Russellian ``at-at'' account of motion. 
So also (implicitly) does Lewis' discussion.   

  So the idea of the proposal is that the difference in these properties 
amounts to a difference in the `local arrangement of qualities' as demanded 
by Humean supervenience. Thus Lewis (1999, p. 211) begins by approvingly 
quoting Robinson, suggesting we should
\begin{quote} 
\ldots see the collection of qualities characteristic of the occupation of 
space by matter as in some sense jointly self-propagating; the fact of 
matter occupying space is itself causally responsible ... for the matter 
going on occupying space in the near neighbourhood immediately thereafter. 
... [The posited vectors] figure causally in determining the direction of 
propagation of [themselves as well as] other material properties. (Robinson 
1989, p. 406-407.) 
\end{quote}
Lewis then goes on to formulate the proposal more formally, as a putative 
law that partially specifies a vector field $V$. The specification is 
partial, both in (i) being admitted to be a ``first approximation'', and 
(ii) specifying only the direction but not the length of the vector at each 
point. But (ii) hardly matters: it will be obvious that Robinson and Lewis 
could frame their proposal entirely in terms of postulating a timelike 
direction field (i.e. a specification at each point of continuous matter of 
a timelike direction), rather than a vector field. But I shall follow them 
and talk of a vector field.\\
\indent In giving this formulation, Lewis' aim is partly to avoid various 
objections or limitations. In particular, the formulation should not invoke 
either persistence or causation, since  these are meant to supervene on the 
local arrangement of qualities, taken of course as including facts about the 
vector field $V$. Thus the formulation is to avoid circularity objections 
that had been urged by Zimmerman (1998) against some related proposals.\\
\indent So in particular: the vector field $V$ cannot simply be the 
instantaneous (four-dimensional) velocity (Russellian, not Tooleyan!) of the 
matter at the point in question.  For $V$ is to contribute to an analysis of 
(or at least to a supervenience basis for) persistence and thereby of 
velocity.\\
\indent Similarly, since Lewis agrees that causation is crucial to 
persistence (`the most important sort of glue that unites the successive 
stages of a persisting thing is causal glue': 1999, p. 210), causation 
cannot be invoked in the course of specifying the vector field $V$. 

Lewis proposes that (for a world with continuous space and time), the 
specification of $V$ `might go something like this':
\begin{quote}
Let $p$ be any spacetime point, and let $t$ be any smooth timelike 
trajectory through spacetime with $p$ as its final limit point. Let each 
point of $t$ before $p$ be occupied by matter with its vector [i.e. vector 
of the vector field $V$] pointing in the direction of $t$ at that point. [So 
in the jargon of modern geometry, $t$ is an integral curve of $V$.] Then, 
{\em ceteris paribus}, there will be matter also at $p$. (1999, p. 211.)
\end{quote}
Here, the `{\em ceteris paribus}' clause is to allow for the fact that the 
point-sized bit of matter might cease to exist before $p$, because of 
`destructive forces or self-destructive tendencies' (ibid.).\\
\indent Lewis also stresses that this proposal is to be read as a law of 
succession, not of causation. This means, I take it, that the `Then, {\em 
ceteris paribus}' is to be read as a material conditional.

\subsubsection{Assessment}\label{sssec;LRassessed}
I think Lewis' proposal fails. After saying why, I will broach the more 
general (and I think, more important) issue of how plausible is an extreme 
{\em pointillisme} like Lewis' Humean supervenience. This will return us to 
the discussion in Sections 4.2.2.C and 4.2.2.D.

\paragraph{4.3.2.A The vector field remains 
unspecified}\label{332A:unspecified}
I claim Lewis' proposal is too weak: it does not go far enough to specify 
$V$. For it only says, of any timelike open curve that is an integral curve 
of $V$, that the future end-point $p$ of this curve will, {\em ceteris 
paribus}, have matter at it.\\
\indent But every suitably smooth vector field $U$ defined on a open region 
$R$ of spacetime has integral curves throughout $R$; (which are timelike, by 
definition, if $U$ is). (To be precise: `suitably smooth' is none too 
demanding: all we need is that $U$ be $C^1$, i.e. the partial derivatives of 
its components exist and are  continuous.) So suppose Lewis stipulates, that 
the field $V$ is to be timelike and $C^1$ on an open set $R$ which is its 
domain of definition (say, the spatiotemporal region occupied by continuous 
matter): which (``giving rope'') we can assume to be a legitimate, in 
particular non-circular, stipulation. Then his proposal says that, {\em 
ceteris paribus}, every point $p \in R$ has matter at it.\\
\indent But that claim hardly helps to distinguish $V$ from the countless 
other (timelike continuous) vector fields $U$. For however exactly one 
interprets `{\em ceteris paribus}', the claim is surely true of $p$ 
regardless of the integral  curve one considers it as lying on. So the claim 
about $p$ does not constrain the vector field.  Indeed, if Lewis sets out to 
specify $V$ on the spatiotemporal region occupied by continuous matter, the 
claim is thereby assumed to be true for all  $p$ in the region, regardless 
of vector fields. So again, we have said nothing to distinguish $V$ from the 
countless other vector fields $U$.\\
\indent Agreed, Lewis puts forward his proposal as a ``first approximation'' 
to specifying $V$. But so far as I can see, his discussion doesn't contain 
any ingredients which would, for continuous matter, help distinguish $V$ 
from other vector fields.\footnote{Nor can I guess how I might have 
misinterpreted Lewis' proposal. The situation is puzzling: and not just 
because Lewis thought so clearly, and my objection is obvious. Also, the 
objection is analogous to what Lewis himself says (p. 210) against the naive 
idea that $V$ should point in the direction of perfect qualitative 
similarity: viz. that `in non-particulate homogeneous matter, ... lines of 
qualitative similarity run every which way'.}

Zimmerman (1999) makes a somewhat similar objection to Lewis' proposal. But 
his exact intent is not clear to me.\\
\indent He maintains  that in some seemingly possible cases of continuous 
matter, Lewis' proposal does not specify a unique vector field $V$---indeed 
hardly constrains $V$ at all.  He says (p. 214 para 1 and 2) that in 
possible worlds with a  physics of the sort Descartes might have envisaged, 
i.e. where there is {\em nowhere} any vacuum, and only {\em one} kind of 
(continuous homogeneous) stuff fills all of space: `{\em every} vector field 
will satisfy [Lewis'] law.'\\
\indent Thus Zimmerman assumes that:\\
\indent \indent (i): the worlds with which he is concerned are wholly filled 
with the one kind of stuff; and\\
\indent \indent (ii): these worlds are thus filled as a matter of law, not 
happenstance (in the jargon: as a matter of physical or nomic necessity).\\
\indent He also says (p. 214-5) that he needs to assume (i) and (ii) in 
order to criticize Lewis' proposal, together with obvious modifications of 
it which allow for different types (``colours'') of continuous matter. That 
is: Zimmerman thinks Lewis' proposal works, or could be modified to work, 
for worlds in which:\\
\indent (i') continuous matter  does not fill all of space and-or comes in 
various types; or\\
\indent (ii') continuous matter of just one type fills all of space, but 
only as a matter of happenstance. 

\indent In view of my own objection, I do not understand why Zimmerman feels 
he needs to assume (i) and (ii) in order to object to Lewis. He does not 
explicitly say why he does so. {\em Maybe} it is to block some Lewisian 
rejoinder, that would better specify $V$, by adding constraints of either or 
both of two kinds:\\
\indent (i''): constraints about the spatiotemporal relations of the 
continuous  matter  in a bounded volume (say, one of our discs) to other 
matter outside the volume.\\
\indent (ii''): constraints about the nomic or modal properties of matter.\\
\indent \indent But it remains unclear how the details of (i'') and (ii'') 
might go. 

To sum up: For all I can see, my objection, that $V$ is not distinguished 
from countless other vector fields, applies to Lewis' proposal (and thereby: 
the spirit of Zimmerman's objection also applies) for the case that Lewis 
intended it---i.e. the discs of the original RDA.

\paragraph{4.3.2.B What price Humean supervenience? Doubts about intrinsic 
vectorial properties}\label{332B:whatpricehs}
I announced my denial of {\em pointillisme}, and so my antipathy to Humean 
supervenience, already in Section \ref{sssec;vsStrtfwd}. And in Sections 
4.2.2.C and 4.2.2.D, I developed this by arguing that temporal 
extrinsicality was after all ``not so bad''. More specifically, I argued 
that velocity was almost intrinsic, and that we could even define a 
quantity, {\em welocity}, that in a sense avoids velocity's presupposition 
of persistence.

I am afraid these doctrines would not appeal to Lewis! He would probably be 
unimpressed by velocity's being almost intrinsic. For Humean supervenience 
is so central to his metaphysical system that he sets considerable store by 
intrinsicality. So he would probably say that as regards failing to be 
intrinsic, a miss is as good (i.e. bad!) as a mile.

\indent Similarly, I am not confident that he would welcome welocity. I 
agree that he might be ``envious''  of its being well-defined ({\em modulo} 
the freedom to assign referents associated with the $\varepsilon$ operator), 
since his own proposal, the intrinsic vector $V$, is yet to be successfully 
defined.\\
\indent I also agree that in one important respect, welocity fits Lewis' 
conception of intrinsicality. Namely, on Lewis' conception, intrinsicality 
is not hyperintensional: that is, necessarily co-extensive properties are 
alike in being intrinsic, or not. (The reason for this lies in Lewis' 
proposal for how to analyse intrinsicality. Both his preferred analysis (and 
a fallback analysis, in (Langton and Lewis 1998)) take an intrinsic property 
to be one that does not differ between duplicate objects---where duplication 
is defined, in both analyses, as sharing a certain elite minority of 
properties. Clearly, any analysis with these features will imply that 
necessarily co-extensive properties are alike in being intrinsic, or not.)\\
\indent Now, the welocity of an object $o$ is intuitively (albeit 
hypothetically!) ``about'' $o$'s  positions at other times; so that someone 
who construes intrinsicality as hyperintensional may want to argue that 
welocity is extrinsic, or temporally extrinsic (if they use that notion).  
Conversely, the idea that welocity is intrinsic   goes with a conception of 
intrinsicality as not hyperintensional---such as Lewis'.

 Nevertheless, I am not confident Lewis would welcome welocity, just because 
there are two general obstacles to connecting it to his framework:---\\
\indent\indent (a): Lewis talks only of intrinsic and extrinsic properties, 
not (as I have) of {\em temporally} and {\em spatially} intrinsic or 
extrinsic properties; and \\
\indent\indent (b): Lewis' proposed analysis of intrinsicality (and the 
Langton-and-Lewis fallback proposal) is cast in terms of his very general 
metaphysical system, using notions like `natural property' and `possible 
world'. This makes it a delicate matter to classify everyday, or even 
technical scientific, properties in terms of his proposal. So in particular, 
I am unsure whether my welocity counts as intrinsic for Lewis.\\
\indent (I add, in Lewis' defence: I think this ``gap'' between his 
metaphysical categories and the properties we know explains some of the 
counterexamples brought against his proposal. But I also add, against him: 
the ``gap'' makes considerable trouble for his overall Humean project 
(Mainwood 2003); and these counterexamples, together with other 
considerations, also suggest that there is no single intrinsic-extrinsic 
distinction; (cf. Humberstone (1996) and (broadly following him) Weatherson  
2002).)

But it would take us too far afield to further compare  my anti-{\em 
pointillisme} with Lewis' Humean supervenience, even if we considered only  
the topic of velocity. I must leave further discussion of these issues to my 
2004. To advertise that discussion, and to emphasise that controversy 
surrounds even basic questions about the intrinsic-extrinsic distinction in 
application to scientific properties, I end this Subsection by pointing out 
that the very first claim of the Robinson-Lewis proposal, viz. (i) of 
Section \ref{sssec:LRproposal}:\\
\indent (i): a vectorial property at a point can be an intrinsic property of 
that point\\
has been doubted in the metaphysical literature---and even by Robinson and 
(the earlier) Lewis themselves! (Lewis endorsed (i) in what I called the 
`second phase' reply to the RDA; (1994, p. 474).)

\indent I said in Section \ref{sssec:LRproposal} that Robinson `floated' the 
Robinson-Lewis proposal, precisely because he did not endorse it. His 
anxiety concerns the directionality of a vector. He writes: `Direction seems 
to me an inherently relational matter' (1989,p. 408). He supports this with 
the following argument, for which he credits Lewis; (so Lewis seems to have 
come round to believing (i), that vectors can represent intrinsic properties 
of points, sometime between ca. 1988 and writing his (1994, p. 474)).\\
\indent The argument has two premises:\\
\indent (a): A vector quantity could not be instantiated in a 
zero-dimensional world consisting of a single point; (though since 
arbitrarily close points define a direction, there is of course no lower 
limit to the ``size'' of a world in which a point instantiates a vector 
quantity). But it also seems that:\\
\indent (b): Since a point in an extended world that instantiates a vector 
quantity is indeed a {\em point}, it could have a duplicate that existed on 
its own, i.e. was the only object in its world.\\
\indent Taken together, (a) and (b) imply that duplicate points might differ 
in their vectorial properties; so that (at least on a Lewisan approach to 
the intrinsic-extrinsic distinction, according to which intrinsic properties 
are those shared by duplicates) such properties are   extrinsic.\\
\indent Nor is Robinson  alone in worrying that vectors could only represent 
extrinsic  properties of a point. Cf. also: Bricker (1993); Zimmerman (1998, 
p. 277-278; mentioned in Section \ref{sssec;againstvel}'s discussion of 
Tooleyan velocities (footnote 13); and Black (2000, p. 103), who holds that 
vectors can only represent intrinsic  properties in a {\em flat} manifold, 
i.e. roughly, a manifold in which there is a unique preferred way to compare 
vectors located at different points.\\
\indent To sum up: even among authors squarely within contemporary 
metaphysics' approach to the intrinsic-extrinsic distinction, step (i) of 
the Robinson-Lewis proposal remains controversial.

\subsection{Functionally defining persistence and laws: 
Sider}\label{sssec;Sider}
I will describe Sider's reply to the RDA (2001, p. 230-236) in some detail, 
as it is in effect my ``fallback position'': if my own reply failed, I would 
endorse an analogue of his (Section \ref{sssec:detailclassl}).\\
\indent It also combines several of the themes we have introduced. For 
example, it is close to Section 4.2.2.B's idea of simultaneously defining 
velocity and persistence. More important, it is an interesting example of 
combining the two kinds of reply, (Appealing Differences) and (No 
Difference), in the way mentioned in Section \ref{sssec;tworeplies}. First, 
Sider appeals to non-obvious differences that other perdurantist  replies 
have not appealed to; (`non-obvious' because they are differences in the 
discs' environments, not in the discs themselves). But second, if these 
differences are ``imagined away'', along with the more obvious differences 
(like oblateness) that the RDA imagines away, then Sider turns to the (No 
Difference) reply: he ``bites the bullet'' and says that in such a world, 
there is no distinction between the discs. This second part of Sider's 
position is also interesting. For he does not turn to the (No Difference) 
reply merely as a matter of his philosophical  judgment, or ``intuition'': 
nor even as a matter of scientific methodology (as Callender argues). Rather 
it follows from Sider's {\em theory} of how the perdurantist should   go 
about defining persistence,  that in such a world there will be no 
distinction. 

Sider develops his position from the following  two independent 
components.\\
\indent (i): He notes the ``logical circle'' of the laws (of dynamics) and 
persistence. That is: the laws concern persisting objects, and so use the 
notion of persistence. But for perdurantists, persistence, i.e. the relation 
between stages of an ordinary persisting object, has a causal or nomic 
component---and Sider is happy to have it  be nomic, so that the notion of 
persistence presupposes the laws. In the face of this logical circle, Sider 
proposes the now-familiar tactic: Ramsey-Lewis simultaneous functional 
definition, so as to simultaneously specify persistence  and the laws from a 
single body of doctrine.\\
\indent (ii): He adopts the Mill-Ramsey-Lewis best-system theory of laws of 
nature (Lewis 1973, Section 3.3, p. 72-77).

\indent \indent Given these components, Sider's position follows swiftly. He 
writes (using `genidentity', where I have hitherto used `persistence', for 
the relation between stages (temporal parts) of an ordinary persisting 
object---which he calls `continuants'):
\begin{quote}
Consider various ways of grouping stages together into physical continuants. 
Relative to any such way, there are candidate laws of dynamics. The correct 
grouping into physical continuants is that grouping that results in the best 
candidate set of laws of dynamics; the correct laws are the members of this 
candidate set.\\
\indent More carefully. Any law of dynamics is a statement restricted to 
physical continuants, which may be rewritten in terms of the predicate 
`genidentity' as follows: `for any maximal genidentity-interrelated sum $x$ 
...'. Let $S$ be any axiomatization of any candidate set of laws of nature. 
Let $S$(genidentity) be the result of rewriting any dynamical laws in $S$ in 
terms of the genidentity  predicate. Where $G$ is any two-pace predicate 
variable, let $S(G)$ be the result of replacing all occurrences of 
`genidentity' in  $S$(genidentity) with $G$. ... relative to any assignment 
of a two-place relation {\bf G} to the variable  $G$, we can evaluate the 
strength [JNB: and the simplicity] of the resulting system $S({\bf G})$. We 
now define the best system and genidentity at once: they are the pair 
$\langle S({\bf G}), {\bf G} \rangle$, where {\bf G} is a two-place relation 
over stages and $S({\bf G})$ is the system that achieves the best 
combination of strength and simplicity.\\
\indent [The idea is that] we must look {\em globally}, across the entire 
world, to find what assignment yields the best  candidate  laws of dynamics. 
Thus although the states of a spinning disk may qualitatively match those of 
a stationary disk, what is going on elsewhere in the world may result in 
differences of rotation. [JNB: `result in' here means `imply' not cause] 
Suppose, for example, that  a stationary disk with a small hole is impacted 
by an object that seamlessly lodges itself in the hole, resulting in a 
perfectly homoegenous disk ... Suppose further that, in the possible  world 
in question, collisions generally result in transfer of momentum. The best 
simultaneous assignment of genidentity and laws of dynamics will then be one 
according to which this disk is spinning, for a pair of a genidentity 
assignment and set of laws on which the disk does not spin will not contain 
exceptionless laws governing the transfer of momentum. Now suppose further 
that, elsewhere in the same world, a disk with an empty niche had initially 
been spinning, and that a perfectly fitting object moving opposite to the 
direction of rotation collided with the disk. If the speeds and masses are 
appropriate, the best  assignment of genidentity and laws will have the 
result  that this second disk is stationary after the collision. The present 
view, therefore, allows the possibility of differences in rotation between 
homogeneous disks without appealing to non-Humean quantities.   (p. 230-231) 
\end{quote}

So Sider's leading idea is to have both the notion of genidentity  and the 
laws of dynamics ``get established'' in unproblematic cases, and then 
``projected'' to the problematic cases involving continuous  homogeneous 
matter. As he says:
\begin{quote}
it is crucial that the world contain plenty of unproblematic cases {\em not} 
involving uniform homogeneous matter. Once a certain candidate pair of laws 
and  genidentity gets its foothold in these unproblematic cases, it can then 
be projected into the problematic cases involving homogeneous objects, for 
this projection increases the strength of the candidate laws and does not 
decrease their simplicity. (p. 233)
\end{quote}

So Sider goes on to admit (in effect by way of concession to the argument of 
Zimmerman (1999), which I discussed in Section 4.3.2.A) that his view cannot
\begin{quote}
distinguish states of rotation in cases where there is not enough else going 
on in the world to give candidate pairs of genidentity and laws a foothold. 
In a world that contains only a homogeneous disk, the facts will not be 
sufficiently rich to allow one candidate pair to win out; there will 
therefore be no unique facts about genidentity, no unique spacetime worm 
that counts as a given spatial part of the disk, and no fact of the matter  
whether the disk spins or rotates. (p. 233-234)
\end{quote}
Sider then argues (p. 234-236) that he can ``bite this bullet''. (Though 
Sider does not explicitly discuss Zimmerman's space-filling homogeneous 
fluid, he would no doubt also bite the bullet in this case.) That is to say, 
in terms of Section \ref{sssec;tworeplies}'s two kinds of reply: for such a 
world, he adopts the (No Difference) reply; where, as I noted above, this is 
not just a matter of his philosophical  judgment, but follows from his {\em 
theory} of how to go about defining persistence---functional definition, and 
the best-system theory of laws.

 (Sider also argues that he can similarly bite another ``more general'' 
bullet, that arises from his overall strategy of defining both laws of 
dynamics and persistence by looking `{\em globally}, across the entire 
world'. Namely: the bullet that according to his account, whether or not a 
disc is rotating is a very extrinsic matter---i.e. it depends on what goes 
on in spacetime external to the disc. Sider, a good Humean, says he can 
accept this: indeed, for much the same  reasons that a Humean about 
causation accepts that a singular causal fact, say $c$ causes $e$, is 
extrinsic to the two relata $c$ and $e$.)

In metaphysics, I am an aspiring Humean: to that extent, I like Sider's 
position. But the endurantist  will no doubt reply that Sider's 
bullet-biting amounts to conceding the force of the RDA: `even Sider's 
version of perdurantism, with its  sophistication about the account of 
persistence invoking the laws of mechanics, cannot secure facts of 
persistence in the troublesome cases considered by the RDA'. In other words: 
Sider's (No Difference) verdict suggests that after all, there is at best a 
stalemate.\\
\indent From Section \ref{sec:soln} onwards, I will argue that the 
perdurantist  can do  better than this stalemate. But before turning to 
that, we need to consider ...

\section{Describing rotation}\label{ssec;describerot} 
So much, for the moment, for metaphysics!  I now return to Section 
\ref{sssec;persesptlraised}'s demand that the advocate of the RDA (indeed 
all parties to the dispute) should state and justify their claims about 
spatiotemporal structure---i.e. the claims they need to make, in order that  
statements of rotation {\em make sense}. I begin by stressing the need for 
precision (Section \ref{ssec;needprecise}). Then I report how physics 
rigorously describes states of rotation (Sections \ref{131needconnect} and 
\ref{132connectmetrot}), and review how these technicalities bear on the 
endurantism-perdurantism debate (Section \ref{133Allowboth}). This yields 
two of the paper's three main conclusions:\\
\indent (i): the RDA can be formulated more strongly than is usually 
recognized: it is not necessary to ``imagine away'' the dynamical effects of 
rotation (Section \ref{134Goodbadunnecy}); but\\
\indent (ii): in general relativity, the RDA fails (even in the strengthened 
version), because of frame-dragging (Section \ref{432:RDAfailsGR}).

\subsection{The need for precision}\label{ssec;needprecise}
As I argued in Section \ref{sssec;persesptlraised}: to get a grip on the two 
possibilities that the RDA urges on us, it is certainly not enough to just 
draw or visually imagine the contrasting diagrams, with straight and helical 
worldlines. For such diagrams implicitly assume that the 
rotation/non-rotation distinction is defined in terms of a space of 
persisting spatial points; and the question arises what account either 
party, endurantist or perdurantist, can give of such points---or of whatever 
(maybe more technical) notions they need, or choose, to use so as to 
describe rotation.

Besides, this question is also brought out by Callender's (No Difference) 
reply in Section \ref{sssec;tworeplies}. Callender's claim that the 
(Stat)/(Rot) distinction is as spurious as that between (Up) and (Down)  is 
essentially the claim that, pending some further account, the 
straight/helical contrast for a diagram's worldlines can be dismissed as an 
artefact of the diagram: I can change  coordinate  system to make what I 
drew as straight (helical) be now drawn as helical (straight), just as I can 
change coordinates to make an arrow drawn pointing upward get drawn as 
downward.

The general point here is that diagrams can carry implicit assumptions or 
connotations that a certain distinction makes sense (aka: `is physically 
significant/real')---and that one can propose, or hope to have, a theory of 
motion in which that distinction is in fact denied. 

This is a familiar point in the philosophy of geometry. A standard simple 
example, much like the up/down one, is the description   of 3-dimensional 
Euclidean space with cartesian coordinates, i.e. as $\mathR^3$. The diagram 
of the three axes suggests a distinguished point, and three distinguished 
directions: a connotation we immediately do away with by emphasising how we 
can equally well choose coordinate systems with other origins and-or 
axis-directions.\footnote{This leads into a large mathematical  subject, 
which goes back to Klein: articulating the geometric structure of a space by 
singling out a class of coordinate systems that gives its structure an 
especially simple expression, and stating the group structure of this 
class.}

Indeed, this sort of rectilinear example is very relevant to the RDA. For 
the RDA can be---and sometimes has been---developed using, instead of discs 
(or spheres, cylinders etc.) and rotation: rivers of homogeneous continuous 
matter undergoing an  homogeneous  steady flow, i.e. with the velocities  of 
all the point-sized bits of matter being the same as each other, and 
constant in time.\\
\indent Thus the argument against perdurantism would be that the 
perdurantist apparently cannot distinguish between the river being 
stationary and flowing steadily. (But most authors in the RDA literature who 
mention rivers do not confine themselves to steady homogeneous flow: they 
gesture at the endless variety of possible flows, with all sorts of eddies, 
which allegedly all ``look the same'' to the perdurantist or Humean.\\
\indent And the argument prompts the now-familiar two kinds of reply. 
(Appealing Differences): Can the perdurantist distinguish the cases by 
appealing to, for example, motion relative to the river bank? And if not, 
say because the river is the only thing in the world (say, filling all 
space), can she appeal to instantaneous velocity or causation? On the other 
hand, (No Difference): can the perdurantist deny that there is a 
distinction?

So in fact our topic in this Section is, not just how is rotation rigorously 
described, but: how is all motion, even rectilinear  motion, rigorously 
described? But I shall emphasise rotation, since:\\
 \indent (i) the RDA literature does so, and:\\
\indent (ii) nowadays, the rejection of absolute space makes an argument, 
based on the stationary vs. irrotationally steadily flowing river, look 
weak. That is: if the river is ``lonely'', the only thing in the world, then 
the (No Difference) reply seems convincing. So rotation seems to give the 
endurantist their best chance of making trouble for perdurantism.

The rigorous description of motion, and especially  rotation, in modern 
geometry and physics   is a very large and subtle subject. But to assess the 
RDA we can fortunately  make do with some simple points.  I  start in 
Section \ref{131needconnect}, simply and traditionally, by discussing how 
Newton argued for persisting spatial points, i.e. an absolute space, by 
appealing to the dynamical effects of rotation. This will lead  to Section 
\ref{132connectmetrot}'s  summary of some aspects of the modern {\em 
kinematical} description of rotation.

\subsection{Motion needs a connection}\label{131needconnect}
In his bucket and globes thought-experiments, Newton appealed to the 
dynamical effects of rotation to argue that the theory of motion needed to  
postulate an absolute space of persisting spatial points. (At least this is 
the usual reading: but for subtleties and controversy, cf. e.g. Rynasiewicz 
1995, Mainwood 2004.) So at first sight, it seems that both   the 
endurantist and perdurantist might hope to appeal to, or adapt, Newton's 
argument so as to give an account of the persistence of spatial  points.\\
\indent (Here I assume that for the advocate of the RDA, this does not 
conflict with the fact that the RDA imagines away such dynamical effects. 
The idea is that the advocate  follows Newton in arguing for the {\em 
actual} existence of absolute space; but then says that for the purposes of 
the argument, the dynamical effects can be imagined away---and that this is 
not so ``unlike the actual world'' as to let the perdurantist off the hook 
of having to distinguish the cases of (Stat) and (Rot).)

But Newton's arguments (and their ilk)  can be resisted. There are two 
points here, of which the second is more important for us.\\
\indent (i): Those inclined to relationism about space and time (like 
Leibniz and Mach) will say that the correct account of space must be based 
on relations between material bodies---and that therefore for a ``lonely'' 
disc, i.e. a disc alone in the universe, there can be no distinction between 
rotation and non-rotation. In effect, Leibniz and Mach hoped to  develop a 
mechanics in which Newton's arguments would fail, because the mechanics 
would  contain a law that vetoes Newton's putative possibilities in which 
the total material content of the universe rotates: the law would  require 
that the total angular momentum of the universe be zero. Such a relational 
mechanics has now been developed, especially by Barbour et al.; (for 
discussion and references, cf. Earman 1989, p. 27-30, 92-96, Belot 2000, p. 
570-574, 580-582, Pooley and Brown 2002, Butterfield 2002 296-311). But I 
set these theories aside in what follows.

\indent (ii) Nowadays, it is clear that, even apart from alternative 
relationist mechanics, Newton's arguments fail in one precise sense. That 
is: all now agree that:\\
\indent \indent (a) Though (relationism apart) Newton was justified in 
inferring from the dynamical effects of rotation that acceleration had an  
absolute (i.e. coordinate-independent) physical significance;\\
\indent \indent (b) And though Newton was justified, within the mathematics 
of his time, in inferring that absolute   acceleration could only make sense 
if there was also absolute velocity (since acceleration seems to be ``just''  
the time-derivative of velocity), and thereby also absolute position (since 
velocity  seems to be ``just''  the time-derivative of position);\\
\indent \indent (c) Nevertheless, modern mathematics enables us to make 
sense of absolute acceleration (and its quantitative measures), and so of 
the contrast between straight and helical worldlines, {\em without} an 
absolute space ({\em and} without having a notion of absolute velocity).

 The idea in (c) is mathematically  subtle: it only became clear in the 
1920s (in the work of Weyl, Cartan etc.) after relativity theory prompted 
physicists to think in terms of spacetime concepts. But I will  not need to 
develop it in detail (cf. e.g. Sklar 1974 pp. 202-206, Earman 1989 p.33). 
Here it suffices  to say that we    postulate a geometric structure on 
spacetime called an {\em affine connection} (for short: connection), which  
essentially defines a notion of straightness, and thereby notions of amounts 
of curvature, for curves  in spacetime. Applied to timelike curves, these 
are notions of unacceleratedness, and amounts of acceleration. A spacetime 
that is non-relativistic (has a notion of absolute simultaneity) and is 
equipped with such a connection---but is not equipped with a notion of 
absolute space, that induces the connection---is called {\em neo-Newtonian} 
or {\em Galilean}.

So to sum up (ii):  we {\em can} make sense of absolute acceleration without 
absolute rest or absolute velocity. Although a connection {\em can} be 
legitimately defined by a notion of absolute rest, as Newton in effect did, 
it  is a logically weaker idea than a notion of absolute rest, and so can be 
postulated directly---without the rest. Besides, relativistic theories (both 
special and general)  also have a connection in  just this way (without 
absolute rest): they differ in that they {\em also} lack absolute 
simultaneity.

\subsection{Connections, metric and rotation}\label{132connectmetrot}
So let us ask: how is motion, and in particular  rotation, described using a 
connection? I shall summarize the answer in three Subsections. The first 
introduces the ingredients needed for describing motion; the second gives 
more details about the description of rotation; and the third reports some 
subtleties of general relativity. 

\subsubsection{Common ingredients}\label{421commoningredts}
The first thing to say is that  most (but not all!) of the  ingredients for 
describing motion are {\em the same} in  most spacetime  theories: both 
non-relativistic (with and without absolute space: Newtonian and 
neo-Newtonian) and relativistic (special and general). (Again, I set aside 
the relational theories of Barbour et al.) But we will see in Section 
\ref{423rotnGR} that general relativity has some very special features.\\
\indent The foremost ``common ingredient'' is that all these theories  
describe rotation  by invoking two types of mathematical structure, which 
mesh in an appropriate way.\\
\indent  The first type is relatively familiar: it is metrical structure, 
which we can think of as primarily assigning a length to curves in 
spacetime. In relativistic theories, there is a single notion of length for 
all  curves: a spatiotemporal metric. In  non-relativistic theories, there 
are two notions of length---spatial length for spacelike curves, and 
temporal length for timelike curves: so there is a spatial metric and a 
temporal metric. For both kinds of theory, I will speak of `a spatial metric 
and a temporal metric'.\\
\indent The second type of structure  is the connection, which gives a  
standard of straightness, and numerical degrees of curvature, for an 
arbitrary curve in spacetime, and so in particular for the worldline of a 
point-particle, or point-sized bit of matter in a continuous body.\\
\indent The meshing required between the two types of structure is called 
{\em compatibility}. (In fact, in relativistic theories (whether special or 
general) any spatiotemporal  metric  has a unique compatible connection; but 
in non-relativistic theories, the two metrics (spatial and temporal) do not 
fix a unique compatible connection.)\\
\indent So to sum up: Only once we have in hand a spatial metric, temporal  
metric and a compatible connection, does the judgment that a disc is 
rotating---that its matter's worldlines are ``helical rather than 
straight''---even {\em make sense}.\footnote{Or rather, this is true once we 
set aside relational mechanics. More precisely: only with these structures 
can one make sense of the judgment that the disc is rotating, irrespective 
of its relations to other bodies---and in particular, if it is lonely.}

\subsubsection{Details: the rotation tensor}\label{422omega}
This Subsection and the next  spell out some details about how the metrics 
and connection give a framework for describing rotation. This Subsection 
makes two points which are in common between the theories; the next makes 
points which are specific to general relativity.

\indent (a) {\em Acceleration of a single particle}:\\
 For a single worldline, i.e. the worldline of a single point-particle, the 
connection defines at each point along the worldline a (four-dimensional) 
acceleration of the point-particle. Using the metrics, one can also define 
the more familiar three-dimensional acceleration. This point  holds good in 
both non-relativistic and relativistic theories.

(b) {\em Rotation, local and non-local}:\\
 Though one can define in these theories  the  rotation of one 
point-particle relative to another, this notion is not usually treated in 
the textbooks: (and general relativity holds some surprise about it---cf. 
(b) of Section \ref{423rotnGR}). Nor is the notion of a swarm of 
point-particles rotating about another particle (or about a spatial point) 
treated in the textbooks. In fact, they concentrate on the case of most 
relevance to the RDA: the case where we are given a {\em congruence} of 
timelike curves, i.e. a continuously infinite collection of worldlines whose 
points of intersection with a (possibly finite) spacelike slice completely 
fill the slice. (So the worldlines might be  given as the integral curves of 
the 4-velocity vector field of some continuous matter.) And for this case, 
the textbooks define a {\em local} notion of rotation.\\
\indent That is: the metrics and compatible connection together define at 
each point in the congruence a {\em rotation tensor}, usually symbolized as 
$\bf \omega$, which gives a quantitative measure of the speed and direction 
of rotation (of the congruence) in an arbitrarily small neighbourhood of 
that point. Roughly speaking, $\bf \omega$ at a point in spacetime   encodes 
how an observer located there sees the limitingly close worldlines of the 
congruence swirling around her. For us, this construction is important in 
two main ways.\\
\indent \indent (i): The construction of $\bf \omega$ proceeds in much the 
same way in the different theories; (for more details, cf. e.g. Misner et 
al. 1973, p. 566; Dixon 1978, p. 121-128, 140-145, 163-166; Wald 1984, p. 
216-218). \\
\indent \indent (ii): The construction is a {\em robust} local limit of 
other {\em  non-local} definitions of rotation. By this, I mean the 
following; (I thank David Malament for explaining this).  There are various 
intuitively compelling (and experimentally realizable) criteria for whether 
an extended object, such as a disc, is rotating; but as one considers 
smaller and smaller discs, the verdicts of these various criteria as to 
whether a given disc is rotating {\em converge} on the verdict given by the 
rotation tensor (i.e. by whether or not $\bf \omega = 0$). So the local 
notion given by the rotation tensor is a common ``robust'' limit of the 
other criteria. Besides, this is so in all the different theories.\\
\indent \indent \indent To give an example: one such criterion supposes that 
an observer  at the centre of the disc bolts a telescope to a water-bucket 
and then continually observes a light source fixed on the edge of the disc: 
the disc is judged to be rotating iff the water-surface is concave. Another 
criterion supposes that a light source fixed on the edge of the disc sends 
light-signals right  around  the edge of the disc, in both directions, and 
asks whether the two signals arrive back simultaneously: the disc is judged 
to be rotating iff there is a difference.\\
\indent In (c) of Section \ref{423rotnGR}, I will discuss how  in general 
relativity, these criteria (and others) can {\em disagree} in their verdicts 
about whether an extended  disc is rotating. But for the moment, I just make 
the more ``positive'' point that as the disc shrinks in size, all these 
criteria must (in {\em all} the theories) tend towards agreeing with each 
other---and with the mathematical condition that the rotation tensor $\bf 
\omega$ at the centre of the disc is non-zero.\footnote{By the way: to 
define, not the rotation tensor, but  merely the qualitative distinction, 
rotating vs. non-rotating,  one does not need all of Section 
\ref{421commoningredts}'s ingredients, metrics and compatible connection. 
One needs only a {\em conformal structure}: which is, roughly speaking, a 
structure in which angles are meaningful but lengths are not.  A conformal 
structure can be encoded by an equivalence class of metrics, with 
equivalence classes  $[g_0] : = \{g: g = \Omega g_0\}$; where $\Omega$ is a 
positive smooth $\mathR$-valued function on spacetime, and  where this 
structure is to be again compatible with the connection (where `compatible' 
is spelt out by adapting the usual relativistic and non-relativistic 
conditions for a given metric by an existential quantifier over the 
equivalence class).}

\subsubsection{Rotation in general relativity}\label{423rotnGR}
I turn to report three points which indicate the subtlety of rotation in 
general relativity. The first point is standard material in the physics 
textbooks: but worth reporting since, as we shall see, it implies that the 
RDA fails in general relativity. The second and third points are specialist 
knowledge: striking results by Malament (2002, 2003).

(a) {\em Frame-dragging}:
According to general relativity, there is an (amazing) physical effect of 
rotation, understood in Section \ref{422omega}'s sense that ${\bf \omega} 
\neq 0$), on spacetime itself.
Namely, a rotating body distorts its nearby spacetime geometry; or as it is 
more vividly put, the rotating body ``drags'' the inertial frames in its 
vicinity (hence the name `frame-dragging'). That is: test particles falling 
freely under gravity near a body move differently, according to whether the 
body is rotating (and in what sense, and how fast)---they ``feel'' not just 
the mass of the body, but also its state of rotation.  (Cf. Misner et al. 
1973, p. 699, 879, 1117.)\\
\indent The theory of this effect goes back to 1918 (by Thirring and Lense). 
The  effect is numerically minuscule, even when the rotating body is very 
massive, e.g. the earth. Yet the dragging of frames by the rotating earth 
may soon be detected.\footnote{Namely, by tiny supercooled gyroscopes in an 
orbiting satellite, recently launched. Some numbers make  vivid how 
ambitious, and delicate, is this experiment (called `Gravity Probe B').  
After a year in orbit, the drag on a gyroscope will be 42 milliarc-seconds, 
which is the angle subtended by a metre-stick at a distance of 3000 miles, 
or the thickness of a sheet of paper at a distance of a mile. To prevent 
this minuscule effect being masked by random thermal motions, the gyroscope 
must be cooled to very close to absolute zero; and then one has to measure 
the effect by radio contact with the satellite a year after its launch.  No 
wonder the experiment has been designed over some thirty years!}      

(b) {\em Relative rotation of two particles}\\
 In this and the next point, I report two much less well-known 
peculiarities---indeed surprises---of rotation in general relativity. But 
they are only needed briefly later (in (B) of Section \ref{432:RDAfailsGR}); 
so the reader can skip to the summary of this Subsection.\\
\indent For a pair of worldlines, $X$ and $Y$ say, in either a 
non-relativistic or a relativistic theory, one can define a ``direction from 
$X$ to $Y$'' (and {\em vice versa}), and its ``rate of change'', and thereby 
define the ``angular velocity of $Y$ relative to $X$''. Besides, the 
physical ideas behind these definitions are natural, and  similar, in 
non-relativistic and relativistic theories: e.g. in relativity theory, the 
direction from $X$ to $Y$ is  given by the direction of the tube of a 
telescope held by an observer on $X$ who continually observes $Y$.  (And one 
can extend these definitions so as to talk about a collection of worldlines 
(particles) $Y_1, Y_2, \dots$ rotating relative to a given particle $X$.)\\
\indent But even with just two worldlines, general  relativity  surprises 
us. In non-relativistic theories, the defined notions have the expected 
properties: in particular, the angular velocities of $Y$ relative to $X$, 
call it $\omega_{XY}$, and of $X$ relative to $Y$, $\omega_{YX}$, are equal. 
But this is {\em not} so in  general relativity. In this theory, $Y$ can be 
non-rotating relative to $X$, i.e. $\omega_{XY} = 0$, while $X$ rotates 
relative to $Y$, i.e. $\omega_{YX} \neq 0$; and this can be so while the 
distance between $X$ and $Y$ (in any reasonable sense of `distance') remains 
constant, and is as small as you care to demand! (For details, cf. Malament 
2003.)

(c) {\em Conflicting criteria of rotation}\\
 Finally, general  relativity  also holds considerable surprises about the 
RDA's case: the rotation of a disc (or a sphere or hoop). One surprise is 
that different intuitively compelling (and experimentally realizable) 
criteria for whether a disc is rotating can give different verdicts---not in 
all spacetimes, but in some. Thus recall the two examples from (b) of 
Section \ref{422omega}. The first asks if the water-surface in the bucket at 
the centre of the disc is concave; the second asks if light-signals 
circumnavigating the disc in opposite directions arrive back at different 
times.   These criteria will match in their verdicts for any disc in a 
non-relativistic spacetime, or in the Minkowski spacetime of special 
relativity. Besides,   in any general relativistic spacetime, they must tend 
to agreement with each other, and with whether the rotation tensor is 
non-zero,  for smaller and smaller discs (as discussed in (b):(ii) of 
Section \ref{422omega}). But for a disc of given size, there are general 
relativistic spacetimes in which the verdicts will {\em differ}; e.g. Kerr 
spacetime (Malament 2002).\\
\indent Indeed, more is true: {\em any} criterion of rotation for a disc 
must violate {\em some} intuitively compelling condition in {\em some} 
general relativistic spacetime! More precisely: Malament (2002) shows that 
any criterion of rotation (in the very weak sense of a binary 
classification, for any disc in any state of motion, as to whether it 
qualifies as rotating), that agrees with the water-surface criterion (and so 
with the rotation tensor criterion: ${\bf \omega} \neq 0$?) in the limit of 
smaller and smaller discs, must violate another compelling condition, when 
it is  applied to a spacetime like the Kerr spacetime. (This other condition 
is roughly: if a disc $d_1$ is not rotating, and $d_2$ is rigidly attached 
to $d_1$ in the sense that the distance between any two point-sized bits of 
matter in $d_1$ and $d_2$ is constant over time, then $d_2$ is also not 
rotating.)

To sum up Sections \ref{131needconnect} and \ref{132connectmetrot}:--- To 
make sense of rotation, it is by no means enough to draw  straight vs. 
helical worldlines. One needs a considerable body of theory: specifically, 
spatial and temporal  metrical structure, and a compatible connection 
(whether or not induced {\em a la} Newton by a notion of absolute rest). 
With this equipment, one can define (in much the same way in the different 
theories) a robust local notion of rotation, expressed by the rotation 
tensor. But in general relativity, rotation has  complex and even 
counter-intuitive features, especially as regards the rotation of extended 
bodies---like a disc.
  
I believe this technical material yields two significant conclusions about 
the RDA, which I will  develop in Sections \ref{134Goodbadunnecy} and 
\ref{432:RDAfailsGR}. But first (Section \ref{133Allowboth}), I need to 
connect this material to the endurantism-perdurantism debate in general, by 
stating some familiar general assumptions about the bearing of physical 
theories on metaphysical theses.

\subsection{The endurantism-perdurantism debate in the light of 
physics}\label{133Allowboth}
I began this Section by recalling Section \ref{sssec;persesptlraised}'s 
demand that both endurantist and perdurantist should state and   justify the 
claims they need to make about spatiotemporal structure. We have now seen 
that these claims are technical; and that physics always formulates them  in 
terms of equipping a  manifold of spacetime points with various mathematical 
structures. So even apart from the RDA, the question arises whether the 
endurantist and perdurantist have ``equal rights'' to these claims. 

This is  a large question. I will lay out some of its aspects, but not 
pursue them in detail. My reason is that I want to give endurantism some 
rope. That is: since I want to give the RDA against perdurantism as good a 
run as possible, I will give endurantism the benefit of the doubt about its 
right to these claims. 

\indent (i) {\em Traditional associations}:\\
Our question has two traditional associations:\\
\indent\indent (a):  Perdurantists have traditionally argued that their 
position fits much better than does endurantism with the description of 
matter, space and time in modern physics; and in particular, with the 
description  in spacetime theories like relativity theory. On the other 
hand, endurantists often  distinguished the conceptual schemes of physics 
and everyday thought, and took their position to be about the latter.\\
\indent\indent (b): The endurantism-perdurantism debate is also 
traditionally aligned with the debate whether there is objective ``temporal 
becoming'', as against the ``tenseless'' or ``block universe'' theory of 
time being true. Again, modern physics, especially relativity theory, has 
been taken to support the tenseless view, and thereby perdurantism. 

\indent (ii) {\em Today's debate}: \\
But nowadays, endurantism's traditional associations, as sketched in (a) and 
(b) of (i), are broken.\\
\indent As regards (a), many endurantists are ``scientific realists'', and 
even substantivalists about spacetime. They believe that successful 
scientific theories like relativity theory, literally construed, are 
approximately true; and even that spacetime points  are {\em bona fide} 
objects bearing the properties and relations represented by mathematical 
structures like metrics and connection.\\
\indent As regards (b), the tenseless view is nowadays often called 
`eternalism'---and many endurantists endorse it. (On the other hand, the 
currently most popular version of becoming seems to be presentism, the 
doctrine that only the present exists---which is no doubt at least as hard 
to reconcile with relativity's denial of absolute simultaneity, as are other 
versions of temporal becoming.)\\
\indent I think the breaking of these associations reflects both: the rise 
of philosophical naturalism (cf.  Section \ref{ssec;naturalism}); and (more 
contentiously!), the difficulty of defending (or even making sense of!) the 
idea of temporal becoming. In any case, the upshot is that nowadays, the 
endurantist is likely to claim ``full rights'' to the technical claims of 
modern spacetime theories, just as much as perdurantist does. (For more 
discussion of the current standing of both traditional associations, (a) and 
(b), cf. Butterfield 2004 and Sider 2001 pp. 75-76, 110-119.) 

\indent (iii) {\em Three questions}: \\
Accordingly, I think that our large question breaks down, at least nowadays, 
into the following three questions, (A) to (C). The first two questions,  I 
propose to set aside, since they are independent of the endurantism vs. 
perdurantism debate.

\indent \indent (A): The first pertains to general philosophy of science. It 
is the question: should one be (a) a scientific realist about the 
theoretical claims of modern spacetime theories, especially general 
relativity, or (b) some sort of instrumentalist (maybe constructive 
empiricist) about them?  So in setting this question  aside, I shall in 
effect speak like a scientific realist: which is anyway the widespread 
practice of much current discussion of endurantism and perdurantism---cf. 
(ii) above.

\indent \indent (B): The second question pertains to the philosophy of 
(chrono)geometry. Even if we are scientific realists, and even 
substantivalists, there is a further question about how we should interpret 
spacetime points' properties and relations as represented by e.g. metrics 
and connection. One view is that they are in some strong sense independent 
of the physics of matter (and radiation), at least in theories where the 
metrics and connection are not dynamical. (I think this is the dominant view 
among substantivalists.) An alternative view is that these properties and 
relations {\em are} dependent on  the physics of matter: for they are a way 
of compendiously representing some features, especially invariances, of the 
dynamical equations governing matter. (This is the view---at least as I read 
them!---of Brown and Pooley; cf.  Brown and Pooley 2001, 2004.) Again, I 
shall not pursue this question; so I shall in effect speak like a 
substantivalist---which is anyway widespread in current discussion.

\indent \indent (C): So: setting (A) and (B) set aside, and assuming 
substantivalism, and (as in (i)) that the perdurantist thereby has ``full 
rights'' to the technical claims of modern spacetime theories, our question 
becomes: does an endurantist have equal rights to them?\\
\indent Note that even with these assumptions in place, one can envisage two 
versions of endurantism. The first version is committed to spacetime points 
and sets of them (and their properties and relations), but does not accept 
spacetime regions as spatiotemporally located objects (say as mereological 
fusions of points). Rather, regions are to be treated as sets of points; and 
sets are ``abstract'' in at least the sense that they are not located in 
spacetime---and therefore in no sense persist. Since such regions, taken to 
be spatiotemporal  objects, would surely persist by perduring, not enduring, 
this version's denial  that regions are spatiotemporal  objects enables it 
to hold, not only that ordinary material objects endure, but also that {\em 
no} spatiotemporal object perdures.\\
\indent  On the other hand, a second version of endurantism accepts 
spacetime  regions as perduring objects, and so is what I will call a {\em 
mixed view}:  some objects endure, but others perdure. (These others include 
at least spacetime regions, but maybe also other  objects which one might 
call `events', like wars or meals; (for more discussion, cf. Butterfield 
2004).)

\indent So our question is whether either of these versions of endurantism 
has as much right to the technical claims of modern spacetime theories as 
the perdurantist does.\\
\indent  A proper answer to this question would have to investigate two main 
topics:\\
\indent \indent [i]: whether there are problems about the mixed view (i.e. 
the mixed view for spacetime points---nevermind wars and meals); and\\
\indent \indent  [ii]: whether relativity theory makes problems for 
endurantism (as it certainly does for temporal becoming);\\
\indent and then decide whether any such problems could be solved.

As I announced above,  here I want to give the RDA against perdurantism as 
good a run as possible, and so I will simply {\em assume} that (setting 
aside the RDA) there are no such problems---that the answer to both [i] and 
[ii] is `No'.\\
\indent This assumption can be partly defended by appealing to a  formal 
equivalence between the ways that endurantism and perdurantism  describe the 
motions of point-particles and continua (in both non-relativistic and 
relativistic spacetimes). The idea of the equivalence is that:\\
\indent (a): an endurantist will represent the motion of a point-particle, 
or a point-sized bit of matter in a continuum, by a single function $q: t 
\mapsto q(t) \in {\cal M}$, mapping times at which it exists to locations in 
a manifold $\cal M$ (either space or spacetime); while\\
\indent (b): the perdurantist will use a collection of functions, labelled 
by time-intervals that together cover the object's lifetime; for example, if 
it exists throughout the closed time-interval $[a,b]$, there might be a 
function $q_{[a,b]}: t \in [a,b] \mapsto q_{[a,b]}(t) \in {\cal M}$.\\
\indent I develop this equivalence (including extending it to spatially 
extended objects), and relate it to both [i] and [ii] in Butterfield 2004 
and 2004a. But I should also note that this equivalence gives only a partial 
defence of the assumption, that the answer to both [i] and [ii] is `No'. For 
the equivalence is formal; and formal equivalences are liable to be broken 
by philosophical considerations.  (For more discussion, cf. e.g. Balashov 
1999, 2000 for [ii]; and Sider 2001 p. 110-119 for [i], and p. 79-87 for 
[ii].)

\subsection{Dynamical effects revisited}\label{134Goodbadunnecy}
So much by way of general connections between the physics of rotation and 
the endurantism-perdurantism debate. I now return to the RDA, and to 
arguing, in this Subsection and the next, for two main conclusions.
  
In this Subsection, I return to the accompaniments of rotation, especially 
dynamical effects like oblateness. Since Section 
\ref{sssec;accompanimtsrotn} (and especially since Section  
\ref{sssec;tworeplies}) I have assumed, along with the metaphysical 
literature, that the RDA's advocates need to justify ``imagining away'' any 
accompaniments that the perdurantist might latch on to as marking the 
distinction between the discs, (Stat) and (Rot). 

The physics of rotation---the material in Sections \ref{131needconnect} and 
\ref{132connectmetrot}---yields three main points about this assumption, 
which I develop in the next three Subsections. In short:---\\ 
\indent (i): One main theme of Sections \ref{131needconnect} and 
\ref{132connectmetrot} supports the practice by the RDA's advocates, of 
ignoring such accompaniments (Section \ref{451rotnkinematic}).\\
\indent (ii): But the advocates are lucky---it is an undeserved 
victory---since their {\em avowed} reasons for ignoring such accompaniments, 
especially  dynamical effects, are {\em worse} than what these Subsections 
provide (Section \ref{452cheapimaginatn}).\\
\indent (iii): But in any case, the RDA can be developed very effectively,  
{\em without} imagining away all such accompaniments (Section 
\ref{453RDAwitheffects}).   

\subsubsection{Rotation is kinematic}\label{451rotnkinematic}
We saw in Section  \ref{422omega} that rotation---and its quantitative 
measures, given by the rotation tensor locally, and by various criteria 
non-locally---is definable in a wholly {\em kinematic} way: i.e. without 
mention of dynamics. In less abstract terms, it is definable in terms of 
acceleration, without mention of forces (or more generally, the causes and 
effects of motion).

\indent Agreed, it was perhaps Newton's greatest insight to couple 
acceleration and force, viz. in his second law of motion ${\bf F} = m{\bf 
a}$. But that is a nomic, not logical connection. Force could instead be 
coupled to velocity, i.e. the first derivative of position (an 
``Aristotelian mechanics''), or to a higher derivative than the second: such 
alternative schemes change physical behaviour enormously, but are logically 
coherent.

\indent (Indeed, in a framework in which force is coupled to velocity, but 
which is otherwise as close to classical mechanics as possible, a body's 
future motion would be determined by its position, and the force acting on 
it. This would imply, I take it, that:---\\
\indent \indent (a): Metaphysicians like Tooley, Bigelow and Pargetter would 
feel no temptation to introduce a heterodox intrinsic  notion of velocity to 
act as cause and {\em explanans} of future positions. \\
\indent \indent (b): More importantly for us: the RDA would have much less 
bite, at least for a ``naturalistic'' perdurantist, who is willing to let 
her account of perdurance depend on the actual laws. For now ingredients 
that the RDA agrees to be available to the perdurantist (more generally, the 
Humean), viz. position and forces, {\em would be} enough to determine future 
positions. I will return to this in the context of quantum theory (Section 
\ref{511 Unitarity}).

\indent  More radically,  rotation  makes sense without any  forces at 
all---nevermind how force couples to kinematic quantities. To talk in terms 
of possible worlds: there are  worlds with a spacetime manifold, spatial and 
temporal metrics and compatible connection, and a congruence of timelike 
curves representing continuous matter---again, nevermind the forces. A pair 
of these worlds can match in countless ways and yet differ  as to whether 
the matter is rotating, in say the usual local sense, i.e. at some  given 
point in their common spacetime. Just suppose that in one world the rotation 
tensor is zero at the point, while in the other it is non-zero. 

I take this as evidence that  perdurantism should strive to accommodate the 
distinction between these possibilities.\\
\indent I do not claim that it is conclusive evidence. Some perdurantists 
such as Callender (Section \ref{sssec;tworeplies}) will still prefer the `No 
Difference' reply to the RDA. That is: they will say that worlds with no 
dynamics are so unlike the actual world, that perdurantists have no 
responsibility to distinguish rotation and non-rotation within them (cf. 
Callender 2001, p. 38). \\
\indent But I do not need to  resolve this dispute between myself and 
fellow-perdurantists. For all perdurantists can agree to the more important 
conclusions in the following Subsections.

\subsubsection{Beware of rigidity}\label{452cheapimaginatn}
On the other hand, I think that advocates of the RDA have often had worse 
reasons than than that just given, for insisting that the perdurantist 
should  distinguish (Stat) and (Rot) even without any of the usual 
accompaniments of rotation. I will not try to catalogue people's errors, but 
will focus on one prevalent reason. (Parts of this Subsection's critique 
will carry over to versions of the RDA that  use a homogeneous fluid, rather 
than a rigid solid.) 

This reason is the belief that it is entirely straightforward to ``imagine 
away'' the accompaniments, since  one only needs to stipulate that the discs 
are perfectly rigid. This implies in particular  that the rotating disc will 
not be oblate---and so the RDA will be posed, to the consternation of the 
perdurantist.

This reason is defective in two ways. First: to say these two words 
`perfectly rigid', so ``trippingly off the tongue'', is to forget that 
within the theories of classical continuum  physics, perfect rigidity  is a 
very strong  idealization---it violates central principles of these 
theories.\\
\indent To take our example:  what in fact would happen when a  (classical, 
continuous, homogeneous) stationary disc is given a push at its edge to make 
it rotate, is very complicated. A disturbance would  travel outward  (at the 
speed of sound for the disc's material) from the place where the push is 
applied,  leading to a  complex process that  settled down so that the whole 
disc rotated approximately uniformly, with internal cohesive forces exerting 
the required centripetal forces on parts of the disc. (In the actual quantum 
world, this description is a very good approximation for solid discs,  the 
cohesive  forces being electromagnetic forces between atoms. But I am here 
just assuming  a  classical continuum treatment.) Without going into further 
details, this is enough to bring out that assuming perfect rigidity requires 
that the disc's cohesive forces should respond ``infinitely quickly'' to 
distorting influences. More precisely, it amounts to vetoing any account of 
how the whole disc is set in motion as a consequence of the motions of the 
parts. (In physics jargon: it vetoes any  constitutive theory.)\footnote{Two 
incidental remarks about rigidity. (1): There is also the worry that perfect 
rigidity violates relativity's prohibition on faster-than-light signals.  
But in fact, relativistic theories allow  generalized notions of rigidity: 
for a philosopher's introduction, cf. Earman (1989, Chapter 5.5, pp. 
98-101).  (2): Among Bigelow and Pargetter's arguments for their heterodox 
account of instantaneous velocity, as not always a limit of average 
velocities, is a thought-experiment involving perfectly rigid spheres (1989, 
pp. 292-293, 1990, pp. 67-68). As it happens, I disagree with their 
argument, but I will not go into details: as I have said (starting in 
Section \ref{ssec;intrpropvelies}), I am not convinced by such heterodox 
accounts of velocity---and my reply to the RDA does not need them.}

Second: it is {\em not} true that perfect rigidity gets rid of all the 
actual technical accompaniments of rotation. For not all such accompaniments 
are kinematically manifested, i.e. associated with changes in shape or size,  
like   oblateness. There are also forces and energies that {\em would} be 
present in a perfectly rigid rotating disc. (In physics jargon: some 
dynamical  effects of rotation involve stress rather than strain.)  There 
will be cohesive forces throughout the disc's interior which would be absent 
if the disc were stationary: besides, the disc's energy is greater---which 
means in relativity that its {\em mass} is greater. Though such 
accompaniments are more technical, less common-sensical, than being oblate,  
that is  no reason to think the perdurantist is less able to appeal to them, 
so as to distinguish (Stat) and (Rot).\footnote{At least: it is only a 
reason if we take the endurantism-perdurantism debate as entirely a matter 
of  analyzing everyday concepts. In particular, the RDA cannot just consider 
an oblate rotating disc and a non-rotating one moulded so as to be congruent 
to it (as proposed by Hawley 2001, p. 83-84).} So it seems the RDA will also 
need to imagine away these ``kinematically hidden'' accompaniments.

Finally, I note that this critique of just assuming perfect rigidity leads 
us back to Section \ref{sssec;vsStrtfwd}'s theme, that classical mechanics 
is more subtle and problematic than philosophers usually recognize.\\
\indent More specifically, I think that a  traditional  philosophical  view, 
that forces are unobservable, underlies the second defect above, i.e. the 
allegation that the perdurantist can appeal only to accompaniments of 
rotation that are kinematically manifested. I cannot here go into details 
about why this view is wrong. Suffice it to say that I think it mainly 
arises from either or both of:\\
\indent \indent (a): an overly strong empiricism, that the only physical 
quantities to which we have empirical access are some small handful, no 
doubt  including  length, time, mass and charge---but excluding force;\\
\indent \indent (b): a distortion of  Hertz' research program in the 
foundations of classical mechanics. Thus Hertz proposed to explain forces in 
terms of cyclic microscopic variables; (Lanczos (1986, Section V.5, p. 
130-132) explains the idea). But had he succeeded, forces would {\em not} 
have been rendered unobservable.
  
\subsubsection{An improved RDA---allowing dynamical 
effects}\label{453RDAwitheffects}
I said just now that it seems the RDA will  need to  imagine away  
accompaniments involving stress, as well as those involving  strain. But in 
fact, not so: it only seems so. More precisely: the RDA can be developed, 
and be a powerful argument  against perdurantism, {\em without}  imagining 
away any accompaniments of rotation. We only have to change the example a 
bit, so that two  possibilities, apparently distinct on account of 
worldlines, match exactly in such accompaniments.\footnote{Paul Mainwood and 
David Wallace devised the following formulation in a seminar  in autumn 
2003. The idea of exploiting the distinction between two senses of rotation, 
so as to avoid having to  imagine away the usual accompaniments, had already 
been briefly advocated by Dean Zimmerman (1998, p. 268-269), crediting an 
anonymous referee. But beware: Zimmerman's discussion can be read as placing 
each disc in a separate possible world---in which case it fails, as 
explained in (ii) below. Zimmerman kindly points out (personal 
communication) that this was not his intention; so that his formulation of 
the RDA is essentially the same as that invented by Mainwood and Wallace. 
For novelty and precision, I present theirs. My thanks to them and 
Zimmerman.} 

Thus the endurantist challenges the perdurantist to distinguish the 
possibilities:---\\
\indent (Same): Two perfectly circular discs, $d_1$ and $d_2$, both made of 
continuous homogeneous matter and lying in the same spatial plane---but 
otherwise as different as you please from one another---spin in the same 
sense (i.e. both clockwise as seen from one side of the plane, and so 
anti-clockwise as seen from the other side).\\
\indent (Different): Two discs, $d'_1$ and $d'_2$, match $d_1$ and $d_2$ 
{\em respectively} in all respects (at all times); except that $d'_1$ and 
$d'_2$ spin in opposite senses relative to one another.

The idea is that all the usual accompaniments  (stress as well as strain: 
forces and energies as well as distortion) match between $d_1$ and $d'_1$; 
and similarly between $d_2$ and $d'_2$. So there is no need to imagine them 
away, in order to challenge the perdurantist.\\
\indent  Nor is there any need for discs {\em within} one of the possible 
worlds to match in any respect, except being perfectly circular, made of 
continuous homogeneous matter, lying in the same spatial plane---and for 
(Same), spinning with the same sense.

Four comments, in descending order of importance, by way of clarifying this 
formulation of the RDA:---\\
\indent \indent (i): Intuitively, (Different)  describes equally well  two 
distinct possibilities:  one in which $d'_1$ spins in the same sense as both 
$d_1$ and $d_2$; and the other in which instead,  $d'_2$ shares their common 
sense of rotation. This contrast of course  depends on there being a 
fiducial spatial direction in common between the  possibilities. I agree 
that this idea is perfectly coherent: though I emphasise, as Callender did 
with his pseudo-distinction between the arrow's states (Up) and (Down) 
(Section \ref{sssec;tworeplies}), that the direction needs to be specified 
by something salient in the environment, such as a local gravitational field 
giving one an up-down distinction, on pain of its being a distinction 
without a difference, i.e. a spurious distinction---an artefact of a 
diagram, or of our visual imagination. So: given such a specification, 
(Different) indeed represents two possibilities. No matter: to challenge the 
perdurantist,  the RDA can simply consider  either one of them. \\
\indent \indent (ii): (This follows on from (i).) The danger of making a 
distinction without a difference also crops up in another way. As mentioned 
in footnote 29, a formulation of the RDA in terms of distinguishing two 
senses of rotation (and thereby keeping the usual accompaniments) has been 
urged before, by Zimmerman (1998, p. 268-9). But Zimmerman's brief 
discussion can be read as   challenging  the perdurantist to distinguish 
between a disc rotating clockwise, alone in its possible  world, and a 
duplicate disc rotating counterclockwise at the same rate, alone in {\em 
its} world. And this formulation fails for the reason emphasised in (i): the 
clockwise-counterclockwise distinction  assumes a fiducial spatial direction 
in common between the possibilities, which for these ``lonely'' discs is a 
spurious distinction. (Callender (2001, p. 32, 36-7) seems to read, and 
object to, Zimmerman in this way.) Our formulation above  avoids this 
difficulty by considering two discs in each possible world, so that we  need 
only {\em intra}-world comparisons of the sense of rotation.\\
\indent \indent (iii): The possibilities can be modified in various ways. In 
particular, to secure the needed intra-world comparisons of sense of 
rotation, we do not need two discs. (Same) could instead contain just one 
disc, rotating in the same sense as a curved arrow drawn on a sheet of  
paper lying beside it; (Different) would then similarly contain a single 
disc rotating contrary to the sense of another curved arrow drawn on an 
adjacent sheet of paper.\\
\indent \indent (iv): As in the original RDA, at the end of Section 
\ref{sssec;track}: we can also allow the discs' properties to change over 
time, and to vary in a circularly symmetric way---provided of course that 
they do so in a suitably matching way.

Finally, a  comment about how the perdurantist should  reply to this version 
of the RDA. I shall develop my own reply in the next Subsection and 
subsequent Sections. Here I just report my guess about how Sider (Section 
\ref{sssec;Sider}) would reply to this version of the RDA; (which of course, 
he does not discuss).\\
\indent I think Sider would bite the bullet, just as he did in the ``lonely 
disc'' worlds he considered. That is: he would say that in a sufficiently 
simple two-disc world, there need be no unique facts about persistence, and 
so no fact of the matter about whether the two discs' senses of rotation 
match; (and similarly for the analogous world with one disc and a curved 
arrow drawn on paper).

\subsection{The RDA fails in general relativity}\label{432:RDAfailsGR}
So much by way of expounding some implications of the physics of 
rotation---the material in Sections \ref{131needconnect} and 
\ref{132connectmetrot}---for the RDA literature's usual assumption that the 
RDA needs to imagine away the actual accompaniments of rotation.\\
\indent But beware: the last Subsection's punchline---that this assumption 
is unnecessary, that the RDA can keep all the usual accompaniments---is {\em 
fragile}. This improved RDA, and the original version, both fail in the 
context of general relativity, because of the dragging of inertial frames 
around rotating bodies  (cf. (a) of Section \ref{423rotnGR}). 

That is: in general relativity, the trajectory of a test-particle falling 
towards a massive body depends on whether (and how) the body is rotating: 
the rotating mass ``drags'', albeit very slightly, the inertial frames in 
its vicinity (Misner et al. 1973 pp. 699, 879, 1117). This  frame-dragging 
means that  the RDA   fails in the sense that, in the usual version, the 
inertial frames (the worldlines of test particles) are dragged around the 
rotating disc (Rot), but not around (Stat); and in Section 
\ref{453RDAwitheffects}'s version, there cannot be the perfect match in 
rotation's accompaniments both between $d_1$ and $d'_1$ and between $d_2$ 
and $d'_2$, since the dragging of inertial frames around a rotating body is 
different, for different senses of rotation. In short: the RDA fails because 
frame-dragging represents an appealing difference, to which a ``sufficiently 
naturalist'' perdurantist can appeal so as to answer the challenge of 
distinguishing the possibilities.\footnote{This argument against the RDA, in 
its usual version, is due to Callender (2001, p. 38); it is part of his `No 
Difference' reply.}

Before asking how the advocate of the RDA might respond, it is worth making 
two comments. \\
\indent\indent  (A): First it is worth listing---as a partial review of the 
story so far---the five main ideas that have led to this conclusion. This 
``cast, in order of appearance'' is:\\
\indent (i): the idea that since a (consistent and precise) physical theory 
specifies a set of solutions (in philosophers' jargon: possible worlds), the 
RDA could hold in one such theory and fail in another; (cf.  (2A) in Section 
\ref{ssec;naturalism}); \\
\indent (ii): the idea that classical mechanics is subtle and problematic, 
and leads to relativity theory and quantum theory; so that there is good 
reason to consider general relativity as the setting of the RDA (even apart 
from its being our best guess about space, time and matter); (cf. especially 
the qualms about action-at-a-distance in Newtonian gravity, in Section 
\ref{sssec;vsStrtfwd});\\
\indent (iii): the idea that the perdurantist can reply to the RDA by 
finding differences between the possibilities, (Stat) and (Rot), that she 
can  appeal to; (Appealing Differences) in Section \ref{sssec;tworeplies};\\
\indent (iv): the dragging of inertial frames around rotating bodies in 
general relativity (cf. (a) of Section \ref{423rotnGR});\\
\indent (v): the idea that the endurantist will want to accept, as much as 
the perdurantist, the technical claims of general relativity's description 
of spacetime (presumably by being a scientific realist, maybe even some form 
of substantivalist); and this will involve a largely literal (though perhaps 
not a substantivalist) construal  of general relativity's ascription of a 
dynamical geometry to otherwise empty spacetime; (Section 
\ref{133Allowboth}).

(B): The RDA failing in general relativity does {\em not} mean there is no 
more to say about the endurantism-perdurantism debate in the context of 
general relativity. As always in philosophy, there is plenty to explore! In 
particular, the subtleties of rotation in general relativity (witness 
Malament's results in (b) and (c) of Section \ref{423rotnGR}) return us to 
question [ii] at end of Section \ref{133Allowboth}. That is: can endurantism 
and perdurantism really make equally good sense of all these subtleties?\\
\indent If not, so much the worse for whichever party cannot make sense of 
them. For once you have gone beyond traditional conceptual analysis to the  
extent of considering general relativity, it would surely be {\em ad hoc} to 
rule out of court whichever of the general relativistic spacetimes, such as 
the Kerr spacetime, you have trouble making sense of. That is, you cannot 
just declare that the spacetime  represents a world which is ``so unlike 
ours'' as to make considerations based on it irrelevant to the 
endurantism-perdurantism debate. (In any case, the spacetime in question 
might not be ``exotic'': assuming scientific realism, it might describe, or 
approximately describe, our world.) 

I turn to the question how the advocate of the RDA can respond. Could she 
improve the argument's thought-experiment so as to allow for frame-dragging, 
in the kind of way that (Different) and (Same) improve on (Stat) and (Rot) 
by allowing for the usual accompaniments of rotation? Perhaps, but I do not 
see how.\\
\indent On the other hand, the endurantist has two  lines of reply, even if 
she cannot thus improve the thought-experiment. Both return us to some 
questions  raised before.\\
\indent First, she might emphasise that in developing the RDA for general 
relativity (in the usual, or Section \ref{453RDAwitheffects}'s, version) she 
can stipulate that the discs are ``lonely'', i.e. that there are to be no 
test-particles travelling the dragged worldlines. Does this stipulation make 
the difference to which the perdurantist  appeals---viz. whether the frames 
are dragged, and if so,  how---counterfactual? The answer depends on the 
interpretation of general relativity. Roughly speaking, a substantivalist 
will answer `No', since they take the metrical structure of spacetime to be 
real and occurrent: it is not just an encoding of how suitable bodies would 
behave. But the endurantist may argue that she can accept general 
relativity, and so develop the RDA for it, without being a substantivalist 
in this sense; (cf.  Section \ref{133Allowboth}). On the other hand, even if 
we accept that the difference is counterfactual, perhaps the perdurantist 
can still appeal to it: (cf. Section \ref{sssec;accompanimtsrotn}).\\
\indent  The second reply is the obvious one about philosophical method. 
Surely no philosophical account of persistence should be ``so far gone'' in 
naturalism as to depend on general relativity: it should be able to 
accommodate continuous matter in classical and special relativistic 
spacetimes (cf. (2.B) in Section \ref{ssec;naturalism}). And for these 
cases, the RDA remains unrefuted, at least in Section 
\ref{453RDAwitheffects}'s improved version.

I think the second reply has force. But in Sections \ref{psa;perdmwouttears} 
and \ref{sssec:Appealbeyondclassl}, I will argue that the perdurantist can 
meet the challenge of defeating the RDA even outside general relativity: in 
short, by accepting only non-instantaneous temporal parts. Besides, this 
version of perdurantism is supported by some heterodox proposals about the 
intrinsic-extrinsic distinction among properties: proposals which are 
themselves supported by some features of classical and quantum  physics.

\section{Replying to the RDA}\label{sec:soln}
\subsection{Two replies}\label{sssec:detailclassl}
We have seen in Section \ref{ssec;describerot} how the RDA fails in general 
relativity---but, so far, looks good in classical mechanics. But only `so 
far'! From now on, I will develop two replies to the RDA: the first in this 
Section (Section \ref{sssec:detailclassl}), and the second in Sections 
\ref{psa;perdmwouttears} and \ref{sssec:Appealbeyondclassl}.\\
\indent The first reply is an analogue of Sider's (Section 
\ref{sssec;Sider}); the main difference being that it is  less 
metaphysically committed and closer to the detail of empirical enquiry. But 
like Sider's reply, it will be a combination of (Appealing Differences) and 
(No Difference). In particular, it bites the bullet as Sider does, in that 
it denies the distinction between the discs, in sufficiently simple worlds. 
The endurantist advocate of the RDA will of course see this bullet-biting as 
conceding victory to the RDA, or at best as forcing a stalemate.\\
\indent  This situation will prompt the second reply, which is my preferred 
reply. It is a version of (Appealing Differences), not (No Difference), 
given by a modest version of perdurantism which accepts only 
non-instantaneous temporal parts.

\subsection{Some details of persistence within classical 
mechanics}\label{sssec:detailclassl}
\subsubsection{Comparison with Sider}\label{152A:CompSider}
I can present this reply to the RDA most clearly, by first stating the 
similarities and differences between it and Sider's  reply. There are three 
similarities.\\
\indent (i) Like him, I envisage that the perdurantist  appeals to a wide 
``web of belief'' to yield a perdurantist  definition  of persistence.\\
\indent (ii) Like him, I allow this web to involve technicalities, so that 
the perdurantist  account of persistence  is ``naturalistic'' (Section 
\ref{ssec;naturalism}): in particular, the account  is not a conceptual 
analysis of the sort traditionally sought by metaphysicians. \\
\indent So the idea is that the perdurantist  defines persistence by 
appealing to the various relations that the notion has to other notions, 
including technical ones. Like Sider, I will discuss these relations only 
for classical mechanics; (in particular, I will not need to distinguish 
between using a Newtonian and a special relativistic spacetime).\\
\indent (iii) Like Sider, this reply will bite the bullet in that it denies 
the distinction between the discs, for cases using sufficiently simple 
possible  worlds, such as worlds with lonely homogeneous discs or with 
space-filling homogeneous fluids.

\indent On the other hand, there are two main differences from Sider.\\
\indent (i) I will give more details about the theories of mechanics. For 
example, while Sider talks of `the laws of dynamics' (or `laws of nature'), 
I will be more specific, e.g. distinguishing point-particles and continua.\\
\indent (ii) I will be less metaphysically committed. In particular:\\
\indent \indent (a) I will not be committed to the Mill-Ramsey-Lewis 
best-system  theory of laws; nor to any other specific way of selecting the 
true ``laws of dynamics'' from a list of rival candidates. \\
\indent  \indent (b) Nor will I be committed to the use of Ramsey-Lewis 
simultaneous functional definition; i.e. to the assumption that  the body of 
doctrine selected by the method  in (a)---the doctrine dubbed $S({\bf G})$ 
in Sider's notation---is logically strong enough to uniquely specify $\bf 
G$.\footnote{I think this debatable assumption of the Ramsey-Lewis technique 
tends to be forgotten: though not by Sider, nor of course by Lewis. Indeed, 
Lewis recognized that here lurk deep difficulties for his theory  of natural 
properties; cf. his (2004).}

\indent So the broad idea of this position is that by exploiting details 
about mechanics, the perdurantist can claim to define persistence, without 
having to be committed to Sider's philosophical  methods (a) and (b).

A side-remark. One can of course take the details of this reply as ``filling 
in'' Sider's position, e.g. as giving details about how the best-system 
theory of laws applies to mechanics, rather than as providing  an 
alternative to Sider's position. That is, in terms of Sider's notation (cf. 
Section \ref{sssec;Sider}): Sider could endorse the details in the next 
Subsection as exactly {\em what makes} the system  $S({\bf G})$ `the system 
that achieves the best combination of strength and simplicity' (his p. 231).

\subsubsection{Persistence and the web of belief}\label{152B:SiderCheap?}
So let us join Sider in imagining a world governed by classical mechanics: a 
world that is not too simple, but has the sort of variety and complexity of 
motions of objects that we see in the actual macroscopic world.\\
\indent So here, we are to set aside: (i) the fact that the actual world is 
quantum; and (ii) Section \ref{ssec;therelevcephys}'s misgivings about how 
problematic the ontology of classical mechanics taken on its own is. (There 
is  no reason to think (i) and (ii) prevent a classical mechanical world 
having a variety and complexity of motions similar to that of the actual 
macroscopic world.)\\
\indent We are also to set aside the question how in this imagined world  
the laws of mechanics earn the name of laws: in particular, it need not be 
(though it could be) by the best-system analysis of lawhood.

\indent My task is to sketch (in more detail than Sider does) the 
``functional role'' of persistence,  the web of belief in which it is 
embedded. To do so, I shall proceed in three stages.\\
\indent  (1) {\em Kinds of Object}: First, I shall be more specific about 
the various kinds of object that are described by classical mechanics:  I 
distinguish four kinds.\\
\indent  (2) {\em Persistence}: Then I discuss the persistence of these 
objects. This is a matter of applying familiar factors, qualitative 
similarity and causation, to the four kinds of objects.\\
\indent  (3) {\em Establishing the laws}: Then I sketch how the behaviour of 
these persisting objects  could underpin  the laws of mechanics, both as 
true generalizations and as laws: even without using Ramsey-Lewis 
simultaneous definition.\footnote{I will not need to distinguish the various 
formulations of classical mechanics, in particular Newtonian, Lagrangian and 
Hamiltonian: one can think throughout just of Newton's second law ${\bf F} = 
m{\bf a}$.}\\
\indent But I stress at the outset that, as is now familiar from Quine's 
``web'' metaphor: ideas and propositions about each of these three stages of 
course influence, in various complex ways, ideas and propositions about the 
others. We do {\em not} first fix the objects, then discover or decide their 
conditions of persistence (diachronic criteria of identity), and finally 
establish what laws they obey. Rather, what counts as an object is 
influenced by the convenience of having different putative criteria of 
identity agree in their verdicts. And the laws of mechanics (or candidate 
laws) can contribute to determining persistence. For example, we might judge 
that $o$ at $t$ is the same persisting object as $o'$ at $t'$, despite some 
contrary evidence (e.g. insufficient qualitative similarity), just because 
the laws of mechanics prescribe for $o$ at $t$ (together with its velocity) 
a future trajectory that at $t'$ passes through where $o'$ then is. (This 
mutual influence between (1), (2) and (3) of course  illustrates my 
similarities to Sider: both of us appeal  naturalistically to a complex web 
of belief.)

(1) {\em Kinds of Object}:\\
As mentioned in Section \ref{ssec;therelevcephys}, the main distinction 
among objects which classical mechanics makes is the distinction between:\\
\indent (a): point-particles, i.e. extensionless point-masses moving through 
empty space (and so interacting by action-at-a-distance forces); \\  
\indent (b): continua, i.e. bodies whose entire volume is filled with 
matter.

Mathematically, this difference is in the first place one of finitude vs. 
infinity:---\\
\indent According to (a), a system consists of a finite number of 
point-particles, so that the system's state is given by finitely many real 
numbers. (In fact, one needs six for each point-particle: three for its 
position in space, given by, say, coordinates in a cartesian coordinate 
system; and three for the components of its momentum.) So (a) conceives an 
extended macroscopic body, whether solid, liquid or gas---a brick, or a 
sample of water or air---as a swarm of a gigantic number of point-particles. 
\\
\indent On the other hand, according to (b), a system---even a single small 
rigid body like a marble---consists of continuum-many point-sized bits of 
matter, one at each spatial  point in the volume occupied by the body: so 
these bits of matter are truly ``cheek by jowl'' to one another! So we 
expect the system to be described by continuously many real numbers: indeed, 
``six times continuously many'', to specify the position and momentum of 
each point-sized bit of matter. (I say `we expect', because this is a 
simplification, albeit a harmless one in the present discussion. That is,  
as I stressed in (1) of Section \ref{sssec;vsStrtfwd}: the {\em 
pointilliste} particles-in-motion picture is wrong. Classical mechanics in 
fact describes continua not point-by-point but in terms of the states of the 
countless arbitrary (in general, overlapping) sub-regions of the body.)

\indent Accordingly, the systems treated {\em {\`{a}} la} (a) and (b) are 
often called, respectively, `finite-dimensional' and `infinite-dimensional' 
systems: or for short, `finite' and `infinite' systems. 
Broadly speaking, finite systems are in principle  simpler since they are 
described by ordinary differential equations, while infinite systems require 
us to use partial differential equations, which are considerably more subtle 
and complicated, both in theory and in practice.

 But even finite systems can be very complicated---in various senses, which 
I will not need to distinguish. All I need to say here is that even for a 
finite system, i.e. a swarm of point-particles, the number of real numbers 
needed to specify the state (called the number of {\em degrees of freedom}) 
might well be, though finite, intractably large. But fortunately, mechanics 
has various strategies for reducing such daunting numbers to something 
manageable.\\
\indent The paradigm example of such a strategy is to assume that a body is 
{\em rigid}. Indeed, it is a strategy that applies both to finite systems 
and continua---in both cases, enormously reducing the number of degrees of 
freedom one needs to consider.   The idea of rigidity is not just the 
everyday vague idea of being solid, like a brick: it is the precise 
assumption that all the distances between the body's smallest 
constituents---whether they are point-particles in a swarm, or point-sized 
bits of matter ``cheek by jowl'' filling a continuum---are constant in 
time.\footnote{This assumption implies that the positions of all the 
point-particles, or point-sized bits of matter, is fixed once we fix the 
position of just {\em three} of them. For imagine: if the tips of three of 
your fingers were placed at  three given positions within a rigid brick, and 
someone specified the exact positions of the finger-tips, then they would 
have implicitly specified the positions of all the brick's constituent 
parts. Similarly, rigidity reduces enormously how many momenta one needs to 
specify, in order to specify a state, i.e. in order to implicitly specify 
all the momenta.}

Putting together the finite vs. infinite contrast and the rigid vs. flexible 
contrast,  we get a distinction between four kinds of object described by 
classical mechanics:\\
\indent (i) a point-particle, i.e. an extensionless point-mass; \\
\indent  (ii) an extended object (in physics jargon:  body) that is small 
and rigid enough that both its internal structure (whether a swarm of 
point-particles or a continuum) and its orientation can be ignored, so that 
it can be successfully modelled as a point-particle;\\ 
\indent (iii): an  object that, though extended, is rigid enough that its 
internal structure (whether a swarm of point-particles or a continuum)  can 
be ignored, so that it can be successfully modelled as a rigid body; (if it 
is {\em also}  small  enough that its  orientation can be ignored, so that 
it can be  modelled as a point-particle, we revert to case (ii)); \\
\indent (iv): an extended object  that is both large and flexible enough 
(especially: a fluid) that its internal structure (whether a swarm of 
point-particles or a continuum)  {\em cannot} be ignored, and it cannot be 
modelled as a rigid body.\\
\indent  To sum up: objects of kinds (i)-(iii) either are, or can be 
modelled as, finite systems. But objects of kind (iv) cannot be, and are 
thereby in general  the hardest objects to model successfully using 
classical mechanics.

(2): {\em Persistence}\\
In saying these objects are `point-particles', or `successfully modelled as 
point-particles', `successfully modelled as rigid bodies' etc.,  I have 
implicitly assumed that they persist. But what ingredients should enter into 
the definition  of persistence? Or less ambitiously: into its supervenience 
basis, or its functional role?\\
\indent We have already briefly discussed the two main factors that the 
philosophical literature (especially in the tradition of conceptual analyis) 
considers. Namely:\\
\indent [i] qualitative similarity (cf. {\em Follow} in Section 
\ref{ssec;theRDA});\\
\indent [ii]  causation (cf. Section \ref{sssec;appealcause}).\\ 
Of course, how these factors might figure in a definition, or the functional 
role, of persistence is a large and controversial subject. It is even 
controversial if, {\em pace} my denials in Sections \ref{sssec;vsStrtfwd} 
and \ref{sssec;vsbracket}, we make the assumptions ({\em Straightforward}) 
and ({\em Bracket}), i.e. we assume that the interpretation of classical 
mechanics, or more generally classical physics,  is straightforward and does 
not lead in to the relativity and quantum theories. As discussed in Section 
\ref{ssec;therelevcephys}, that is what the philosophical  literature about 
persistence  usually assumes; and in the present context, i.e. for the sort 
of classical mechanical possible  world that we are presently (following 
Sider) envisaging, I of course agree that it is fair enough. 

I will not review, let alone try to settle, this controversy. Indeed, even 
if the philosophical literature settled this controversy to its own 
satisfaction, work would remain for me (and Sider): since for us the 
contribution of the laws of mechanics to the definition (or supervenience 
basis, or functional role) of persistence would remain to be spelt out. So 
here I will confine myself to sketching in just a bit more detail, the 
factors [i] and [ii], and how they apply to the objects of classical 
mechanics. (I discuss the controversy, also from an endurantist perspective, 
more fully  elsewhere: 2004, 2004a). 

 I think many, even most, philosophers on both sides of the 
endurantism-perdurantism debate agree that for most objects, the definition 
of persistence (i.e. of perdurance as a relation between temporal parts,  or 
for endurantists, of objects' criteria of identity) will invoke one or both 
of the  factors, [i] and [ii] (which also might well overlap).

[i]: Qualitative similarity concerns whether the object at the two times (or 
in perdurantist terms: the two stages) has suitably similar qualitative 
properties. Here, `suitably similar' is to be read flexibly. It is to allow 
for:\\
\indent \indent(i) only a tiny minority of properties counting in the 
comparison;\\
\indent \indent(ii) considerable change in the object's properties, provided 
the change is ``continuous''; i.e. provided the object goes through some 
kind of chain of small changes.

[ii]: Causation concerns whether the state of the object at the later time 
(or the later stage) is suitably causally related by the earlier state or 
stage. Here again `suitably causally related' is to be read flexibly. It is 
to allow for:\\
\indent \indent (i) various rival doctrines about causation---including even 
the special variety, `immanent causation', that some philosophers believe 
underpins persistence (Section \ref{sssec;appealcause});\\
\indent \indent (ii) a suitable chain of states or stages linked by 
causation.
 
It is notoriously difficult to go beyond this vague consensus to give a 
precise definition of persistence (a precise criterion of identity): even if 
we allow the definition/criterion to vary from one sort of object to 
another.\\
\indent Agreed, in {\em practice}, we have little difficulty with the 
macroscopic objects of everyday life (Austin's `medium-sized dry goods', and 
analogous `wet goods', such as organisms, rivers, etc.), for the simple 
reason that for the vast majority of such objects, various relatively simple 
definitions of persistence or criteria of identity which one might propose, 
based on the ideas of [i] and [ii], agree in their verdicts. The different 
proposed criteria are ``convergent''. In other words, such an object 
typically has many observable qualitative properties, and the changes in 
these properties are slow and-or rare enough, and-or linked sufficiently 
systematically (causally) to other events, that the various proposals  based 
on  [i] and [ii] yield the very same judgments of persistence over time.\\
\indent So it takes strange cases to tease apart the proposals' verdicts. 
Hence the  tradition within conceptual analysis of considering puzzle cases, 
as a tactic to help us formulate a definition/criterion that covers all 
(logically or metaphysically) possible cases.\\
\indent The philosophy of personal identity provides the most obvious 
examples: viz. puzzle cases where the verdicts of proposals based on bodily 
properties, and those based on psychological properties, differ. (This is so 
even if we consider only one of [i] and [ii]. For [i], the trajectories  
determined by ``tracking''  bodily properties (albeit slowly changing) and 
by ``tracking''  psychological properties (albeit slowly changing) diverge. 
For [ii], the causal chain of bodily states diverges from that for 
psychological states.) 
  
As I said, I will here duck out of the project of trying to go beyond the 
vague consensus above. But by way of justifying my doing so, I note two 
points. First, I believe this project is independent (at least, to a large 
extent) both of  the endurantism-perdurantism debate, and of whether 
classical mechanics is true. (I think this claim is uncontroversial: by and 
large, the literature addressing this project sets aside these two issues.)  

Second, recall (from the start of this Subsection) that my present task is 
to sketch the functional role of persistence in a classical mechanical world 
with the sort of variety and complexity of motions of objects that we see in 
the actual macroscopic world. And for that task, it seems legitimate to 
assume that the actual fact just mentioned---that in the actual world, most 
macroscopic objects change their observable qualitative properties in a slow 
and-or rare and-or orderly enough way that various putative criteria of 
identity agree---``carries over'' to the envisaged classical mechanical 
world. In particular, I see nothing in my denials of ({\em Straightforward}) 
and ({\em Bracket}) (Sections \ref{sssec;vsStrtfwd} and 
\ref{sssec;vsbracket}) to prevent such a widespread agreement of criteria of 
identity in a classical mechanical world.

Assuming this widespread agreement of criteria of identity, it is 
straightforward to sketch how  the ideas [i] and [ii] would  apply in {\em 
practice} to the four kinds of object distinguished in (1) above. In other 
words, we can see how the practice of physics in the envisaged world would 
be able to ignore, at least in large measure, puzzle cases and conundrums 
about persistence (including the RDA), just as it does in the actual world. 
Being myself wary of appealing to causation (cf. end of Section 
\ref{sssec;appealcause}), I will only consider applying [i], i.e. 
qualitative similarity. (For some more discussion of causation's role in the 
definition of persistence, cf. my (2004a).)

As discussed in Section \ref{ssec;theRDA}, qualitative similarity   works 
well  when applied to bodies of kind (i): point-particles each with a 
continuous spacetime trajectory (worldline), moving either in a void or in a 
continuous fluid with suitably different properties---a different 
``colour'', or made of different ``stuff'', than the point-particle. For 
however exactly we define `maximum qualitative similarity', there will no 
doubt be,  starting at a point-particle at $t_0$, a unique timelike curve of 
qualitative  similarity passing through it: the worldline of the particle. 
(Indeed, for the case of a void we could dispense with qualitative 
similarity, and have the {\em definiens} refer just to spacetime points' 
property  of being occupied by matter.)

Similarly for the other kinds (ii), (iii) and (iv). Again, qualitative 
similarity will in practice work well, at least for most cases.\\
\indent For an object of kind (ii) is small and rigid enough to be modelled 
as a point-particle: i.e. a successful physical description of it does not 
need to keep track of its spatial parts.\\
\indent And though a successful physical description of an object of kind 
(iii) {\em does} need to keep track of the object's spatial parts, the 
assumption of rigidity makes this a vastly easier task than it otherwise 
would be. In particular, qualitative similarity will again work well, 
provided there is some property or other (such as colour, density, 
temperature ...) that (a) varies sufficiently across the object's different 
spatial parts, and (b) for each such part changes over time 
slowly/rarely/systematically enough, to enable ``tracking'' of the spatial 
part. Broadly speaking, qualitative similarity will only fail in the 
limiting case of continuous matter that is utterly homogeneous as regards 
all properties: i.e. the case of the RDA.\\
\indent For kind (iv), objects large and flexible enough that they must be 
treated as non-rigid, the assumption of rigidity is unavailable, and 
tracking the parts of objects will be correspondingly harder. But again, we 
can expect qualitative similarity to work well, at least in practice, for 
objects with properties that vary spatially, and change over time in an 
orderly enough way: for cases unlike that of the RDA.\\
\indent  (Of course, for all four kinds, there might be no {\em single} 
definition of persistence (criterion of identity) in terms of qualitative 
similarity for the whole kind: the kind might be divided into subsets, each 
with their own definition/criterion.)

To sum up this discussion:--- I have argued it is legitimate to assume that 
in the envisaged classical mechanical world:\\
\indent (i) the account of persistence, for each of our four kinds of  
object, will appeal to the same factors, qualitative similarity and 
causation, that most philosophers actually appeal to; and \\
\indent (ii) objects change properties  slowly and/or regularly enough that 
in {\em practice}, various proposed criteria of identity agree, and 
conundrums about persistence like the RDA do not arise.

(3): {\em Obeying the laws of mechanics}\\
Finally, I discuss what it means in the envisaged classical mechanical 
world, for our four kinds of persisting object to `obey the laws of 
mechanics'. I shall discuss in order, (a) `obeys'  and (b) `laws', urging 
that for my purposes both can be left vague.

\indent  (a): That a kind of object obeys a certain mechanical 
generalization (in particular Newton's second law, ${\bf F} = m{\bf 
a}$)\footnote{Again, I do not need to distinguish the various formulations 
of classical mechanics; cf. footnote 32.} should  not  mean merely that for 
each object in the kind there is some possible schedule of forces exerted on 
the object, such that were they exerted, the object's motion would satisfy 
the generalization. That would be far too weak: in particular Newton's 
second law  could be obeyed in a spurious and {\em ad hoc} way by each 
object having a schedule of forces tailor-made to describe its motion, no 
matter how  peculiar it might be.\\
\indent  Rather, it is to mean, roughly, 
that there is some overall assignment of the forces exerted at each time on 
each object (and for an object of kind (iii) or (iv): its spatial parts) 
that\\
\indent \indent (i) is derived from some general principles or formulas 
applying to all objects of the kind (or at least all of a broad sub-kind, 
e.g. among point-particles, the electrically charged ones), and\\
\indent \indent (ii) makes the generalization satisfied by the object's 
motion.\\
\indent But for present purposes, we do not need to be more precise than (i) 
and (ii). In particular, the principles or formulas in (i) surely need not 
be  require using only forces familiar from the actual practice, and 
macroscopic success, of classical mechanics: e.g. gravitational forces taken 
as fixed by Newton's inverse-square law. I think it would also be too much 
to require the forces to have more abstract features familiar from actual 
classical mechanics, such as being two-body, rather than many-body, in 
nature.

\indent (b): So much by way of sketching how  the laws of mechanics could be 
{\em true} in the envisaged world (with (2) giving an account  of the 
persistence of (1)'s four kinds of object). I turn to their being {\em 
laws}.\\
\indent Though as an aspiring Humean, I am attracted to the best-system 
analysis of the notion of a law of nature, it is clear that the discussion 
above does not need the notion, or this analysis of it. We can make do with 
the theory-relative notion of a law {\em of mechanics} (or more generally, 
of a given physical  theory); and though this notion might be explicated by 
some theory-relative version of the best-system analysis, it need not be.\\
\indent Indeed, I think that even if you are sceptical of any general 
explication of law, even a theory-relative one, you are likely to accept 
that in the envisaged world, the laws of mechanics earn the name of `law' if 
anything does. After all, recall requirements (i) and (ii) in (a) above, 
that there be a principled overall assignment of the forces exerted on 
objects that their various motions satisfy: what else need you require of 
laws of mechanics?

\indent  To sum up my stages (1) to (3):--- We have seen how a possible  
classical mechanical world could contain various kinds of object, and  
sustain a notion of persistence for them such that they satisfy classical 
mechanical laws. Besides, this need not involve specifying first the 
objects, then the account of their persistence, then the laws. Rather, the 
three stages can influence each other: in particular, the account of 
persistence can invoke the laws (cf. comments just before (1)). Nor need it 
involve Ramsey-Lewis simultaneous definition or the best-system analysis of 
laws.

{\em Coda: Biting the bullet}\\
  Finally, I admit that  that my stages (1)-(3) lead to the same  
bullet-biting which Sider admits he must do for the case of a lonely 
homogeneous disc; and which he must also do for my preferred version of the 
RDA, using two discs or a disc and a sheet of paper, and for Zimmerman's 
space-filling homogeneous fluid. That is, stages (1)-(3) lead to saying that 
in such cases, there is (No Difference) between the two putative cases; (cf. 
the discussion at the end of Section \ref{sssec;Sider}).\\
\indent This means the endurantist  will reply to me, as they did to Sider, 
that this amounts to conceding the force of the RDA: `even this version of 
perdurantism, with its sophisticated appeal to the web of belief, cannot 
secure facts of persistence  in the troublesome cases considered by the 
RDA'. This suggests that again, there is after all at best a  stalemate 
between the endurantist and this sort of perdurantist. 

\indent But I think the perdurantist {\em can} do better than this. They can 
secure facts of persistence in the troublesome cases such as lonely discs 
and space-filling fluids---by adopting the position in the next two Sections 
...

\section{Perdurantism without tears: the classical 
case}\label{psa;perdmwouttears}
I turn to my second, and favoured, reply to the RDA. It meshes with Section 
\ref{ssec;therelevcephys}'s rejection of the widespread assumptions ({\em 
Straightforward}) and ({\em Bracket}). That is: it fits  my claims that:\\
\indent (i) classical mechanics  is subtle and problematic and\\
\indent (ii) classical mechanics leads to relativity and the quantum.
 
More specifically:--- As to claim (i), this reply uses that claim's 
rejection of {\em pointillisme} to say that the perdurantist can take 
objects to have only temporally extended i.e. non-instantaneous temporal 
parts. As we shall see, this makes the perdurantist's account of persistence 
non-reductive: it uses notions which presuppose persistence (cf. (1) in 
Section \ref{ssec;naturalism}).  But for reasons already discussed 
(especially Section 4.2.2.C), the account is only ``slightly'' 
non-reductive: the presupposition of persistence is ``mild'' in the way that 
Section 4.2.2.C maintained velocity's presupposition of persistence was 
mild. In any case, the perdurantist who accepts only non-instantaneous 
temporal parts has an (Appealing Differences) reply to the RDA: that is, she 
can appeal to differences between the two discs. Furthermore, I maintain 
that  non-instantaneous temporal parts can do the various jobs, within the 
endurantism-perdurantism debate, that the perdurantist demands of temporal 
parts. 

All this, I will argue in this Section. Section 
\ref{sssec:Appealbeyondclassl} will pick up on claim (ii) above.  It will 
support this non-{\em pointilliste} version of perdurantism by considering 
how classical mechanical objects ``emerge'' from quantum theory. (This 
argument will also suggest augmenting discussion of the intrinsic-extrinsic 
distinction with a new idea: relativizing the distinction to bodies of 
doctrine, such as scientific theories.)

\subsection{Rejecting instantaneous temporal parts}\label{psa;vetoinsts}
At the end of Section \ref{ssec;intrpropvelies}, I said I would take it that 
both average and instantaneous velocity presuppose the notion of 
persistence, and are extrinsic properties. But when we consider a 
non-instantaneous temporal part, the second point needs to be qualified.\\
\indent For one of the part's constituent pieces of matter having a certain 
worldline segment within the part is surely an intrinsic property of the 
part. And similarly for lesser, i.e. logically weaker, information than the 
entire worldline segment. For example, that a constituent piece of matter 
has a certain average velocity over a time-interval ``within'' the temporal  
part is intrinsic to the part: notwithstanding the fact that average 
velocity presupposes the notion of persistence. Similarly for instantaneous 
velocity at a time ``within'' the temporal part. \\
\indent At least, these properties are intrinsic  to the part, {\em modulo} 
the topic I set aside in Section \ref{131needconnect}, viz. how to justify 
the appeal to persisting spatial points, and a spatial metric, that is 
needed for the idea of the distance traversed by the persisting object.\\
\indent This situation returns us to the terminology of {\em temporally 
intrinsic} properties which I introduced in (A) of Section 4.2.2.C. Roughly 
speaking, these are properties whose possession by an object $o$ at a time 
implies nothing about matters of fact (especially about $o$) at other times 
(though it may imply propositions about other places). Thus the fact that 
one of a non-instantaneous temporal part's constituent pieces of matter, 
$o$, has a certain instantaneous  velocity at a time $t$ ``within'' the part 
corresponds to a temporally intrinsic property of the part, though the 
velocity is temporally  extrinsic for $o$ at the {\em instant} $t$.

\indent The above points are of course independent of whether matter is 
atomic or continuous. The piece of matter can be a point-particle or a 
point-sized bit of matter in a continuum. (Indeed, the qualification could 
be stated in the very same words for an extended piece of matter, provided 
it was small enough for us to model it as point-like, i.e. having a 
worldline, and a single velocity: but I can focus on unextended pieces of 
matter.)\\
\indent To sum up: a non-instantaneous temporal part has a rich set of 
intrinsic, or at least temporally intrinsic, properties concerning the 
worldline-segments and average and instantaneous velocities, during the 
part,  of its constituent pieces of matter. 

Now consider a version of perdurantism that accepts only non-instantaneous 
temporal parts. (I will not discuss the pre-history of this proposal in 
authors like Whitehead: for details cf. Grattan-Guinness (2002). But I will 
soon discuss whether it should  accept all such parts, i.e. parts with an 
arbitrarily short, though non-zero, temporal extent.)\\
\indent Since such parts have a rich set of intrinsic properties, the 
prospects for the perdurantist project of defining persistence (or providing 
a supervenience-basis for it, or at least some non-reductive account of it) 
look a great deal better than for a {\em pointilliste}  version of 
perdurantism accepting only instantaneous parts (or accepting also extended 
parts, yet requiring persistence to supervene on the intrinsic properties of 
instantaneous parts, as in Lewis' Humean supervenience). For with these rich 
sets of properties, there are so many more ingredients which one could use 
in the {\em definiens} of persistence (or more generally, in the account of 
persistence). More precisely: the perdurantist's prospects are a great deal 
better, provided their definition or account of persistence  can 
legitimately refer to these intrinsic  properties of non-instantaneous 
temporal parts.\\
\indent In the rest of this paper,  I will endorse this version of 
perdurantism, both in general and as a reply to the RDA (both the usual 
formulation and Section \ref{453RDAwitheffects}'s stronger one). 

The reply it affords to the RDA is as follows. The worldline segments, 
average velocities and instantaneous velocities of point-sized bits of 
matter within a homogeneous disc provide intrinsic properties of the disc's 
temporal parts. Assuming that the perdurantist can appeal to these intrinsic 
properties---an assumption I will discuss in Section 
\ref{psa:noninststplyextrpropies}---she can certainly distinguish the discs. 
Indeed, with these intrinsic properties to hand, she may well have no more 
of a problem about her project of defining persistence, for the parts of a 
perfectly circular homogeneous disc, than for the parts of an inhomogeneous 
one. There are two aspects to this, which we can call `kinematical' and  
`dynamical'.

\paragraph{7.1.A ``Kinematics''}\label{71AKinics}
First, the perdurantist can appeal to the mathematical fact that every 
suitably smooth vector field $U$ defined on a open region $R$ of spacetime 
has integral curves throughout $R$: curves  which are timelike, by 
definition, if $U$ is timelike. (I mentioned this when discussing Lewis' 
proposal in Section 4.3.2.A. To be precise: `suitably smooth' requires only 
that $U$ be $C^1$, i.e. the partial derivatives of its components exist and 
are  continuous.) So the idea is that the intrinsic properties of a 
non-instantaneous temporal part of a classical continuum specify the vector 
field $U$, of instantaneous velocities (to be precise: 4-velocities) of the 
point-sized bits of matter, on the spacetime region $R$ of the part. $U$ 
then specifies integral curves, i.e. the worldlines within $R$ of the bits 
of matter. Besides, by considering a set of such non-instantaneous parts 
that ``cover'' the entire period for which a given bit of matter exists, its 
entire worldline can be reconstructed.

\indent There are two points to make about this proposal; of which the 
second will lead us to ``dynamics''.\\
\indent (1): Agreed, this proposal seems at first sight a cheat, a case of 
theft over honest toil. But I am for the moment just {\em assuming} that the 
perdurantist can appeal to intrinsic properties of non-instantaneous parts, 
even though some of them involve the notion of persistence: postponing 
discussion to Section \ref{psa:noninststplyextrpropies}. And rest assured, I 
will there admit that this assumption makes this kind of perdurantism  
``non-reductive''. (This assumption also marks the difference from Lewis' 
proposal in Paragraph 4.3.2.A: lacking the assumption, Lewis had trouble 
specifying the vector field $U$.)

\indent (2): The idea of reconstructing an object's entire worldline by 
concatenating segments (each lying in one of a ``covering'' set of 
non-instantaneous temporal parts) returns us to the formal equivalence I 
mentioned at the end of Section \ref{133Allowboth}. That equivalence had the 
perdurantist represent the location in spacetime ${\cal M}$ of a 
point-particle, or a point-sized bit of matter in a continuum, by a 
collection of functions, labelled by time-intervals that together cover the 
object's lifetime; for example, if it exists throughout the closed 
time-interval $[a,b]$, there might be a function $q_{[a,b]}: t \in [a,b] 
\mapsto q_{[a,b]}(t) \in {\cal M}$. Indeed, one can show how to reconstruct 
worldlines from such functions, even for a point-sized bit of matter in an 
utterly homogeneous continuum, provided the functions' domains are 
non-degenerate time-intervals, i.e. {\em not} singleton sets of times. (For 
details, cf. Butterfield 2004a, Section 3.)\\
\indent This reconstruction of worldlines from a collection of functions 
raises two points. First, I admit again that it seems at first sight a case 
of theft over honest toil: the perdurantist reconstructs worldlines from 
functions that involve the notion of persistence. Here I  again refer to 
Section \ref{psa:noninststplyextrpropies}'s discussion.\\
\indent Second, this reconstruction of worldlines is ``kinematical''. It 
uses no information about the properties of the moving matter, in particular  
the causes of its motion (``dynamics''): it simply invokes a set of 
functions that immediately specify worldline-segments. So it is natural to 
ask whether our kind of perdurantist can give an account of persistence that 
in some way appeals to (i) the properties of the moving matter, or (ii) the 
causes of its motion. I already reviewed in (2) of Section 
\ref{152B:SiderCheap?} how appealing to (i) would work in practice for 
inhomogeneous matter in a classical mechanical world: and the rejection of 
instantaneous temporal parts obviously does not affect that appeal. But our 
perdurantist can also appeal to (ii), at least if she is a ``naturalist''. 
This leads to ``dynamics''.\footnote{This ``kinematics-dynamics'' contrast 
exemplifies two more general contrasts in the philosophy of identity 
(discussed in in my 2004a, Section 4.1) which I call (i) `ontic-epistemic' 
and (ii) `conceptual-empirical'. (i) concerns whether the criterion or 
account of identity specifies the ``constitutive facts'' of persistence, or 
our grounds---everyday or technical, occasional or systematic---for 
judgments of persistence. (ii) concerns whether the criterion or account 
eschews the concepts and results of empirical theories, e.g. physical 
theories, or is willing to invoke them.}   

\paragraph{7.1.B ``Dynamics''}\label{71BDynics} Our perdurantist can indeed 
appeal to dynamics. That is: if she is sufficiently ``naturalist'' that she 
is willing to appeal to the laws of motion, then in a classical mechanical 
world, the definition of persistence can ``piggy-back'' on the determinism 
of those laws. (Cf. my endorsement of Quine's ``web'' metaphor just before 
(1) of Section \ref{152B:SiderCheap?}: the laws of mechanics can contribute 
to determining persistence.)\\
\indent That is: in common cases, the classical  laws (above all, Newton's 
second law, that Force = mass $\times$ acceleration) fully determine the 
motion of a point-particle, or a point-sized piece of matter in a continuum, 
over an interval of time $[t_1,t_2]$, in terms of its initial position and 
velocity at $t_1$ and the regime of forces on it during $[t_1,t_2]$: all of 
which the perdurantist  can take to be given by intrinsic properties of a 
temporal part that begins before $t_1$ and ends after $t_2$.\\
\indent Agreed, that is rough speaking: hence my `in common cases'. For 
accuracy, I should note some of the subtleties, in particular the threat to 
determinism from solutions in which some quantities become infinite within a 
finite period of time after the initial time $t_1$. For point-particles, 
such solutions are known to exist even if we veto collisions (cf. (2) of 
Section \ref{sssec;vsStrtfwd}); for a popular account of this, cf. Diacu and 
Holmes (1996, Chapter 3). For continua, whether there are such solutions is 
a deep open question: witness the fact that one of the Clay Institute's 
million-dollar Millennium Prizes is for a proof or disproof of the rigorous 
existence for all times of solutions of the equations that govern a 
classical fluid, i.e. the Navier-Stokes equations.\footnote{For a popular 
account, cf. Devlin (2002, Chapter 4); for a monograph discussion of what is 
known about the simpler case of a perfect fluid (Euler's equations), cf. 
Section 4.4 and Example 5.5.8  of Abraham and Marsden (1978)---thanks to 
Gordon Belot for this reference.}\\
\indent But for present purposes, I can discount these subtleties: here it 
is enough to suggest that a naturalist perdurantist can go about defining 
persistence in terms of integrating the equations of motion.

So much by way of replying to the RDA. But I do need to linger on defending 
this version of perdurantism, especially the assumption that the 
perdurantist {\em can} appeal to the non-instantaneous parts' intrinsic 
properties.  I will defend this perdurantism in four stages. The first two 
stages are metaphysical: I expound them in the next two Subsections. The 
third and fourth stages will return us to the philosophy of physics, and 
will each involve a  proposal about the intrinsic-extrinsic distinction 
among properties. The third stage (Section \ref{velocityrevisit}) just 
appeals to what Section \ref{322CDegreeExtry} already argued for, concerning 
the classical mechanical description of motion: that velocity is hardly 
extrinsic. The fourth stage, in Section \ref{sssec:Appealbeyondclassl}, 
concerns quantum theory.

\subsection{Intrinsic properties of non-instantaneous temporal 
parts}\label{psa:noninststplyextrpropies}
Intrinsic properties of non-instantaneous temporal parts raise three issues; 
which I address in three Subsections.

\subsubsection{Can the perdurantist appeal to them?}\label{721Discs}
I claim that the perdurantist can legitimately appeal to these parts' 
intrinsic properties, even though some of them involve the notion of 
persistence. Does this mean that my sort of perdurantist just gives up on 
the project of defining persistence (or at least providing a supervenience 
basis for it) in terms that do not presuppose it? Agreed, giving up need not 
spell defeat for perdurantism. For a non-reductive perdurantism of the sort 
mentioned in (1) of Section \ref{ssec;naturalism} might have various 
merits---and merits that are not undermined by accepting only 
non-instantaneous temporal parts. (I will support this in Section 
\ref{psa:noninstsdojob}.) But {\em does} my sort of perdurantist give up?

 Yes and No! Yes, in that she aims to give {\em some} account of 
persistence, yet is willing to have the account invoke notions that 
presuppose persistence; in particular, instantaneous velocity.\\
\indent But also, No: for reasons hinted at in Section 7.1.B's discussion of 
persistence ``piggy-backing'' on the laws of motion. That is: my sort of 
perdurantist need not assume persistence as a primitive---or that 
persistence  is somehow satisfactorily defined (or accounted for, say with a 
supervenience thesis)---for some specific set of parts: say, a set that 
covers the lifetime of the persisting object in question, or a set 
containing all those temporal  parts  with a  temporal extent (lifetime) 
less than some bound. She can perfectly well pursue the project of defining, 
or accounting for, persistence as a relation between {\em any} two 
non-instantaneous parts (including any two sub-parts of any given 
non-instantaneous part).\\
\indent And even if the perdurantist accepts all such parts, so that there 
are parts with arbitrarily short, though non-zero, temporal extents, I 
maintain that this need not involve a vicious regress of endlessly deferred 
definitions or accounts of persistence. For the account may, for 
time-intervals less than some amount, become suitably ``uniform'', i.e. with 
no substantive variations for shorter times. In short: it can be ``turtles 
all the way down'', provided that below a certain level, the turtles are all 
the same. Of course, this is in effect what happens in an account of 
persistence that piggy-backs on the classical deterministic laws of motion, 
determining future and past positions in terms of present position and 
instantaneous velocity (or momentum).

\subsubsection{Temporal intrinsicality at an instant is  
rare}\label{722bjps}
 I turn to a general point 
 about the sorts of property  invoked in an account of persistence: a point 
that applies to both endurantist and perdurantist, and to accounts of 
criteria of identity for specific kinds of object, e.g. persons, where there 
are issues, e.g. about the weighing of diverse factors such as bodily and 
psychological similarity, absent from the highly general 
endurantism-perdurantism debate.\\
\indent The point is simply that almost no properties are temporally 
intrinsic to their instance  {\em at an instant}. That is: almost all 
properties require features of their instance not only at a single instant, 
but also at other (albeit perhaps  close) times. So an account of 
persistence, or a criterion of identity for a specific kind of object, needs 
must appeal to temporally extrinsic properties; (though the other times 
involved may be close to the given one).

\indent Unfortunately, this fact is obscured in most philosophical  
discussion of persistence (at least in the tradition of conceptual 
analysis). This discussion focusses on the idea of giving an account of, or 
criterion for,  $o$ at time $t$ being the same persisting object (maybe of a 
specific kind, e.g. person) as $o'$ at $t'$, that invokes everyday  
properties. As discussed in (2) {\em Persistence}, in Section 
\ref{152B:SiderCheap?}, the idea is almost always that the object(s) (in 
perdurantist terms: the two temporal parts)  need to be:\\
\indent \indent [i] suitably similar as to these properties: where `suitably 
similar' allows considerable change provided there is some kind of chain of 
small changes; and-or\\
\indent \indent [ii] suitably causally related, with the properties being 
the causally relevant ones (in other jargon: part of the specification of 
the object's causal state); where again there can be a suitable chain of 
stages or states linked by causation.\\
\indent So far, so good: I have no objection to searching for this sort of 
account or criterion, nor to its invoking everyday properties in ways [i] 
and-or [ii]. But the locution `at time $t$', and the focus on everyday 
properties, makes philosophers often choose as their examples observational 
properties, i.e. properties which can be ascribed ``at a glance'': be they 
``everyday-taxonomic'' like `is a rock/leaf/chair' or ``purely sensory'' 
like `is red/hot'. And since they can be ascribed at a glance, philosophers 
are tempted to think they are  temporally intrinsic in the strong sense of 
requiring something of their instance only for a instant.\footnote{All 
parties can agree that among {\em non}-observational everyday properties, 
most are temporally extrinsic; indeed they often require features  at other 
times of objects other than their instance: for example, being married 
requires a spouse at a past wedding, and no intervening divorce or death.}\\
\indent And {\em that} is false. We are very gross creatures: our perceptual 
apparatus is insensitive to such properties. Rather, the process of 
perception ``averages'', in myriadly complex (and often adaptive) ways, over 
the instant-by-instant  properties of not only the object but also the 
medium, and our  perceptual apparatus itself. So any observational property 
is temporally extrinsic at an instant: it demands features of its instance 
over a time-interval of at least about one twentieth of a second---and in 
general a very complex, open-ended and vague array of features, to boot.

When we set aside conceptual analysis and everyday properties, and consider 
the properties of technical science, in particular physical theories, the 
same conclusion holds good: most properties are temporally extrinsic at an 
instant (though as emphasised, they may well be intrinsic to a 
non-instantaneous temporal part). Thus most of the hundred-odd physical 
quantities that get an entry in a physics dictionary are clearly temporally 
extrinsic at an instant. I have already mentioned velocity: obviously 
momentum, angular momentum and kinetic energy are temporally extrinsic for 
the same reason. Many other quantities, such as temperature, conductivity 
(thermal and electrical), permeability and permittivity, depend for their 
definition (as well as their value) on collective phenomena that require a 
process or  situation  to last longer than an instant (though perhaps much 
less than a second).

\indent I admit that within classical physics, three familiar quantities 
{\em are} good candidates for being temporally intrinsic even to an instant: 
viz. position, mass and electric charge. Besides, for a point-particle: 
these also seem to be spatially intrinsic at a spatial point, not just for 
an extended spatial region. At least, this is so {\em modulo} the topic I 
set aside in Section \ref{131needconnect}, about the basis of spatial 
geometry: that is to say, a ``relationist'' about spatial geometry would no 
doubt object to the claim that the position of a point-particle is spatially 
intrinsic to a point.\footnote{Beware! For a point-sized bit of matter in a 
continuum, the trio of position, mass-density and charge-density {\em seem} 
to be not only temporally intrinsic, but also spatially intrinsic---provided 
we can interpret the densities (i) as defining mass and charge through 
integration, rather than (ii) being themselves defined from the masses and 
charges of finite volumes, by taking the limit of smaller and smaller 
volumes. But in fact, we cannot interpret the densities like this: (i) 
fails, and we need (ii)---another mark against {\em pointillisme}, in my 
view (2004).}\\
\indent I also admit that  this trio seeming to  be intrinsic---taken 
together with the great success of classical physics in reducing much of the 
behaviour of large complex objects  to the classical mechanics and 
electrodynamics of postulated tiny components, whether point-particles or 
point-sized bits of matter in a continuum (``micro-reductionism'')---has 
undoubtedly been one strong reason, perhaps the main reason, for the   
prevalence in philosophy  of {\em pointilliste}  doctrines like Lewis' 
Humean supervenience.\\
\indent Of course, the RDA is precisely an argument that such doctrines come 
to grief on the topic of persistence.\footnote{Philosophers tend to forget 
that they also have trouble in physics. The classical mechanics and 
electrodynamics of point-particles and continua have considerable conceptual 
tensions, some of which are aggravated by a {\em pointilliste} picture; cf. 
Section \ref{sssec;vsStrtfwd}.} And my present point is that the rarity of 
temporal intrinsicality at an instant supports my proposal to be 
perdurantist without being {\em pointilliste}---and so to block the RDA.

\subsubsection{A better reason for temporal intrinsicality}\label{723bjps}
Finally, an incidental point. Philosophers discussing persistence have 
another reason to focus on temporally intrinsic properties, in addition to 
the erroneous tendency to think observational properties are temporally 
intrinsic to an instant. I admit that it is a better reason. But it is a 
reason only for properties temporally intrinsic for shortish intervals, up 
to about a second: not for the stronger notion of temporal intrinsicality at 
an instant---which is the target of my anti-{\em pointilliste} campaign. In 
short, the reason is that a property that is temporally intrinsic for a 
longish interval is liable to be useless in a criterion of identity.\\
\indent In detail: All parties (both endurantists and perdurantists) can 
agree that an account of persistence, or a criterion of identity, had better 
not invoke a property that requires some feature of its instance within a 
period of time similar to the time-scale over which the account or criterion 
is to be applied. For doing so is liable to make the criterion hard or even 
impossible to apply. Thus suppose an account of the conditions under which 
$o$ at time $t$ is the same persisting object (maybe of a specific kind, 
e.g. person) as $o'$ at $t'$, invokes a property $P$: requiring, say, that 
$o$ at $t$ must be $P$ and so must $o'$ at $t'$. (The argument  works 
equally well with other requirements, e.g. that only one of the two need be 
$P$, but that change as regards $P$ is suitably continuous, with some kind 
of chain of small changes.) Then if being $P$ at $t$ requires a feature 
$\phi$ at a time close to $t'$, it may well be hard to apply the account: 
having to ascertain that $\phi$ holds close to $t'$ might entangle one in 
ascertaining whether the persistence claim for $o$ and $o'$ holds.

\subsection{Non-instantaneous parts can do the 
jobs}\label{psa:noninstsdojob}
I turn to the second stage of my defence of perdurantism without 
instantaneous temporal parts. I claim that, by and large, non-instantaneous 
temporal parts do the various jobs, within the endurantism-perdurantism 
debate, that the perdurantist  demands of temporal parts, just as well as 
instantaneous temporal parts. More precisely: this is so once the 
perdurantist ``just says No'' to the siren-calls of {\em pointillisme}. Of 
course, I cannot here discuss all these jobs: I will  make do with three 
short comments. The first comment is general, and will be illustrated by the 
second and third, which concern particular jobs temporal parts are invoked 
to do.

\paragraph{7.3.A Humean supervenience revisited}\label{73AHSrevisit} The 
first comment is an offer of a peace-pipe to the neo-Humean. She envisages 
the world as ``loose and separate'', a succession of ``distinct 
existences'': ``just one darned thing after another''. My version of 
perdurantism can agree, in that it might well accept {\em all} 
non-instantaneous temporal parts, no matter how short-lived: my veto is only 
against utterly instantaneous parts.\\
\indent Besides, my perdurantist can echo Lewis' Humean supervenience, by 
making some claim along the lines that all the facts supervene on the 
temporally local facts; i.e. the facts specified by the intrinsic (if you 
like: temporally and spatially intrinsic) properties of all the 
non-instantaneous temporal parts. To state this echo more precisely: she can 
claim that for any covering of spacetime ${\cal M}$ by a family ${\cal F}$ 
of non-instantaneous temporal parts (no matter how short-lived some or all 
of the parts may be), all the facts supervene on the intrinsic properties of 
elements of ${\cal F}$. (Here, `covering' is understood in mathematicians' 
usual sense: a set ${\cal M}$ is covered by a family ${\cal F}$ of sets iff 
${\cal M} \subseteq \cup {\cal F}$; and similarly if $\cal M$ and the 
elements of $\cal F$ are treated not as sets, but as say mereological 
fusions.)   \\
\indent So the only aspect of Lewis' Humean supervenience that my 
perdurantist  needs to deny is the {\em pointilliste} idea that all the 
facts supervene on the intrinsic properties of spacetime points (or  of 
spatially extended instants of time, i.e. spacelike surfaces). I think 
neo-Humeans should find this a price worth paying: having all the facts 
supervene on the intrinsic properties of all the non-instantaneous temporal 
parts should be enough to satisfy a Humean's ambition to have the ``global'' 
supervene on the ``local''.\footnote{Agreed: since these parts in general 
overlap, the ``fundamental description of the world'', given by the infinite 
conjunction of all (the ascriptions of) the intrinsic properties of all such 
parts, is highly redundant. But I say: no worries. After all, the exact 
spatial analogue occurs in continuum classical mechanics: to describe a 
continuum, this theory needs---{\em not} the infinite point-by-point 
conjunction of all the properties of points---but the highly redundant 
infinite region-by-region conjunction of all properties of all regions; cf. 
Section \ref{sssec;vsStrtfwd}.}

\paragraph{7.3.B The problem of change}\label{73BChange} The second comment 
concerns the so-called `problem of change'. Perdurantists argue that $o$'s 
changing in respect of a property  $P$ is best understood in terms of one 
temporal part having $P$ and another having $\neg P$. In particular, they 
argue that the endurantist has to understand $P$ (and $\neg P$) as a 
relation  to a time, and that for the case of an intrinsic property $P$ this 
is surely wrong. Hence the problem is also called the `problem of temporary 
intrinsics'; (cf. e.g. Sider 2001 pp. 92-98, Lewis 2002).\\
\indent So far as I can tell, almost all the arguments for the perdurantist 
understanding of change carry over, so as to support my version of 
perdurantism, i.e. perdurantism without instantaneous temporal parts. 
(Admitted: as do almost all the arguments {\em against} the perdurantist 
understanding of change.) The main reason is of course that if within a 
single non-instantaneous  part $o$ there is change in respect of $P$, the 
perdurantist will understand the change in terms of one shorter-lived part 
of $o$ having $P$, and another not---and this need not involve any regress 
(Section \ref{721Discs}). Besides: since temporal intrinsicality at an 
instant is rare (Section \ref{722bjps}), the perdurantist's argument that 
endurantism has trouble with temporary intrinsics is more persuasive as an 
argument  for non-instantaneous temporal parts.\\
\indent But there is one objection; (my thanks to Oliver Pooley). Suppose 
that a temporary intrinsic property such as shape changes continuously over 
time, so that an object $o$ is square for merely an instant: to secure an 
instance of squareness {\em simpliciter} in this scenario,  the perdurantist  
surely needs an instantaneous temporal part.\\
\indent Reply: Given the supposition, this is certainly right. Here I can 
only bite the bullet, by any or all of:\\
\indent \indent (i) dropping the problem of change from the list of jobs my 
non-instantaneous  temporal parts are to  do; or\\
\indent \indent (ii) urging that since temporal intrinsicality at an instant  
is rare (Section \ref{722bjps})  my temporal parts can solve the problem of 
change for the vast majority of temporary intrinsic properties; and besides, 
urging that succeeding with this vast majority should satisfy the neo-Humean  
(cf. the first comment above);  or\\
\indent \indent (iii) adopting a ``mixed'' view, more congenial to {\em 
pointillisme}, that admits instantaneous parts as well as non-instantaneous 
ones, but then argues that it is legitimate to account for persistence (and 
so answer the RDA) by invoking only the non-instantaneous ones, as I have.\\
\indent Of these options, I on the whole prefer reply (ii). But I will not 
in this paper try to choose between these replies: in particular, I will not 
refer again to the mixed view, though I agree it is tenable.   

\paragraph{7.3.C Puzzles of coincidence}\label{73Ccoincidence}
Thirdly, the situation as regards the debate over `puzzles of coincidence' 
is similar to that for the problem of change. The puzzles (reviewed by Sider 
2001, p. 5-10, 141-152) concern such cases as the statue and the clay, or 
the fission and fusion of objects such as amoebae---or even persons. For 
example, after an artist makes on Tuesday a statue out of a lump of clay, 
the statue and clay seem to be the very same object. But they seem to differ 
in their temporally extrinsic  properties (often in this debate called 
`historical properties', e.g. by Sider 2001, p. 5, 142): the statue but not 
the lump was created on Tuesday, the lump but not the statue existed on 
Monday. Perdurantists argue that these puzzles are best understood in terms 
of distinct objects sharing temporal  parts, just as objects can share 
spatial parts (such as two roads having a stretch  in common). \\
\indent Again: so far as I can tell, almost all the arguments for the 
perdurantist understanding of these puzzles carry over, so as to support my  
perdurantism without instantaneous  temporal parts. (As do, I admit, the 
arguments against!). For example, almost all the arguments in Sider's 
critique of endurantist approaches (2001, p. 154-188), and in his advocacy 
of perdurantism (2001, p. 152-153, p. 188-208), carry over.

I said `almost all the arguments' carry over. For there are two wrinkles. 
First, Pooley puts the analogue of his objection in Section 7.3.B. Suppose 
that two objects fuse for merely an instant: here the perdurantist surely 
needs an instantaneous temporal part. I reply: I think this objection is 
weaker than its analogue in Section 7.3.B, because its supposition is more 
of an idealization, more a merely logical or metaphysical possibility, 
rather than part of the content of classical mechanics. That is: classical 
mechanics does describe deformable objects changing shape continuously, as 
the objection in Section 7.3.B requires. But it does {\em not} describe 
instantaneous fusions. Indeed as mentioned in Section \ref{sssec;vsStrtfwd}, 
classical mechanics finds collisions, even of point-particles, 
problematic---let alone fusions and fissions. (There is of course no problem 
about the spatial analogue of instantaneous fusions, i.e. two 3-dimensional 
objects sharing a 2-dimensional part: think of two semi-detached houses!)   

\indent The second wrinkle is that the issue whether to accept instantaneous 
temporal parts {\em does} bear on one significant division {\em within} the 
perdurantist camp. This distinction concerns how the perdurantist treats 
temporal language. The traditional  perdurantist view is that an object of 
ordinary ontology---i.e. a referent of an ordinary term, a subject of 
ordinary predications, an element of ordinary domains of quantification---is 
the whole four-dimensional object, the ``maximal spacetime worm''; (Sider 
calls this the `worm view').  But both Sider (2001, p. 188-208) and Hawley 
(2001, pp. 30-32, 41-64) defend the rival `stage view', that the referents 
of our ordinary terms, subjects of ordinary predications etc. {\em are} the 
temporal parts.\\
\indent This is not the place to assess their arguments for this proposal. 
They concern, for example, counting: the stage view says that at each time 
before an amoeba splits into two, there is one amoeba (the stage), a verdict 
which matches everyday thought and language; but since there are then two 
maximal spacetime worms,  the worm view has to say that there are {\em 
stricto sensu} two amoebae, and explain away everyday thought and language 
by invoking some conventions about counting.\footnote{For this line of 
argument, cf. Sider 2001, pp 152-153, 188-193. But Sider has to admit that 
sometimes we count by maximal spacetime worms, not by stages, as in `Fewer 
than two trillion people have set foot in North America throughout history'. 
He writes (2001, p. 197): `if `person' refers to person stages, this 
sentence will turn out false, since more than two trillion (indeed, 
infinitely many if time is dense) person stages have set foot in North 
America throughout history'.}\\
\indent For my purposes here, it suffices to comment on Sider's position  
that the stages he claims to be the referents of ordinary terms are indeed 
instantaneous---and so do not persist: `no person lasts more than an 
instant' (2001, p. 193)! Sider of course agrees that everyday thought and 
language take: (i) ordinary objects to persist, as in `Ted was once a boy'; 
and (ii) most of their properties to be temporally extrinsic at an instant, 
as in `Ted believes perdurantism is true' (Section  \ref{722bjps}). So he 
goes on to argue that he can accommodate (i) and (ii) with a temporal 
analogue of Lewis' counterpart theory (2001, pp 111-113, 193-198).\\
\indent My comment on Sider's position is now obvious. While I admit that 
temporal counterpart theory is coherent and powerful enough to cope with (i) 
and (ii)---the stage view does not have to be so {\em pointilliste} as 
Sider! That is: one could combine my perdurantism, the rejection of 
instantaneous temporal parts, with the stage view.  Not only do most 
arguments for a perdurantist understanding of the puzzles of coincidence 
carry over and support my perdurantism (as I said above). Also, one could 
combine it with some arguments specifically for the stage view: e.g. a 
version of my perdurantism that denies overlapping parts could retain 
Sider's counting argument for favouring the stage view over the worm view 
... But I leave developing this topic for another occasion.

This concludes my metaphysical defence of my version of perdurantism. I hope 
to have made it plausible, quite apart from its blocking the RDA. But the 
philosophy  of physics has some more support to offer it. In the next 
Subsection, the support comes from the classical description of motion.  In 
Section \ref{sssec:Appealbeyondclassl}, the support comes from quantum 
theory. But these two pieces of support are not ``just technical'': each of 
them involves a novel proposal about the intrinsic-extrinsic distinction 
among properties. 

\subsection{Velocity revisited}\label{velocityrevisit}
My version of perdurantism, without instantaneous temporal parts, meshes 
well with Section \ref{322CDegreeExtry}'s arguments that:\\
\indent (i) it is natural to sub-divide the vast class of extrinsic 
properties---and in particular, temporally extrinsic properties---in terms 
of degrees of extrinsicality; and\\
\indent (ii) in the classical mechanical description of motion, velocity is 
hardly extrinsic. 

I will not rehearse again the details of those arguments. I only need to 
recall  the main idea: that the ``only proposition going beyond the 
instance'' that is implied by an ascription of  velocity (or of other 
derivatives of position) is the proposition that the instance $o$ exists 
throughout some open interval of times, perhaps  very short, around the time 
$t$. This proposition corresponds to an intrinsic property of any  
non-instantaneous temporal part containing $t$, no matter how short: a 
property that is thereby hardly extrinsic to $t$.\\
\indent Obviously, this idea  meshes with two points in previous Subsections 
of this Section:---\\
\indent (a): A perdurantist who rejects instantaneous temporal parts can 
account for persistence: either by fusing segments, perhaps arbitrarily 
short, of worldlines (``kinematics''; Section 7.1.A); or (more 
naturalistically) by ``piggy-backing'' on solving the deterministic 
classical laws of motion, given $o$'s initial position and velocity, and the 
forces on it; (``dynamics''; Section 7.1.B).\\
\indent (b): A perdurantism without instantaneous temporal parts can accept 
all {\em non}-instantaneous parts, no matter how small their temporal 
extent. Besides, if one accepts all such parts one can add a claim that all 
facts supervene on the ``temporally local'' facts, in a strong enough sense 
to satisfy all but the most {\em pointilliste} neo-Humeans; (cf. Section 
7.3.A).

\section{Support from decoherence in quantum 
theory}\label{sssec:Appealbeyondclassl}
\subsection{Classical and quantum: relativizing the intrinsic-extrinsic 
distinction}\label{ssec:relizingied}
As I said in Section \ref{ssec;naturalism}, I am not so far gone in 
naturalism as to 
 just dismiss the RDA on the grounds that matter is in fact atomic. I agree: 
a classical mechanical continuum could exist---prompting the RDA, {\em 
modulo} the above replies. My argument in this Section will instead be that 
the way in which classical mechanical objects (both particles and continua) 
are in fact emergent from the quantum realm provides further support for 
Section \ref{psa;perdmwouttears}'s perdurantism without tears, i.e. without 
instantaneous temporal parts.

This argument will use two new assumptions: one about philosophical method, 
the other about the intrinsic-extrinsic distinction.\\
\indent (1): I will now assume that the interpretation of classical 
mechanics---in particular, our conception of how its objects (both particles 
and continua) persist---should be sensitive to how classical mechanical 
objects in fact ``emerge from the quantum''. I agree that this assumption is 
controversial: why not just interpret each theory on its own, as best you 
can? After all, there is no lack of work: as I have stressed, classical 
mechanics is interpretatively subtle and problematic, even without 
considering the dreaded quantum. But I am not alone in endorsing this 
assumption, even as regards the interpretation of a classical theory being 
sensitive to an ``adjacent'' quantum theory. Thus for Belot (1998, p. 
550-554), it is the main moral of his examination of classical 
electromagnetism and the Aharonov-Bohm effect.

\indent (2): My second assumption is that it is legitimate to relativize the 
intrinsic-extrinsic distinction among properties to a body of doctrine. The 
distinction is of course usually discussed in terms of logical or 
metaphysical possibility: the literature discusses taking a property $P$ to 
be intrinsic iff it is logically or metaphysically possible for an object 
$o$ to have $P$ ``while lonely'', or ``whatever the rest of the world is 
like'', or ... But I now assume that it is legitimate to relativize the 
modality to a body of doctrine, such as a scientific theory $T$. (I will not 
need the metaphysically more ambitious idea of relativizing to the ``laws of 
nature'', or to the laws of nature of some possible world.) Therefore I 
shall  talk, for any such body of doctrine or theory $T$, of {\em nomic} 
intrinsicality and extrinsicality.\\
\indent Unless  $T$ is logically or metaphysically necessary---a case I need  
not consider---the relativized modality will be a restricted one. That is: 
not all logically or metaphysically possible worlds make $T$ true. In 
general, this will strengthen the notion of intrinsicality, and 
correspondingly weaken the notion of extrinsicality---however exactly we 
understand the original intrinsic-extrinsic distinction. That is: nomic 
intrinsicality will imply intrinsicality {\em simpliciter}, and 
extrinsicality {\em simpliciter} will imply nomic extrinsicality. For  
intrinsicality is a matter of ``not implying propositions about the 
instance's environment''; and once we assume a theory $T$ is true, any 
proposition in $T$ can be an implicit premise in an implication---yielding 
more implications. So in general, once we assume $T$, more properties will 
be classified as extrinsic. So extrinsicality {\em simpliciter} implies 
nomic extrinsicality; and {\em vice versa} for intrinsicality. (Similar 
remarks apply to my notions of temporal, and spatial, intrinsicality and 
extrinsicality; and to the case where we consider two theories $T_1$ and 
$T_2$, one implying the other.)\\
\indent In fact, this idea of relativized intrinsicality has surfaced in the 
literature (Humberstone 1996, p. 238); but so far as I know, it has not been 
pursued. I agree that  many a metaphysician will at first sight doubt its 
value, though they will probably accept it as coherent. Thus Humberstone 
writes, after floating the idea of  relativizing intrinsicality to a class 
of possible worlds that match in their laws of nature: `From a suitably 
elevated position [i.e. suitably general philosophical stance], this has an 
element of arbitrariness about it: why not restrict attention to 
worlds---not with the same laws as ours, but---with the same tourism 
statistics for Naples as ours?' (ibid.).\\
\indent But I submit that relativization to (our best guess for) the laws of 
physics has some interest! In any case, I can at least show that in the 
present context, it has the interest of being {\em surprising}. For in 
Section \ref{52Cqmdetails} I will argue that the {\em position}, and even 
the {\em existence}, at a time of an emergent classical object (whether a 
particle or a point-sized piece of matter in a continuum) is {\em 
extrinsic}, relative to the laws of quantum theory.\footnote{Besides, the 
extrinsicality has nothing to do with the possible involvement of other 
objects in defining position, as urged by a relational conception of space 
(set aside since Section \ref{131needconnect}). The extrinsicality is what I 
have called {\em temporal} extrinsicality, rather than spatial; and it 
arises from decoherence.}  

But before arguing for this, I should briefly set aside another way in which 
quantum theory bears on persistence, and apparently on the RDA.

\subsubsection{Unitarity: momentum as temporally intrinsic}\label{511 
Unitarity} 
Quantum theory violates an assumption that the RDA depends on, viz. that 
velocity is not part of the instantaneous state of an object. (This 
assumption, first registered in Section \ref{ssec;intrpropvelies}, led to 
discussing Tooley's heterodox proposal that velocity {\em  should} be part 
of the instantaneous  state.) This assumption is often endorsed in the 
metaphysical literature about space, time and motion, even apart from the 
RDA: for example, Sider (2001, p. 39) says `fixing the properties  and 
relations of present objects will not fix their velocities' (cf. also his p. 
34-35).

\indent The assumption tends to be associated with the fact that in 
classical mechanics, in order to determine an object's  future (and past) 
motion, you need not only its present position and the forces acting on it 
(in the time-interval concerned), but also its present velocity; i.e. the 
fact that classical mechanics' equations of motion are second-order in time. 
For in a theory in which position and forces were enough to determine the 
motion (a theory that is first-order in time), it would be more tempting to 
say that velocity is part of the present instantaneous state. At least, it 
would be as tempting to say this, as that the whole future (and past) 
history of the system is part of the present instantaneous state (because of 
the determinism). Certainly, in such a theory the RDA itself would have much 
less sting for a ``naturalistic'' perdurantist, who is willing to let her 
account of perdurance depend on the actual laws. For in such a case, 
ingredients that the RDA's advocate presumably  agrees to be available to 
the perdurantist, viz. position and forces, {\em are} enough to determine 
future positions.  

\indent But quantum theory violates this assumption.\footnote{As readers who 
are {\em cognoscenti} of quantum theory will have long ago noticed: at least 
by the time that Section \ref{psa;vetoinsts} proposed we could have 
perdurantism without tears, by letting the perdurantist ``have'' velocity, 
and even have their account of persistence ``piggy-back'' on integrating the 
classical equations of motion. Apologies for the delay!} It  {\em is} 
first-order in time.
It combines the position and velocity (better: momentum) aspects into a 
single instantaneous state of a system which, together with the forces 
acting on the system, determines its future (and past) states (setting aside 
controversy about whether there is  a ``collapse of the wave-packet'' on 
measurement).\footnote{Agreed, the Hamiltonian formulation of classical 
mechanics also combines position and momentum in its conception of state, 
and so is first-order in time. But there is a crucial disanalogy: neither of 
the pair, position and momentum, determines the other. (Indeed, the 
formulation is equivalent to the Lagrangian or Newtonian formulation, under 
certain conditions, in particular  taking the phase space to be the 
cotangent bundle of a configuration space.) But in quantum theory, the 
position and momentum  representations each determine the other.}

So it is tempting to say that in quantum theory, velocity and momentum are 
just as intrinsic (or temporally intrinsic) to the system at a time, as is 
position; (Arntzenius (2003, p. 282) says this). A bit more precisely: once 
we are willing to relativize the intrinsic-extrinsic distinction to a 
physical theory (as proposed in (2) above), it is tempting to say this.\\
\indent  Furthermore, just as Section \ref{psa;vetoinsts} proposed that in a 
classical setting, a perdurantist accepting only non-instantaneous parts 
could have their account of persistence ``piggy-back'' on integrating the 
classical equations of motion: so in quantum theory, the perdurantist's 
account of persistence  could  appeal to  integrating the quantum equations 
of motion. (But as the weasel-word `system' hints,  it is controversial how 
to relate persisting objects to quantum systems, even if you know the 
systems' complete histories: cf. the next Subsection).

\indent So be it, say I. But again: I am not so far gone in naturalism about 
persistence---I am loath to just dismiss the RDA on the grounds that quantum 
theory is first-order in time. A theory of persistence should accommodate 
classical continua, and this Subsection's points do not bear directly on how 
it can do so. However, I will now argue that quantum theory has {\em other} 
light to shed on our topic---once we ask the interpretation of classical 
mechanics to take note of how classical mechanical objects emerge through 
decoherence.

\subsection{Position and existence as nomically 
extrinsic}\label{52Cqmdetails}
So let us adopt the idea in (2) of Section \ref{ssec:relizingied}, of nomic 
intrinsicality and extrinsicality. The intrinsic-extrinsic distinction among 
properties is to be relativized to bodies of doctrine---in particular, to 
quantum theory. 

Warning: Choosing logically strong bodies of doctrine can yield odd-sounding  
verdicts of extrinsicality. Given our interest in temporal extrinsicality, 
the  obvious example of this is provided by a deterministic theory. Thus 
suppose you choose to relativize, not just to the deterministic theory 
itself, but to the conjunction of the theory and the regime of forces 
imposed on a system in some time-interval $(a,b)$. This yields the verdict 
that every instantaneous  state\footnote{Since for some philosophers, a 
state is not a property, it is better to say: every property that specifies 
such a state.} is temporally very extrinsic: indeed, about as extrinsic as 
it could be. For given the laws of the theory and the forces imposed, any 
instantaneous state of a system   determines the system's states during 
$(a,b)$. But it sounds wrong to say that every instantaneous state is 
temporally very extrinsic.  \\
\indent The solution of course is to exercise some judgment about what is a 
natural or useful body of doctrine to which to relativize. In our example, 
the theory is presumably such a body of doctrine, but its conjunction with a 
specified regime of forces is not: that is too particular (logically 
strong). More generally, we should allow some distinction between 
``central'' and ``peripheral'' statements (or more generally; features) of 
an ambient body of doctrine, and relativize the intrinsic-extrinsic 
distinction  only to  (the conjunction of) the central ones. That is, only 
the central ones are held fixed in all the nomic possibilities, and so by 
nomic intrinsicality and extrinsicality. Then you may say in the example 
that (maybe part of) the specification of the forces is not central, so that 
instantaneous states are not so very temporally extrinsic. 

Let us now apply this sort of relativization to how classical mechanical 
objects emerge through decoherence. Fortunately for us, although the quantum 
measurement problem remains controversial and there remain many open 
technical questions in the physics of decoherence, we need not address these 
controversies and questions. We can sidestep the measurement problem, and 
manage with only the most basic and best-established features of 
decoherence.\footnote{Bacciagaluppi (2003) is an excellent introduction to 
decoherence for philosophers; for more technical details, Guilini et al. 
(2003), Schlosshauer (2003) are also excellent. By the way, all these 
sources endorse the consensus that decoherence cannot by itself provide the 
solution of the measurement problem, but is an important ingredient in any 
such solution.}

Classical mechanical objects (both particles and continua) are in fact 
transient and approximate patterns in the quantum state of an underlying 
quantum system. They are patterns that emerge from an ubiquitous, continuous 
and very efficient process of decoherence, which continues throughout the 
lifetime of the classical object. Roughly speaking, decoherence is diffusion  
(spreading) in to the quantum  system's environment  of coherence, i.e. of 
the  puzzling interference effects in the  probability distributions that 
are the system's state.\\
\indent To keep things simple, I shall discuss this in terms of the 
elementary quantum theory of particles, not quantum field theory. But I 
should note that:\\
\indent \indent (i): quantum particles are themselves transient and 
approximate patterns in the quantum state of an underlying quantum field or 
fields; for  discussion of this, cf. Wallace (2004, especially Section 
5.2);\\ 
\indent \indent (ii):  decoherence also happens within quantum field theory; 
for a review, cf. Guilini et al. (2003, Chapter 4). 

\indent Here are some details about a well-studied model of a quantum  
particle immersed in an environment (called `quantum Brownian motion'). Take 
as the initial quantum state of a tiny dust-particle (radius $10^{-3}$ cm) 
in air, a superposition of two positions for the centre of mass of the 
particle, with the two positions just $10^{-4}$ cm apart (i.e. a tenth of 
the particle's radius), and with (say) the two positions not moving relative 
to one another. The bombardment of the particle by air molecules is {\em 
very} efficient in diffusing the coherence in to the environment: the 
superposition's interference effects converge to zero like $\exp 
(t/10^{-36}$ sec) and remain small for a very long time ($10^{10}$ years)!\\
\indent This means that {\em very} soon the probabilities for any quantity 
on the particle you care to measure are as if there is an even chance of the 
(centre of mass of the) particle being in the two positions; (i.e. 
probabilities  for quantities other than position are also  given by a 50-50 
mixture corresponding to the two positions). Similarly for other initial 
states: if the initial superposition had the two positions separating from 
each other at say $x$ cm sec$^{-1}$, then after a second, the probabilities 
would be as if there is an even chance of the (centre of mass of the) 
particle being in two positions $x + 10^{-4}$ cm apart. Indeed, more 
generally: it is even possible to deduce the approximate validity of the 
deterministic  classical mechanical  equations of motion of a dust-particle 
from the underlying equations for the quantum system, together with a 
description of the decoherence process. 

So the classical object, ``the dust-particle we see'', corresponds to one of 
these two decohered possibilities (in my example: possibilities for the 
position of the centre of mass).  It is a pattern  in a quantum state, which 
also contains another pattern corresponding to the other possibility. If the 
quantum state were sufficiently different, not only would the classical 
object not have the position and momentum we see: it would not {\em exist}. 
In particular, if the decoherence process did not occur, it would never 
exist; and if the decoherence process  did not continue, it would cease to 
exist. That is: the quantum system would continue to exist, but the 
classical dust-particle would not: it would ``disappear into a quantum 
fog''.\footnote{For more discussion of the idea of classical objects as 
patterns in quantum objects, cf. Wallace (2003).}

I propose that we take these propositions, about how classical objects are 
in fact patterns in a quantum state that are formed because of an ongoing 
process of decoherence, as what I called `central'. After all, they are 
crucial to how such objects are in fact constituted. That is: I propose they 
are to be held fixed in assessing whether a property is nomically intrinsic 
or extrinsic. So they are to be available as implicit premises for 
implications from ascriptions of a property to propositions about the world 
beyond the property's instance.\\
\indent It follows that  an ascription to a classical object such as a 
dust-particle, of a position at $t$ (to be precise: for its centre of mass, 
say), is nomically {\em extrinsic}. (I would say: temporally extrinsic, 
since the implications are about facts at times other than $t$). For the 
ascription (together with the implicit premises) implies the (categorical) 
proposition that the object has a position at all other times in a (very 
short but non-zero!)  interval of times  around $t$. Here, the length of the 
interval is determined by the decoherence process' time-scale. \\
\indent  Similarly, a statement that the object exists at a time is 
nomically {\em extrinsic}. For it implies that the object exists at all 
other times in an interval about as long as the decoherence 
time-scale.\footnote{A point of clarification for quantum {\em aficionados}. 
You might object that since\\
\indent (i) the reduced state density matrix of the dust-particle (strictly: 
of its centre of mass degree of freedom) is nearly diagonal (upto some 
desired level of approximation) in position, {\em at an instant};\\
\indent  it surely follows that:\\
\indent  (ii) the position and existence of the classical particle is {\em 
not} nomically temporally extrinsic.\\
\indent I reply: (i) is of course true, but does not imply (ii). For I am 
taking as ``central'', not just the formalism of reduced states etc., but 
also the physical fact of a decoherence process over time.}  

So far I have only discussed the emergence of a classical particle, such as 
a dust-particle. But the discussion just given carries over to continua, as 
regards both physics and philosophy.\\
\indent Admittedly, there are more technical questions about decoherence in 
quantum fluids that are still open than about quantum Brownian motion, which 
is by now very well-studied. But there is already a good understanding of 
decoherence in quantum  fluids, and so of the emergence of classical 
continua. In short: recent work shows that even in a quantum fluid, where 
there is no clear distinction between system and environment, decoherence 
selects certain quantities (roughly, hydrodynamic variables) as ``behaving 
classically''. Again, one can deduce the approximate validity of the 
classical equations of motion for a fluid. (For details, cf. Halliwell 
(1999) and references given there.) \\
\indent As regards philosophy: I said above that the fact that classical 
mechanical objects (both particles and continua) are in fact emergent from 
the quantum realm should be reflected in the interpretation of classical 
mechanics, and so in a naturalistic theory of persistence. One way to do 
this is now clear. Namely: take the nomic extrinsicality of position at a 
time, and even existence at a time, as favouring the denial of instantaneous 
temporal parts.  Thus decoherence supports the perdurantism without the 
tears of {\em pointillisme} which I defended in Section 
\ref{psa;perdmwouttears}: a naturalistic perdurantist can interpret 
classical mechanics in terms of temporally extended temporal parts---and 
thereby block the RDA.

{\em Acknowledgements}:--- I thank: G. Belot, C. Callender, W. Demopoulos, 
G. Emch, I. Gibson, L. Humberstone, P. Mainwood, A. Oliver, O. Pooley, M. 
Redhead, T. Sattig, S. Saunders, N. Shea, D.Wallace, A. Wilson and D. 
Zimmerman for discussions and comments on previous versions; audiences in 
Florence, Granada, Leeds, London England, London Ontario, Montreal and 
Oxford; and participants at a seminar in autumn 2003.

\section{References}
R. Abraham and J. Marsden (1978), {\em Foundations of Mechanics}, 2nd 
edition, Benjamin/Cummings.\\
D. Armstrong (1980), `Identity through time', in ed. P. van Inwagen, {\em 
Time and Cause}, Dordrecht: Reidel.\\
D. Armstrong (1997), {\em A World of States of Affairs}, Cambridge: 
University Press.\\
F. Arntzenius (2000), `Are there really instantaneous velocities?', {\em The 
Monist} {\bf 83}, pp. 187-208.\\
F. Arntzenius (2003), `An arbitrarily short reply to Sheldon Smith on 
instantaneous velocities', {\em Studies in the History and Philosophy of 
Modern Physics} {\bf 34B}, pp. 281-282.\\
G. Bacciagaluppi (2003), `The role of decoherence in quantum theory', in ed. 
E. Zalta, {The Stanford Encyclopedia of Philosophy}, available at: 
http://plato.stanford.edu.\\
Y. Balashov (1999), `Relativistic objects', {\em Nous} {\bf 33} pp. 
644-662.\\
Y. Balashov (2000), `Persistence and spacetime: philosophical lessons of the 
pole and the barn', {\em The Monist} {\bf 83}, pp. 321-340.\\
G. Belot (1998), `Understanding electromagnetism', {\em British Journal for 
the Philosophy of Science} {\bf 49}, pp 531-555.\\
G. Belot (2000), `Geometry and motion', {\em British Journal for the 
Philosophy of Science} {\bf 51}, pp. 561-595.\\
J. Bigelow and R. Pargetter (1989), `Vectors and Change', {\em British 
Journal for the Philosophy of Science} {\bf 40}, pp 289-306.\\
J. Bigelow and R. Pargetter  (1990), {\em Science and Necessity}, Cambridge 
University Press.\\
R. Black (2000), `Against Quidditism', {\em Australasian Journal of 
Philosophy} {\bf 78}, pp. 87-104.\\
P. Bricker (1993), `The fabric of space: intrinsic vs. extrinsic distance 
relations', in ed.s P.French et al., {\em Midwest Studies in Philosophy} 
{\bf 18}, University of Minnesota Press, pp. 271-294.\\
H. Brown and O. Pooley (2001), `The origin of the spacetime metric: Bell's 
`Lorentzian pedagogy' and its significance in general relativity', in C. 
Callender and N. Huggett (eds) {\em Physics Meets Philosophy at the Planck 
Scale}, Cambridge University Press, pp. 256--72; available at 
http://philsci-archive.pitt.edu/archive/00001385/\\
H. Brown and O. Pooley (2004), `Minkowski spacetime: a glorious non-entity', 
forthcoming in V. Petkov ed., {\em The Ontology of Spacetime}; available at 
http://philsci-archive.pitt.edu/archive/00001661/\\
J. Butterfield (2002), `The end of time?', {\em British Journal for the 
Philosophy of Science} {\bf 53}, pp. 289-330.\\
J. Butterfield (2004), `Classical mechanics is not {\em pointilliste} and 
can be perdurantist', in preparation.\\
J. Butterfield (2004a), `On the persistence of particles', forthcoming in 
{\em Foundations of Physics}. Available at Los Alamos archive: 
http://xxx.soton.ac.uk/abs/physics/0401112; and Pittsburgh archive: 
http://philsci-archive.pitt.edu/archive/00001586/\\
C. Callender (2001), `Humean supervenience and rotating homogeneous matter', 
{\em Mind} {\bf 110},  pp. 25-42.\\
F. Diacu and P. Holmes (1996), {\em Celestial Encounters: the Origins of 
Chaos and Stability}, Princeton University Press.\\
K. Devlin (2002), {\em The Millennium Problems}, London: Granta Books.\\
W. Dixon (1978), {\em Special Relativity}, Cambridge: University Press.\\
P. Dowe (2000), {\em Physical Causation}, Cambridge University Press.\\
J. Earman (1989), {\em World Enough and Space-Time}, Cambridge MA: MIT 
Press.\\ 
I. Grattan-Guinness (2002), `Algebras, Projective Geometry, Mathematical 
Logic and Constructing the World: intersections in the philosophy of 
mathematics of A.N. Whitehead', {\em Historia Mathematica} {\bf 29}, pp. 
427-462.\\
D. Guilini et al. (2003), {\em Decoherence and the Appearance of a Classical 
World in Quantum Theory}, New York: Springer (2nd edition).\\
J. Halliwell (1999), `The Emergence of Hydrodynamic Equations from Quantum 
Theory: A Decoherent Histories Analysis', to appear in Proceedings of the 
4th Peyresq Conference, 1999; available at: 
http://xxx.soton.ac.uk/abs/quant-ph/9912037 \\
S. Haslanger (1994), `Humean supervenience and enduring things', {\em 
Australasian Journal of Philosophy} {\bf 72}, pp. 339-359.\\
K. Hawley (1999), `Persistence and Non-supervenient Relations', {\em Mind} 
{\bf 108}, pp. 53-67.\\
K. Hawley (2001), {\em How Things Persist}, Oxford University Press.\\ 
C. Hitchcock (2003), `Of Humean Bondage', {\em British Journal for the 
Philosophy of Science} {\bf 54}, pp. 1-25.\\
I. Humberstone (1996), `Intrinsic/Extrinsic, {\em Synthese} {\bf 108}, pp. 
205-267.\\
C. Lanczos (1986), {\em The Variational Principles of Mechanics}, Dover 
(reprint of 4th edition, 1970).\\
R. Langton and D. Lewis (1998), `Defining `intrinsic'', {\em Philosophy and 
Phenomenological Research} {\bf 58}, pp. 333-345; reprinted in Lewis' {\em 
Papers in Metaphysics and Epistemology} (1999), Cambridge University 
Press.\\
G. Leibniz (1698), `On Nature itself, Or On the inherent force and actions 
of created things'; pp. 498-508 of L.E. Loemker ed. (1989) {\em Gottfried 
Wilhelm Leibniz: Philosophical Papers and Letters}, Dordrect: Kluwer.\\
J. Levy-Leblond (1995), `Quantum Phenomena at Large', in  ed.s Enrico 
Beltrametti and J. Levy-Leblond, {\em Advances in
Quantum Phenomena}, Plenum Press (Nato Series), pp. 281-295.\\
D. Lewis (1970), `How to define theoretical terms', {\em Journal of 
Philosophy} {\bf 67}, 427-446; reprinted in his (1983). \\
D. Lewis (1973), {\em Counterfactuals}, Oxford: Blackwell.\\
D. Lewis (1983), {\em Philosophical Papers, volume I}, New York: Oxford 
University Press.\\
D. Lewis (1986), {\em Philosophical Papers, volume II}, New York: Oxford 
University Press.\\
D. Lewis (1994), `Humean Supervenience Debugged', {\em Mind} {\bf 103}, pp 
473-490; reprinted in his {\em Papers in Metaphysics and Epistemology} 
(1999), Cambridge University Press pp. 224-247.\\
D Lewis (1999), `Zimmerman and the Spinning sphere', {\em Australasian 
Journal of Philosophy} {\bf 77}, pp. 209-212.\\
D Lewis (2002), `Tensing the copula', {\em Mind} {\bf 111}, pp. 1-13.\\
D. Lewis (2004), `Ramseyan Humility', forthcoming in {\em The Canberra 
Plan}, ed. D. Braddon-Mitchell and R. Nola.\\
E. Lieb (1997), `The Pauli Principle and the Stability of Matter 
from Atoms to Stars' in ed. W. Thirring, {\em The Stability of Matter: from 
Atoms to Stars. Selecta of Elliott H. Lieb}, Springer-Verlag.\\ 
P. Mainwood (2003), {\em Properties, Permutations and Physics}, B.Phil 
thesis, University of Oxford; available at: 
http://users.ox.ac.uk/~univ0767/physics/PPP.pdf\\
P. Mainwood (2004), "Thought Experiments in Galileo and Newton's 
Experimental Philosophy", available at: 
http://users.ox.ac.uk/~univ0767/history/ThoughtExperiments.pdf \\
D. Malament (2002), `A No-Go theorem about rotation in relativity theory', 
in D. Malament ed. {\em Reading Natural Philosophy} Chicago: Open Court; p. 
267-293. \\
D. Malament (2003), `On relative orbital rotation in relativity theory', in  
A. Ashtekar, R.Cohen et al. eds. {\em Revisiting the Foundations of 
Relativistic Physics: Festchrift for John Stachel}, Kluwer Academic; pp. 
175-190.\\ 
C. Misner et al. (1973), {\em Gravitation}, San Francisco: Freeman.\\
J. Norton (2003), `Causation as folk science', {\em Philosophers' Imprint}, 
an e-journal at http://www.philosophersimprint.org/003004/; also at: 
http://philsci-archive.pitt.edu/archive/00001214/\\
G. Oppy (2000), `{\em Humean} supervenience?', {\em Philosophical Studies} 
{\bf 101} pp. 77-105.\\
O. Pooley and H. Brown (2002), `Relationism rehabilitated? I: Classical 
mechanics', {\em British Journal for the Philosophy of Science} {\bf 53} pp. 
183-204.\\ 
D. Robinson (1989), `Matter, Motion and Humean supervenience', {\em 
Australasian Journal of Philosophy} {\bf 67},  pp. 394-409.\\
F. Rohrlich (2000), `Causality and the arrow of classical time', {\em 
Studies in the History and Philosophy of Modern Physics} {\bf 31B}, pp. 
1-13.\\
R. Rynasiewicz (1995), `By their Properties, Causes and Effects: Newton's 
Scholium on Time, Space, Place and Motion; Part I---the Text; Part II---The 
Context', {\em Studies in the History and Philosophy of Science} {\bf 26}, 
pp. 133-154 and 295-321.\\
M. Schlosshauer (2003), `Decoherence, the measurement problem, and 
interpretations of quantum mechanics', available at the Los Alamos physics 
arXive: quant-ph/0312059.\\
S. Shoemaker (1979), `Identity, properties and cauality', {\em Midwest 
Studies in Philosophy} {\bf 4}, pp. 321-342.\\
T. Sider (2001), {\em Four-Dimensionalism}, Oxford University Press.\\
L. Sklar (1974), {\em Space, Time and Spacetime}, University of California 
Press.\\
S. Smith (2003), `Are instantaneous velocities real and really 
instantaneous?', {\em Studies in the History and Philosophy of Modern 
Physics} {\bf 34B}, pp. 261-280.\\ 
P. Teller (2002), `The rotating discs argument and Humean supervenience: 
cutting the Gordian knot', {\em Analysis} {\bf 62}, pp. 205-210.\\
M. Tooley (1988), `In Defence of the Existence of States of Motion', {\em 
Philosophical Topics} {\bf 16}, pp. 225-254.\\
R. Torretti (1999), {\em The Philosophy of Physics}, Cambridge University 
Press.\\
C. Truesdell (1991), {\em A First Course in Rational Continuum Mechanics}, 
volume 1; second edition; Academic Press. \\
D. Wallace (2003), `Everett and structure', {\em Studies in the History and 
Philosophy of Modern Physics} {\bf 34B}, 87-105.\\
D. Wallace (2004), `In defence of naivete: The conceptual status of 
Lagrangian quantum field theory', forthcoming in {\em Synthese};\\
available at: http://philsci-archive.pitt.edu/archive/00000519.\\
R. Wald (1984), {\em General Relativity}, Chicago: University Press.\\
B. Weatherson (2002), `Intrinsic vs extrinsic properties', {\em The Stanford 
Encyclopedia of Philosophy}.\\
M. Wilson (1997), `Reflections on Strings',  in ed.s T. Horowitz and G. 
Massey, {\em Thought Experiments in Science and Philosophy},  University of 
Pittsburgh Press, pp. 193-207. \\
M. Wilson (2000), `The Unreasonable Uncooperativeness of Mathematics in the 
Natural Sciences', {\em The Monist} {\bf 83}, pp 296-314.\\
D. Zimmerman (1995), `Theories of masses and problems of constitution', {\em 
Philosophical Review} {\bf 104}, pp. 53-110.\\
D.Zimmerman (1997), `Immanent causation', {\em Philosophical Perspectives} 
{\bf 11}. pp. 433-471. \\
D. Zimmerman (1998), `Temporal parts and supervenient causation: the 
incompatibility of two Humean doctrines', {\em Australasian Journal of 
Philosophy} {\bf 76}, pp. 265-288.\\
D. Zimmerman (1999), `One really big liquid sphere: reply to Lewis', {\em 
Australasian Journal of Philosophy} {\bf 77}, pp. 213-215.

\end{document}